\documentclass[aps,pra,10pt,a4paper,superscriptaddress,onecolumn,preprintnumbers,floatfix,nofootinbib, showkeys]{revtex4-2}

\usepackage{graphicx}
\usepackage[utf8]{inputenc}
\usepackage{tocloft}
\usepackage{epsf}
\usepackage{bm}
\usepackage{amsmath}
\usepackage{amsfonts}
\usepackage{amssymb}
\usepackage{color}
\usepackage{caption}
\usepackage{subcaption}
\usepackage{booktabs} 
\usepackage{float}
\usepackage{empheq}

\usepackage[citecolor=blue,urlcolor=blue,unicode=true,pdfusetitle, bookmarks=true,bookmarksnumbered=true,bookmarksopen=true,bookmarksopenlevel=3, breaklinks=false,pdfborder={0 0 1},backref=false,colorlinks=true, linkcolor=blue]{hyperref}

\numberwithin{equation}{section}

\makeatletter

\renewcommand*{\p@subsection}{}

\renewcommand*{\p@subsubsection}{}

\renewcommand*{\p@paragraph}{}
\makeatother

\providecommand{\U}[1]{\protect\rule{.1in}{.1in}}

\newcommand{\be}{\begin{equation}}
\newcommand{\ee}{\end{equation}}

\newcommand{\mincir}{\raise
-3.truept\hbox{\rlap{\hbox{$\sim$}}\raise4.truept\hbox{$<$}\ }}
\newcommand{\magcir}{\raise
-3.truept\hbox{\rlap{\hbox{$\sim$}}\raise4.truept\hbox{$>$}\ }}

\begin{document}
\title{II. Non-Linear Interacting Dark Energy: Analytical Solutions and Theoretical Pathologies}
\author{Marcel van der Westhuizen}
\email{marcelvdw007@gmail.com}
\affiliation{Centre for Space Research, North-West University, Potchefstroom 2520, South
Africa}
\author{Amare Abebe}
\email{amare.abebe@nithecs.ac.za}
\affiliation{Centre for Space Research, North-West University, Potchefstroom 2520, South
Africa}
\affiliation{National Institute for Theoretical and Computational Sciences (NITheCS),
South Africa}
\author{Eleonora Di Valentino}
\email{e.divalentino@sheffield.ac.uk}
\affiliation{School of Mathematical and Physical Sciences, University of Sheffield, Hounsfield Road, Sheffield S3 7RH, United Kingdom}

\begin{abstract}
We investigate interacting dark energy (IDE) models with phenomenological, non-linear interaction kernels $Q$, specifically $Q_{1}=3H\delta \left(\frac{\rho_{\rm dm}\rho_{\rm de}}{\rho_{\rm dm}+\rho_{\rm de}}\right)$, $Q_{2}=3H\delta \left(\frac{\rho_{\rm dm}^2}{\rho_{\rm dm}+\rho_{\rm de}}\right)$, and $Q_{3}=3H\delta \left(\frac{\rho_{\rm de}^2}{\rho_{\rm dm}+\rho_{\rm de}}\right)$. Using dynamical system techniques developed in our companion paper on linear kernels, we derive new conditions that ensure positive and well-defined energy densities, as well as criteria to avoid future big rip singularities. We find that for $Q_{1}$, all densities remain positive, while for $Q_{2}$ and $Q_{3}$ negative values of either DM or DE are unavoidable if energy flows from DM to DE. We also show that for $Q_{1}$ and $Q_{2}$ a big rip singularity always arises in the phantom regime $w<-1$, whereas for $Q_{3}$ this fate may be avoided if energy flows from DE to DM. In addition, we provide new exact analytical solutions for $\rho_{\rm dm}$ and $\rho_{\rm de}$ in the cases of $Q_{2}$ and $Q_{3}$, and obtain new expressions for the effective equations of state of DM, DE, the total fluid, and the reconstructed dynamical DE equation of state ($w_{\rm dm}^{\rm eff}$, $w_{\rm de}^{\rm eff}$, $w_{\rm tot}^{\rm eff}$, and $\tilde{w}$). Using these results, we discuss phantom crossings, evaluate how each kernel addresses the coincidence problem, and apply statefinder diagnostics to compare the models. These findings extend the theoretical understanding of non-linear IDE models and provide analytical tools for future observational constraints.
\end{abstract}
\keywords{Cosmology; Interacting Dark Energy; Analytical solutions; Negative Energy; Big Rip}\date{\today}
\maketitle
\date{\today }

\section{Introduction}
In our companion paper titled "I. Linear Interacting Dark Energy: Analytical Solutions and Theoretical Pathologies" \cite{vanderWesthuizen:2025I}, we provided an overview of why interacting dark energy (IDE) models, where non-gravitational energy exchange occurs between dark matter (DM) and dark energy (DE), are relevant to modern cosmology. Briefly, IDE models are relevant as candidates to resolve long-standing issues such as the coincidence problem~\cite{Amendola_2000, Zimdahl_2001, Chimento_2003,  Farrar_2004, Wang_2004, Olivares_2006, Sadjadi_2006,  Quartin_2008,Campo_2009, Caldera_Cabral_2009_DSA, He_2011, delcampo2015interactiondarksector, von_Marttens_2019}, the more recent $H_0$ and $S_8$ tensions~\cite{Kumar:2016zpg, Murgia:2016ccp, Kumar:2017dnp, DiValentino:2017iww, Kumar:2021eev, Pan:2023mie, Benisty:2024lmj, Yang:2020uga, Forconi:2023hsj, Pourtsidou:2016ico, DiValentino:2020vnx, DiValentino:2020leo, Nunes:2021zzi, Yang:2018uae, vonMarttens:2019ixw, Lucca:2020zjb, Zhai:2023yny, Bernui:2023byc, Hoerning:2023hks, Giare:2024ytc, Escamilla:2023shf, vanderWesthuizen:2023hcl, Silva:2024ift, DiValentino:2019ffd, Li:2024qso, Pooya:2024wsq, Halder:2024uao, Castello:2023zjr, Yao:2023jau, Mishra:2023ueo, Nunes:2016dlj, Silva:2025hxw, Zheng_2017, Kumar_2019, Anchordoqui_2021, Pan_2019, Guo_2021, Di_Valentino_2021_H0_review, Gariazzo_2022, Wang_2022, Califano_2023, Pan_2024, Liu_2023, Liu_2024, Sabogal_2025, Kumar_2017, Kumar_2019, Di_Valentino_2020_rhode, Anchordoqui_2021, Kumar_2021, Gariazzo_2022, lucca2021darkenergydarkmatterinteractions, Sabogal_2024, Liu_2023, Liu_2024, Sabogal_2025, yang2025probingcoldnaturedark}, and most recently as a mechanism for the possible dynamical nature of DE suggested by the DESI collaboration~\cite{DESI:2025fii} (see also~\cite{DESI:2024mwx,Cortes:2024lgw,Shlivko:2024llw,Luongo:2024fww,Yin:2024hba,Gialamas:2024lyw,Dinda:2024kjf,Najafi:2024qzm,Wang:2024dka,Ye:2024ywg,Tada:2024znt,Carloni:2024zpl,Chan-GyungPark:2024mlx,DESI:2024kob,Ramadan:2024kmn,Notari:2024rti,Orchard:2024bve,Hernandez-Almada:2024ost,Pourojaghi:2024tmw,Giare:2024gpk,Reboucas:2024smm,Giare:2024ocw,Chan-GyungPark:2024brx,Menci:2024hop,Li:2024qus,Li:2024hrv,Notari:2024zmi,Gao:2024ily,Fikri:2024klc,Jiang:2024xnu,Zheng:2024qzi,Gomez-Valent:2024ejh,RoyChoudhury:2024wri,Lewis:2024cqj,Wolf:2025jlc,Shajib:2025tpd,Giare:2025pzu,Chaussidon:2025npr,Kessler:2025kju,Pang:2025lvh,RoyChoudhury:2025dhe,Scherer:2025esj,Teixeira:2025czm,Specogna:2025guo,Cheng:2025lod,Cheng:2025hug,Ozulker:2025ehg,Lee:2025pzo}). We also presented an analysis of the causes and conditions to avoid two of the most often neglected features that plague IDE models, namely negative energy densities and future big rip singularities. This analysis was performed for the most common phenomenological interaction kernels $Q$, where the interaction that determines the energy exchange is proportional in some linear way to the DM density $Q\propto \rho_{\rm{dm}}$, the DE density $Q\propto\rho_{\rm{de}}$, or some linear combination of the dark components given by the general function $Q=3 H (\delta_{\text{dm}} \rho_{\text{dm}} + \delta_{\text{de}}  \rho_{\text{de}})$, where $\delta_{\text{dm}}$ and $\delta_{\text{de}}$ are constants that determine the dependence of the coupling on DM and DE, respectively.

\begin{figure}
    \centering
    \includegraphics[width=0.9 \linewidth]{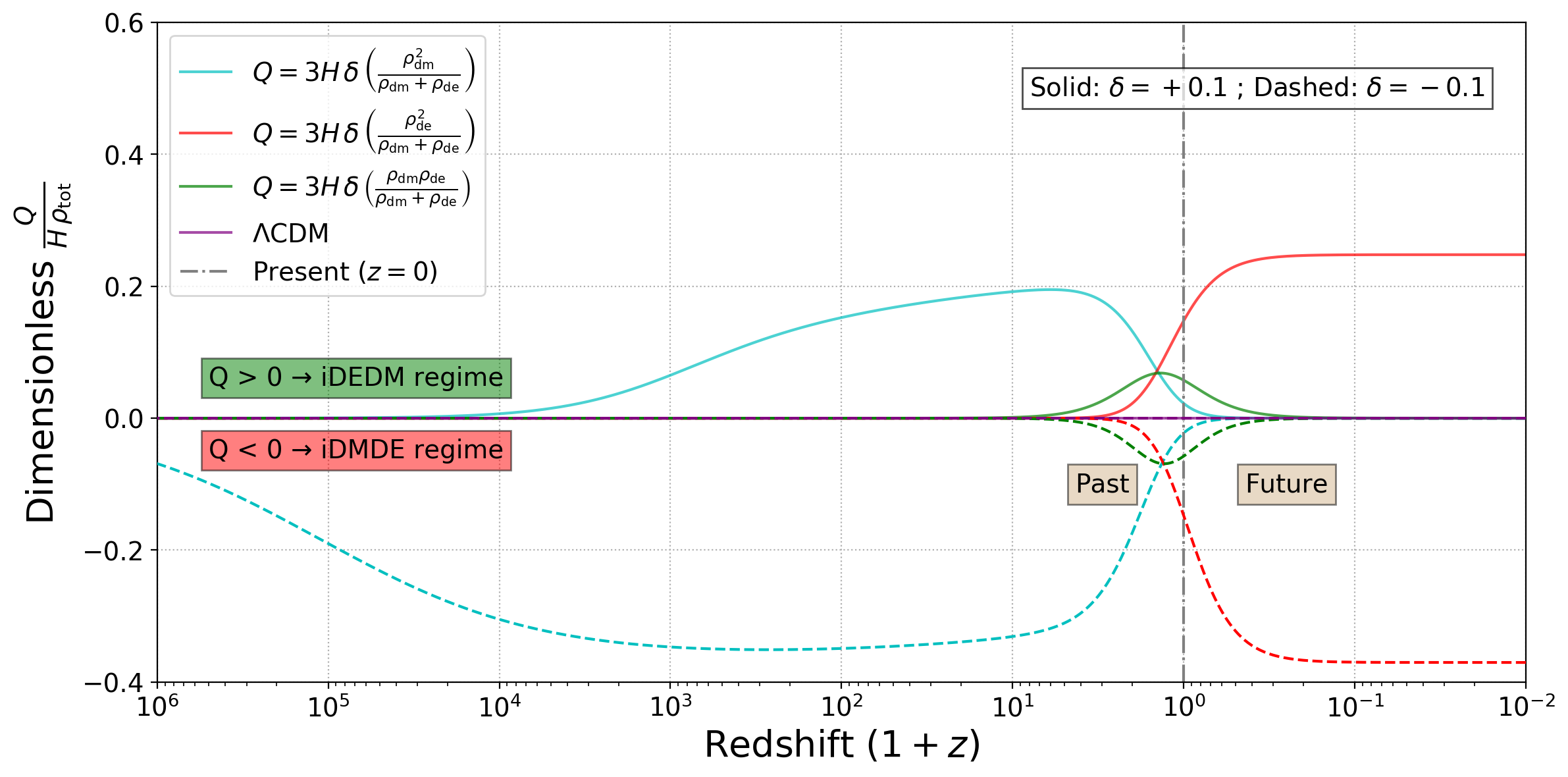}
    \caption{Dimensionless interaction $Q$ versus redshift for three non-linear interaction models, illustrating when the interaction becomes dominant.}
    \label{fig:Q_nonLinear}
\end{figure}

Even though these linear IDE models are the most widely studied, it may be argued that it is more natural for the interaction to be non-linear, and instead proportional to the product of the dark components, such that $Q\propto\rho_{\text{dm}}\rho_{\text{de}}$. The logic behind this is that the interaction rate increases with each of the densities and vanishes if either of the densities is zero ($Q=0$ if $\rho_{\text{dm}}=0$ or $\rho_{\text{de}}=0$), thus preventing energies from crossing the zero boundary and becoming negative, as already noted in Table 1 of our companion paper \cite{vanderWesthuizen:2025I}. This form of interaction is natural and has been used to model both two-body chemical reactions and biological predator-prey systems~\cite{Lip:2010dr, Arevalo:2011hh}. This behavior is shown by the green line in Figure~\ref{fig:Q_nonLinear}, along with two other non-linear interactions that were also studied in~\cite{Arevalo:2011hh}, which served as a foundation for this study. The other two interaction kernels, where $Q\propto\rho_{\rm{dm}}^2$ and $Q\propto\rho_{\rm{de}}^2$, have not been studied extensively, as seen in the brevity of the literature review in Section~\ref{lit}. These two interactions serve as interesting alternatives to the common kernels $Q=3H\delta\rho_{\rm{dm}}$ and $Q=3H\delta\rho_{\rm{de}}$, as they maintain many of the same features while showing subtle differences.

In this paper, we aim to extend the analysis provided in~\cite{Arevalo:2011hh} using the techniques established in our companion paper \cite{vanderWesthuizen:2025I}, while also providing a clear comparison of the differences between linear and non-linear interaction kernels. The results from both papers are summarised in \cite{ vanderWesthuizen:2025III}. The structure of this paper is as follows. 
\begin{itemize}
    \item In Section~\ref{BG_lit}, some background on IDE cosmology is presented. We provide the background equations that will be used throughout this analysis in subsection~\ref{BG_EQ}. Subsection~\ref{lit} presents a summary of previous literature on specific aspects of each of the three non-linear interaction kernels considered in this study.
    \item In Section~\ref{Sec.DSA}, we use dynamical system techniques to study the asymptotic behavior of our models. For all three interactions, we find the critical points and their corresponding eigenvalues to determine the stability of these points. The dynamical system analysis of $Q_{1}=3H \delta \left(\frac{\rho_{\rm{dm}}\rho_{\rm{de}}}{\rho_{\rm{dm}}+\rho_{\rm{de}}} \right)$, $Q_2=3H \delta \left(\frac{\rho_{\rm{dm}}^2}{\rho_{\rm{dm}}+\rho_{\rm{de}}} \right)$ and $Q_3=3H \delta \left(\frac{\rho_{\rm{de}}^2}{\rho_{\rm{dm}}+\rho_{\rm{de}}} \right)$ are presented in subsections~\ref{DSA.Q.dmde},~\ref{DSA.Q.dmdm} and~\ref{DSA.Q.dede}, respectively. We obtain both 3D (Figures~\ref{fig:3D_QNLdmde_phase_portraits},~\ref{fig:3D_QNLdm_phase_portraits} and~\ref{fig:3D_QNLde_phase_portraits}) and 2D phase portraits (Figures~\ref{fig:2D_QNLdm_phase_portraits} and~\ref{fig:2D_QNLde_phase_portraits}), which we use to derive the new conditions listed in Table~\ref{tab:QNLdmde_energy_conditions},~\ref{tab:QNLdm_energy_conditions} and~\ref{tab:QNLde_energy_conditions}, ensuring positive energy densities at all times. The behavior of the background equations at each critical point is provided in Table~\ref{tab:CP_B_QNL_dmde},~\ref{tab:CP_B_QNL_dm} and~\ref{tab:CP_B_QNL_de}, while the stability of the system, described by the doom factor $\textbf{d}$ \eqref{DSA.doom}, is given in Table~\ref{tab:QNLdmde_stability_criteria},~\ref{tab:QNLdm_stability_criteria} and~\ref{tab:QNLde_stability_criteria}, respectively.

    \item In Section~\ref{Background_cosmology}, for each of the three interactions we provide new analytical expressions for cosmological parameters, using the new solutions found for $\rho_{\rm{dm}}$ and $\rho_{\rm{de}}$, which are derived in Appendix~\ref{Appendix_A} and~\ref{Appendix_B}. The results obtained here are consistent with those from the dynamical system analysis in Section~\ref{Sec.DSA}, and in both cases, when $\delta=0$ and $w=-1$, the relevant expressions for the $\Lambda$CDM model are recovered, thus validating both sections. A summary of the derived expressions for each interaction is provided below.
    \begin{itemize}
        \item Section~\ref{analytical.dmde} $Q_{1}=3H \delta \left(\frac{\rho_{\rm{dm}}\rho_{\rm{de}}}{\rho_{\rm{dm}}+\rho_{\rm{de}}} \right)$: The DM and DE densities $\rho_{\text{dm}}$ and  $\rho_{\text{de}}$ \eqref{NLID1_dm_de_BG} were originally provided by~\cite{Arevalo:2011hh}, and their evolution is shown in Figure~\ref{fig:Omega_NLID1}. We find new solutions for the DM-DE equality $z _{\text{(dm=de)}}$ \eqref{NLID1_dm=de_BG}, the DM and DE effective equations of state $w^{\rm{eff}}_{\rm{dm}}$ and $w^{\rm{eff}}_{\rm{de}}$ \eqref{NLID1_omega_eff_dm_de_BG}, the coincidence problem $\zeta$ \eqref{eq:omega_eff_dm_de_NLID1_subbed_past_future} (illustrated in Figure~\ref{fig:CP+omega_dmde_NLID1}), the redshift of the phantom crossing $z_{\text{pc}}$ \eqref{NLID1_phantom_crossing_BG} and its direction \eqref{eq:z_pc_dmde_direction}, the condition that leads to a big rip \eqref{omega_eff_tot_NLID1_BG}, and the time of the big rip  $t_{\text{rip}}$ \eqref{eq:Big_Rip_general_NLID1_BG} (illustrated in Figure~\ref{fig:eos_tot_BR_NLID1}).
        
        \item Section~\ref{analytical.dmdm} $Q_{2}=3H \delta \left(\frac{\rho_{\rm{dm}}^2}{\rho_{\rm{dm}}+\rho_{\rm{de}}} \right)$:  $\rho_{\text{dm}}$ and  $\rho_{\text{de}}$ \eqref{NLID2_dm_de_BG} are new expressions derived for this model, with their evolution shown in Figure~\ref{fig:Omega_NLID1}, alongside the redshift at which DE becomes negative $z_{\text{(de=0)}}$ \eqref{NLID2_de=0_z_BG}. We also find new solutions for $z _{\text {(dm=de)}}$ \eqref{NLID2_dm=de_BG}, $w^{\rm{eff}}_{\rm{dm}}$ and $w^{\rm{eff}}_{\rm{de}}$ \eqref{NLID2_omega_eff_dm_de_BG}, $\zeta$ \eqref{eq:omega_eff_dm_de_NLID2_subbed_past_future} (illustrated in Figure~\ref{fig:CP+omega_dmde_NLID2}), $z_{\text{pc}}$ \eqref{NLID2_phantom_crossing_BG} and its direction \eqref{eq:z_pc_dmdm_direction}, the big rip condition \eqref{omega_eff_tot_NLID2_BG}, and $t_{\text{rip}}$ \eqref{eq:Big_Rip_general_NLID2_BG} (illustrated in Figure~\ref{fig:eos_tot_BR_NLID2}).

        \item Section~\ref{analytical.dede} $Q_{3}=3H \delta \left(\frac{\rho_{\rm{de}}^2}{\rho_{\rm{dm}}+\rho_{\rm{de}}} \right)$:  $\rho_{\text{dm}}$ and  $\rho_{\text{de}}$ \eqref{NLID3_dm_de_BG} are new expressions derived for this model, with their evolution shown in Figure~\ref{fig:Omega_NLID3}, alongside the redshift at which DM becomes negative $z_{\text{(dm=0)}}$ \eqref{NLID3_dm=0_z_BG}. We also find new solutions for $z _{\text{(dm=de)}}$ \eqref{NLID3_dm=de_BG}, $w^{\rm{eff}}_{\rm{dm}}$ and $w^{\rm{eff}}_{\rm{de}}$ \eqref{NLID3_omega_eff_dm_de_BG}, $\zeta$ \eqref{eq:omega_eff_dm_de_NLID3_subbed_past_future} (illustrated in Figure~\ref{fig:CP+omega_dmde_NLID3}), $z_{\text{pc}}$ \eqref{NLID3_phantom_crossing_3} and its direction \eqref{eq:z_pc_dede_direction}, the big rip condition \eqref{omega_eff_tot_BLID3_BG}, and $t_{\text{rip}}$ \eqref{eq:Big_Rip_NLID3_BG} (illustrated in Figure~\ref{fig:eos_tot_BR_NLID3}).
        
    \end{itemize}

    \item In Section~\ref{reconstructed_w}, we show how IDE models can be parameterized as a dynamical DE model, using the reconstructed dynamical DE equation of state $\tilde{w}(z)$. In Appendix~\ref{Appendix_C}, we derive the new simple expression \eqref{wz_general} for $\tilde{w}(z)$, which holds for any IDE model. This is used to obtain $\tilde{w}(z)$ for the three non-linear interactions, given in \eqref{wz_Q_dmde_1}, \eqref{wz_Q_dmdm_1} and \eqref{wz_Q_dede_1}, and which are also plotted alongside  $w^{\rm eff}_{\rm de}$, $w^{\rm eff}_{\rm dm}$, $w^{\rm eff}_{\rm tot}$, and $w$ in Figure~\ref{fig:w_all_Qdmde},~\ref{fig:w_all_Qdmdm} and~\ref{fig:w_all_Qdede}.
    
    \item In Section~\ref{statefinder}, we consider the statefinder diagnostics to help differentiate between the three non-linear interactions studied in Sections~\ref{Sec.DSA} and~\ref{Background_cosmology}. New expressions for the past and future asymptotic behavior of the statefinder diagnostics $r$ and $s$ are given in Table~\ref{tab:Statefinder_future_nonlinear}, and the evolution of these parameters is illustrated in Figure~\ref{fig:qr+sr_NLID}.
    We compare our results with those in the literature and show how these interactions relate to the $\Lambda$CDM model and other DE candidates at early and late times.
    
    \item Finally, in Section~\ref{summary}, we summarize all our main results in a few tables. We provide conditions to avoid parameter-space regions where these solutions are undefined or imaginary (Table~\ref{tab:Com_real}), lead to negative DM or DE densities (Table~\ref{tab:Com_PEC}), or predict future big rip singularities (Table~\ref{tab:Com_AE_BR}). We also show how each model addresses the coincidence problem in both the past and future (Table~\ref{tab:Com_CP}). We then draw conclusions from the results obtained and discuss directions for further research.
\end{itemize}

\section{Background on IDE models} \label{BG_lit}

\subsection{Background equations} \label{BG_EQ}

The background equations used to describe cosmological expansion and the change in the scale factor $a$ throughout this study are the same as those outlined in our companion paper \cite{vanderWesthuizen:2025I}, and are listed below:
\begin{gather} \label{DSA.H}
\begin{split}
H^2&= \left(\frac{\dot{a}}{a} \right)^2 = \frac{8\pi G}{3 } \left(\rho_{\text{r}}+\rho_{\text{bm}}+\rho_{\text{dm}}+\rho_{\text{de}}\right), \\
q &= \left(-\frac{\ddot{a}a}{\dot{a}^2} \right)=    \Omega_{\rm{r}}  + \frac{1}{2} \left( \Omega_{\rm{bm}}+ \Omega_{\rm{dm}} \right)  + \frac{1}{2} \Omega_{\rm{de}}  \left(1 +3 w \right) =   \frac{1}{2}\left(1+3\Omega_{\rm{de}}  w \right),  \\
r_\text{sf}&= \left(\frac{\dddot{a}}{aH^3} \right)=1+\frac{9}{2} \Omega_\text{de} w \left( 1+w^{\rm{eff}}_{\rm{de}}\right), \\
s_\text{sf}&=\frac{r_\text{sf}-1}{3\left(q-\frac{1}{2} \right)}= 1+w^{\rm{eff}}_{\rm{de}}.
\end{split}
\end{gather}

In \eqref{DSA.H}, $H$ is the Hubble parameter (with $\rho_{\text{r}}=\rho_{\text{(r,0)}}a^{-4}$ and $\rho_{\text{bm}}=\rho_{\text{(bm,0)}}a^{-3}$); $q$ is the deceleration parameter; and $r_\text{sf}$ and $s_\text{sf}$ are the statefinder parameters introduced in~\cite{Sahni_2003, Alam_2003}, with their final forms provided in~\cite{Zhang_2006}. To understand how the evolution of DM and DE is affected by their interaction, we may use either of the modified conservation equations below:
\begin{gather} \label{eq:conservation}
\begin{split}
\dot{\rho}_{\text{dm}} + 3H \rho_{\text{dm}} = Q \quad &; \quad  \dot{\rho}_{\text{de}} + 3H (1 + w) \rho_{\text{de}} = -Q,\\
\dot{\rho}_{\text{dm}} + 3H \rho_{\text{dm}}(1 + w_{\rm{dm}}^{\rm{eff}}) = 0 \quad &; \quad  \dot{\rho}_{\text{de}} + 3H (1 + w_{\rm{de}}^{\rm{eff}}) \rho_{\text{de}} = 0.
\end{split}
\end{gather}
In \eqref{eq:conservation}, $Q$ is the interaction function whose sign determines the direction of energy transfer, such that energy flows from DE to DM if $Q>0$ (iDEDM regime), while energy flows from DM to DE if $Q<0$ (iDMDE regime). The effect of the interaction can also be encapsulated using effective equations of state, which provide an equivalent description of fluids without an interaction but with dynamical equations of state. The conservation equations \eqref{eq:conservation} may be rewritten as:
\begin{gather} \label{DSA.H2}
\begin{split}
w^{\rm{eff}}_{\rm{dm}} &= - \frac{Q}{3 H \rho_{\rm{dm}}} \quad ; \quad w^{\rm{eff}}_{\rm{de}} =  w_{\rm{de}} + \frac{Q}{3 H \rho_{\rm{de}}}.
\end{split}
\end{gather}
The effective equations of state determine how the DM and DE fluids dilute with redshift, providing insight into how these models address the coincidence problem. Since the ratio of DM to DE is given by $r= \frac{ {\rho}_{\rm{dm}}}{{\rho}_{\rm{de}}}= r_0 a^{-3(w^{\text{eff}}_{\text{dm}} - w^{\text{eff}}_{\text{de}})}$, we introduce the parameter $\zeta$, which measures how much the coincidence problem is alleviated relative to the $\Lambda$CDM model, where $\zeta_{\Lambda \text{CDM}} = 3$:
\begin{gather} \label{DSA.CP}
\begin{split}
 \zeta_{\rm{IDE}}=3(w^{\text{eff}}_{\text{dm}} - w^{\text{eff}}_{\text{de}}) &\to 
\begin{cases}
|\zeta_{\rm{IDE}}| > 3 & \textit{worsens the coincidence problem}, \\
|\zeta_{\rm{IDE}} |< 3 & \textit{alleviates the coincidence problem}, \\ 
\zeta _{\rm{IDE}}= 0 & \textit{solves the coincidence problem}.
\end{cases} \;
\end{split}
\end{gather}
It is also useful at times to model the entire cosmological system as a single fluid with a dynamical total effective equation of state $w^{\rm{eff}}_{\rm{tot}}$ given by:
\begin{gather} \label{DSA.omega_eff_tot}
\begin{split}
w^{\rm{eff}}_{\rm{tot}} = \frac{P_{\rm{tot}}}{\rho_{\rm{tot}}} = \frac{w_{\rm{r}} \Omega_{\rm{r}}+ w_{\rm{bm}} \Omega_{\rm{bm}}+ w_{\rm{dm}} \Omega_{\rm{dm}} + w_{\rm{de}}\Omega_{\rm{de}}}{\Omega_{\rm{r}}+\Omega_{\rm{bm}}+\Omega_{\rm{dm}}+\Omega_{\rm{de}}}  = \frac{1}{3}\Omega_{\rm{r}} + w\Omega_{\rm{de}}.
\end{split}
\end{gather}
If $w^{\rm{eff}}_{\rm{tot}}<-\frac{1}{3}$, the model will experience accelerated expansion. Additionally, if the asymptotic future is characterized by $w^{\rm{eff}}_{\rm{tot}}<-1$, the universe will encounter a future big rip singularity, as detailed in Section 2.2 of our companion paper \cite{vanderWesthuizen:2025I}. The stability of IDE models is usually determined by requiring a negative sign for the doom factor $\textbf{d}$~\cite{M.B.Gavela_2009}, given for any interaction $Q$ as: 
\begin{gather} \label{DSA.doom}
\begin{split}
\textbf{d}=  \frac{Q}{3H\rho_{\rm{de}}(1+w)}.
\end{split}
\end{gather}
For further discussion on the viability of \eqref{DSA.doom} and alternative approaches, see Section 2.3 of our companion paper on linear IDE models \cite{vanderWesthuizen:2025I}. As in our previous work, we plot all figures using the parameters $H_0=67.4$ km/s/Mpc, $\Omega_{\rm{(r,0)}}=9\times10^{-5}$, $\Omega_{\rm{(bm,0)}}=0.049$, $\Omega_{\rm{(dm,0)}}=0.266$, $\Omega_{\rm{(de,0)}}=0.685$, $w=-1$, and $\delta=\pm0.1$, unless otherwise stated.

\subsection{Literature on each interaction} \label{lit}

The three non-linear interaction kernels that we study have each been investigated to different extents in the literature, though important gaps remain that this work aims to address. In the following, we provide an overview of some of the main contributions associated with each kernel. As in our companion paper, these groupings are not meant to be exhaustive or mutually exclusive, since many works overlap across kernels and often address multiple theoretical or observational aspects simultaneously. Rather, the list below is intended as a guide for researchers interested in specific features of the three non-linear interaction models considered in this study.

\begin{itemize}

\item \textbf{Literature on non-linear IDE model 1:} $\boldsymbol{Q_1=3H\delta \left( \frac{\rho_{\text{dm}}\rho_{\text{de}} } {\rho_{\text{dm}}+\rho_{\text{de}}} \right)}$ —  
Analytical solutions~\cite{Arevalo:2011hh, Chimento_2012, Li_2014, Bolotin_2015}, background cosmology~\cite{Arevalo:2011hh, von_Marttens_2019}, large-scale structure and instabilities~\cite{Li_2014, Li_2023}, dynamical system analysis~\cite{Lip:2010dr, Arevalo:2011hh, paliathanasis2024compartmentalizationcoexistencedarksector}, observational constraints~\cite{He_2008, von_Marttens_2019, Arevalo:2011hh, Aljaf_2021, vanderwesthuizen2025compartmentalizationdarksectoruniverse}, $H_0$ tension~\cite{Yang_2023}, $S_8$ tension~\cite{Yang_2023}, statefinder analysis~\cite{Zhang_2006, carrasco2023discriminatinginteractingdarkenergy}, Bayesian comparison~\cite{Arevalo_2017, Cid_2019}, field theory interpretation~\cite{Pan_2020_Field}, interacting vacuum model~\cite{Sebastianutti_2024}, holographic modelling~\cite{Ma_2010, mazumder2011interactingholographicdarkenergy, Feng_2016}, Chaplygin Gas modelling~\cite{Zhang_2006, Li_2014, Wang_2013}. 

\item \textbf{Literature on non-linear IDE model 2:} $\boldsymbol{Q_2=3H\delta \left( \frac{\rho^2_{\text{dm}} } {\rho_{\text{dm}}+\rho_{\text{de}}} \right)}$ —  
Analytical solutions~\cite{Arevalo:2011hh, Bolotin_2015}, background cosmology~\cite{Arevalo:2011hh, von_Marttens_2019}, statefinder analysis~\cite{carrasco2023discriminatinginteractingdarkenergy}, Bayesian comparison~\cite{Arevalo_2017, Cid_2019}, dynamical system analysis~\cite{Arevalo:2011hh, Rodriguez_Benites_2024}, observational constraints~\cite{von_Marttens_2019, Arevalo:2011hh, Aljaf_2021}. 

\item \textbf{Literature on non-linear IDE model 3:} $\boldsymbol{Q_3=3H\delta \left( \frac{\rho^2_{\text{de}} } {\rho_{\text{dm}}+\rho_{\text{de}}} \right)}$ —  
Analytical solutions~\cite{Arevalo:2011hh, Bolotin_2015}, background cosmology~\cite{von_Marttens_2019, Arevalo:2011hh}, dynamical system analysis~\cite{Arevalo:2011hh}, statefinder analysis~\cite{carrasco2023discriminatinginteractingdarkenergy}, Bayesian comparison~\cite{Arevalo_2017, Cid_2019}, observational constraints~\cite{von_Marttens_2019, Arevalo:2011hh, Aljaf_2021}.
\end{itemize}

\section{Dynamical system analysis} \label{Sec.DSA}

\subsection{Setting up the dynamical system}  \label{Setup.DSA}

For this analysis, we start with the same four-fluid dynamical system used in our companion paper for a general interaction $Q$ \cite{vanderWesthuizen:2025I}:
\begin{gather} \label{DSA.9}
\begin{split}
{\Omega}'_{\rm{de}} &= \Omega_{\rm{de}} \left[ 1-  \Omega_{\rm{bm}} -\Omega_{\rm{dm}} - \Omega_{\rm{de}}  \left(1 -3 w \right) - 3w \right]  - \frac{8 \pi G }{3H^3} Q, \\
{\Omega}'_{\rm{dm}} &= \Omega_{\rm{dm}} \left[ 1-  \Omega_{\rm{bm}} -\Omega_{\rm{dm}} - \Omega_{\rm{de}}  \left(1 -3 w \right)  \right]  + \frac{8 \pi G }{3H^3} Q, \\
{\Omega}'_{\rm{bm}} &= \Omega_{\rm{bm}} \left[ 1-  \Omega_{\rm{bm}} -\Omega_{\rm{dm}} - \Omega_{\rm{de}}  \left(1 -3 w \right)  \right].
\end{split}
\end{gather}
The dynamical system \eqref{DSA.9} was derived in~\cite{vanderWesthuizen:2023hcl}, and reduces to the $\Lambda$CDM case found in~\cite{GR_book} when $Q=0$ and $w=-1$.  
Note that under the flatness assumption the radiation density satisfies $\Omega_{\rm{r}}=1-\Omega_{\rm{bm}}-\Omega_{\rm{dm}} -\Omega_{\rm{de}}$. For this study, we adopt the same two guiding assumptions as in~\cite{V_liviita_2008}, namely that the correct dynamics of the universe do not deviate too drastically from the present description of the $\Lambda$CDM model.
\begin{gather} \label{DSA.10}
\begin{split}
w<0 \quad &\text{(DE has negative pressure)}, \\
\delta<|w| \quad &\text{(interaction strength is not too strong)}.
\end{split}
\end{gather}
Solutions to the dynamical system \eqref{DSA.9} are known as critical points and may be classified into three categories, depending on how trajectories converge or diverge around these points. If the system is perturbed at a critical point and the possible trajectories converge back to that point, then the critical point is classified as a stable node, sink, or future attractor. Conversely, if the trajectories diverge from this point, the critical point is classified as an unstable node, source, or past attractor. If some trajectories converge to the point while others diverge, the critical point is classified as a saddle point. These three types of critical points are determined mathematically by the sign of the eigenvalues $\lambda$ of the Jacobian matrix of the system: stable nodes have all negative eigenvalues $\lambda<0$; unstable sources have all positive eigenvalues $\lambda>0$; while saddle points have a combination of positive and negative eigenvalues.  
For further applications of dynamical system analysis to cosmology, see~\cite{Bahamonde_2018}.

To determine positive energy conditions, we require the following to hold in the 2D projection of the system in the $(\Omega_{\rm{dm}}, \Omega_{\rm{de}})$ plane:
\begin{enumerate}
         \item \underline{Positive critical points}: We require conditions that ensure positive coordinates for each critical point $(\Omega_{\rm{dm}}\ge0 \;, \;\Omega_{\rm{de}}\ge0)$. 
         \item \underline{Positive trajectories}: The phase portraits we obtain have a region bounded by three invariant submanifolds connecting critical points, where $(\Omega_{\rm{dm}}\ge0 \;, \;\Omega_{\rm{de}}\ge0)$ throughout the region. We require constraints on $(\Omega_{\rm{(dm,0)}}, \Omega_{\rm{(de,0)}})$ that ensure trajectories start and remain within this bounded region.
\end{enumerate}
To gain insight into the asymptotic behavior of the system, we also derive expressions for the equations given in Section~\ref{BG_EQ} at each critical point.

\subsection{Dynamical system analysis: Interaction function $Q_1=3H \delta \left(\frac{\rho_{\rm{dm}}\rho_{\rm{de}}}{\rho_{\rm{dm}}+\rho_{\rm{de}}} \right)$} \label{DSA.Q.dmde}

For the interaction $Q=3H \delta \left(\frac{\rho_{\rm{dm}}\rho_{\rm{de}}}{\rho_{\rm{dm}}+\rho_{\rm{de}}} \right)$, the dynamical system (\ref{DSA.9}) becomes:
\begin{gather} \label{DSA.QNLdmde.1}
\begin{split}
{\Omega}'_{\rm{de}} &= \Omega_{\rm{de}} \left[ 1-  \Omega_{\rm{bm}} -\Omega_{\rm{dm}} - \Omega_{\rm{de}}  \left(1 -3 w \right) - 3w \right]  - 3 \delta \left(\frac{\Omega_{\rm{dm}}\Omega_{\rm{de}}}{\Omega_{\rm{dm}} + \Omega_{\rm{de}}} \right), \\
{\Omega}'_{\rm{dm}} &= \Omega_{\rm{dm}} \left[ 1-  \Omega_{\rm{bm}} -\Omega_{\rm{dm}} - \Omega_{\rm{de}}  \left(1 -3 w \right)  \right]  + 3 \delta \left(\frac{\Omega_{\rm{dm}}\Omega_{\rm{de}}}{\Omega_{\rm{dm}} + \Omega_{\rm{de}}} \right), \\
{\Omega}'_{\rm{bm}} &= \Omega_{\rm{bm}} \left[ 1-  \Omega_{\rm{bm}} -\Omega_{\rm{dm}} - \Omega_{\rm{de}}  \left(1 -3 w \right)  \right], \\
\end{split}
\end{gather}
where we used the relation $ \frac{8 \pi G }{3H^2} \rho_i= \Omega_i$. In this case, since DM participates in the interaction while baryonic matter does not, the two fluids evolve differently and cannot be grouped together. From the dynamical system \eqref{DSA.QNLdmde.1} we find: \\ 

\begin{figure}[htbp]
    \centering
    \begin{subfigure}[b]{0.495\linewidth}
        \centering
        \includegraphics[width=\linewidth]{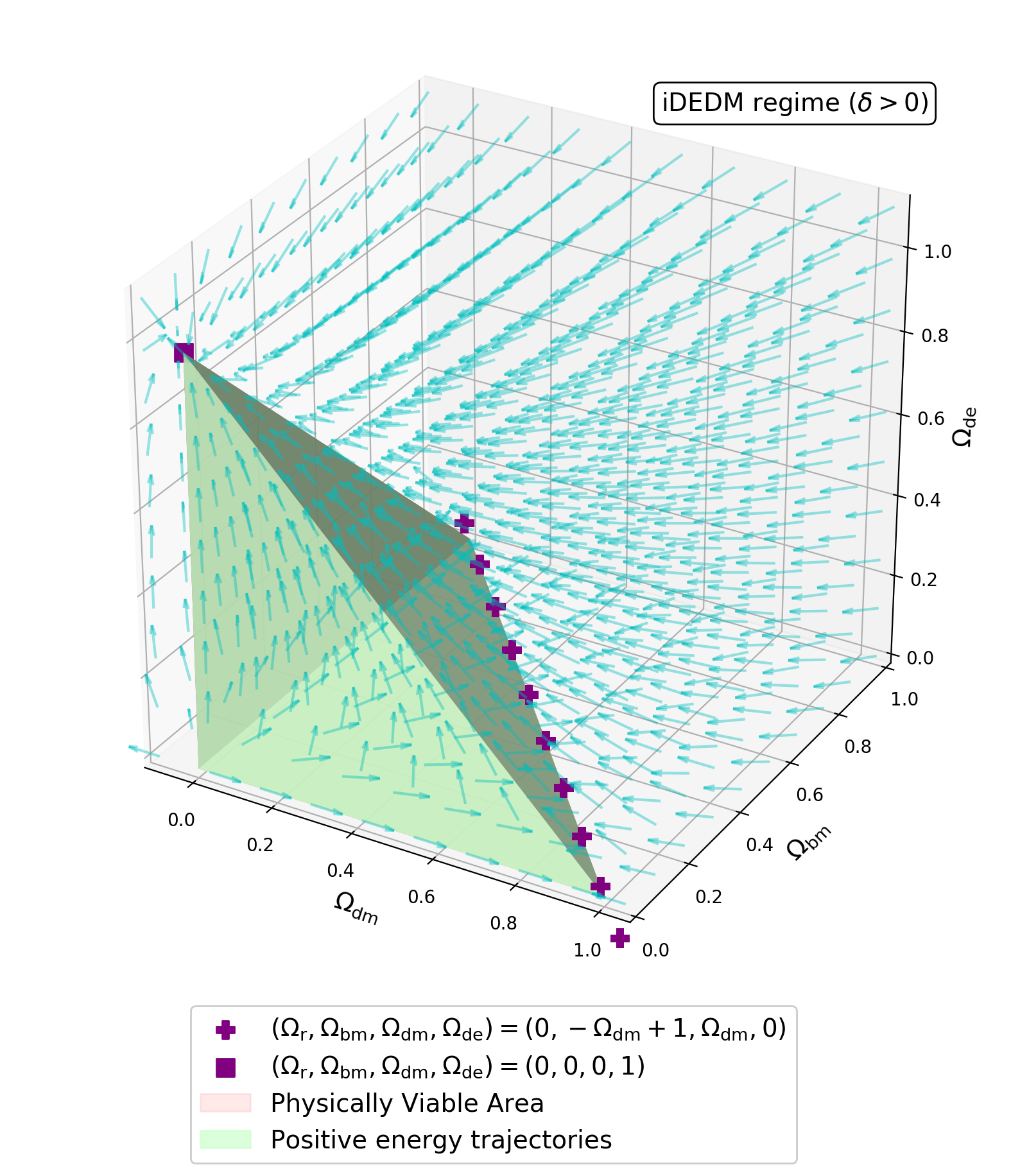}
    \end{subfigure}%
    \hspace{0pt} 
    \begin{subfigure}[b]{0.495\linewidth}
        \centering
        \includegraphics[width=\linewidth]{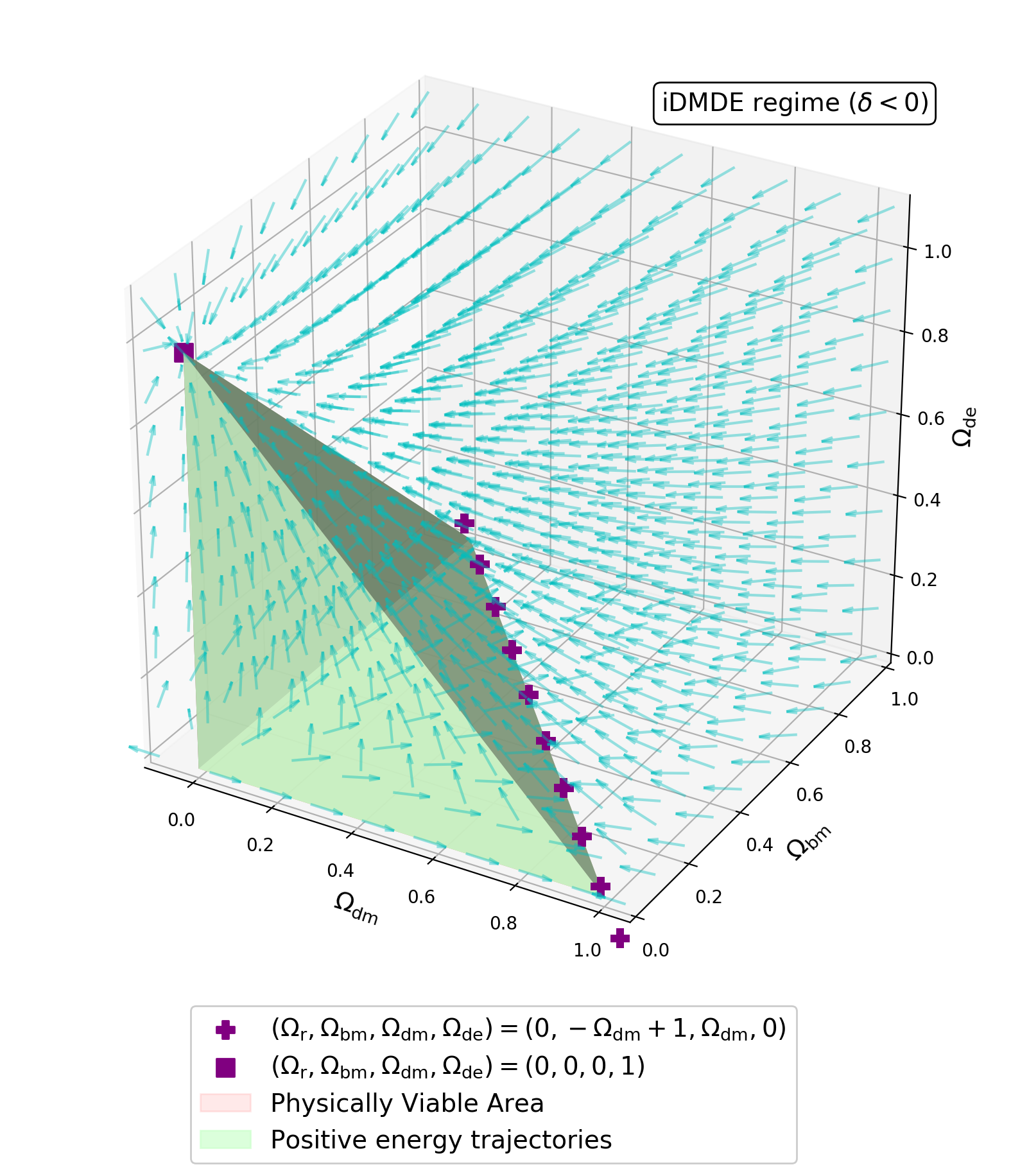}
    \end{subfigure}    
    \caption{3D phase portraits in the iDEDM (left panel, $\delta=+0.1$) and iDMDE (right panel, $\delta=-0.1$) regimes, showing positive energy trajectories at all times for $Q=3H \delta \left(\frac{\rho_{\rm{dm}}\rho_{\rm{de}}}{\rho_{\rm{dm}}+\rho_{\rm{de}}} \right)$.}
    \label{fig:3D_QNLdmde_phase_portraits}
\end{figure}

\underline{Critical Point $P_{\text{m}}$: matter-dominated phase.}
\begin{gather} \label{DSA.QNLdmde.2}
\Omega_{\rm{bm}} = -\Omega_{\rm{dm}}+1, \quad \Omega_{\rm{dm}} = \Omega_{\rm{dm}}, \quad \Omega_{\rm{de}} = 0, \quad \rightarrow \quad \Omega_{\rm{r}} = 0  \quad  \; ; \quad  \lambda = \begin{bmatrix}
0 \\
-1\\
-3 (w+\delta)
\end{bmatrix}.
\end{gather}
The coordinates in \eqref{DSA.QNLdmde.2} correspond to a combination of baryonic and DM domination, i.e. a \textit{matter-dominated phase}, as shown by $\Omega_{\rm{m}}=\Omega_{\rm{bm}}+\Omega_{\rm{dm}}=-\Omega_{\rm{dm}}+1+\Omega_{\rm{dm}}=1$. This means that in the past, DM and baryonic matter behave equivalently and may be grouped together. This critical point is not a single point but rather a line along the axis where the sum of the two components equals one, as seen in Figure~\ref{fig:3D_QNLdmde_phase_portraits}. This behavior is reflected in the eigenvalues obtained.  
The first eigenvalue is zero, indicating a line or manifold consisting of a continuous set of equilibria where the sum of baryonic matter and DM equals one. The second eigenvalue is negative, while the third is positive, since $(\delta+w)<0 \;\Rightarrow\;-3(\delta+w)>0$. This implies that the manifold acts as a saddle point. 
\begin{gather} \label{DSA.QNLdmde.4}
\begin{split}
\underline{\text{Conditions for}} : \text{ Matter-dominated manifold of saddle points} \begin{cases} 
w<0,  \\ 
(\delta+w)<0,  
\end{cases}
\end{split}
\end{gather} 

\underline{Critical Point $P_{\text{de}}$: dark energy-dominated phase.}
\begin{gather} \label{DSA.QNLdmde.5}
\Omega_{\rm{bm}} = 0, \quad \Omega_{\rm{dm}} = 0, \quad \Omega_{\rm{de}} = 1,  \quad \rightarrow \quad \Omega_{\rm{r}} = 0  \quad ; \quad  \lambda = \begin{bmatrix}
3 w -1 \\
3 w \\
3 (w+\delta)
\end{bmatrix}.
\end{gather}
Since we assume $w<0$, the second eigenvalue is negative, $3 w <0$, which also implies that the first eigenvalue is negative, $(3w-1)<0$. For stability, the third eigenvalue would need to be negative as well. However, this requires $(w+\delta)<0$, whereas the text currently states $(w+\delta)>0 \rightarrow 3(w+\delta)>0$, which would make the third eigenvalue positive and the point unstable. This condition therefore determines whether $P_{\text{de}}$ is a stable attractor or not. As all the eigenvalues are negative, this critical point is a stable node (sink) if:
\begin{gather} \label{DSA.QNLdmde.6.5}
\begin{split}
\underline{\text{Conditions for}} : \text{ Dark energy-dominated stable node (sink)} \begin{cases} 
w<0,  \\ 
\delta+w<0,  
\end{cases}
\end{split}
\end{gather}

In Figure~\ref{fig:3D_QNLdmde_phase_portraits}, we see that there is no radiation-dominated unstable node, although most trajectories still originate from coordinates where radiation is the dominant fluid. A purely radiation-dominated origin with $(\Omega_{\rm{r}},  \Omega_{\rm{bm}}, \Omega_{\rm{dm}},\Omega_{\rm{de}} )=(1,0,0,0)$ is not feasible, as this would make the denominator in $Q=3H \delta \left(\frac{\rho_{\rm{dm}}\rho_{\rm{de}}}{\rho_{\rm{dm}}+\rho_{\rm{de}}} \right)$ vanish, which is unphysical. In practice, since the numerator approaches zero before the denominator, the flow lines in Figure~\ref{fig:3D_QNLdmde_phase_portraits} display behavior similar to that of a radiation-dominated origin. This argument applies to all three non-linear IDE models studied in this paper. The presence of an early radiation-dominated era is also evident from the evolution of the analytical solutions for the density parameters in Figures~\ref{fig:Omega_NLID1},~\ref{fig:Omega_NLID2}, and~\ref{fig:Omega_NLID3}.

From the coordinates of the two critical points above, we see that all \textbf{energies are positive at all critical points for any choice of parameters}. Positive energy densities for any parameter choice are also confirmed by the boundary analysis of this model, since there exists an invariant submanifold at both boundaries, $\Omega_{\rm{dm}}=0$ and $\Omega_{\rm{de}}=0$. Substituting $\Omega_{\rm{dm}}=0$ into $\Omega'_{\rm{dm}}$ and $\Omega_{\rm{de}}=0$ into $\Omega'_{\rm{de}}$ in the dynamical system \eqref{DSA.QNLdmde.1} yields $\Omega'_{\rm{dm}}=0$ and $\Omega'_{\rm{de}}=0$, which implies that the flow lines cannot cross into negative $\Omega_{\rm{dm}}$ or $\Omega_{\rm{de}}$. This ensures positive DM and DE densities within the physically viable region. The impact of different parameter choices on DM and DE densities in both the past and future expansions is summarized in Table~\ref{tab:QNLdmde_energy_conditions}, while the behavior of the model at each critical point with respect to various cosmological parameters is given in Table~\ref{tab:CP_B_QNL_dmde}. \\

\begin{table}[h]
    \centering
    \renewcommand{\arraystretch}{1.4} 
    \begin{tabular}{|c|c|c|c|c|c|c|}
        \hline
        \textbf{Conditions} & Energy flow & $\rho_{\text{dm}}$ (Past) & $\rho_{\text{dm}}$ (Future) & $\rho_{\text{de}}$ (Past) & $\rho_{\text{de}}$ (Future) & \textbf{Physical} \\
        \hline         \hline
        $\delta > 0$   & DE $\to$ DM & $+$ & $+$ & $+$ & $+$ & $\checkmark$ \\
        \hline
        $\delta < 0$ & DM $\to$ DE & $+$ & $+$ & $+$ & $+$  & $\checkmark$ \\
        \hline
    \end{tabular}
    \caption{Conditions for positive energy densities throughout cosmic evolution with $w<0$, for $Q=3H \delta \left(\frac{\rho_{\rm{dm}}\rho_{\rm{de}}}{\rho_{\rm{dm}}+\rho_{\rm{de}}} \right)$.}
    \label{tab:QNLdmde_energy_conditions}
\end{table}

\begin{table}[H]
\centering
\renewcommand{\arraystretch}{1.5} 
\setlength{\tabcolsep}{10pt} 
\begin{tabular}{|c|c|c|}
\hline
\textbf{} & \textbf{$P_{\text{m}}$} & \textbf{$P_{\text{de}}$} \\ \hline
\textbf{Class} &  Saddle Point & Stable node (sink) \\ \hline
$\Omega_{\rm{r}}$ & $0$ & $0$ \\ \hline
$\Omega_{\rm{bm}}$& $-\Omega_{\rm{dm}}+1$  & $0$\\ \hline
$\Omega_{\rm{dm}}$  & $\Omega_{\rm{dm}}$  & $0$\\ \hline
$\Omega_{\rm{de}}$ & $0$ & $1$ \\ \hline
$\Omega_{\rm{dm}}\ge0$  & $\forall \delta$  &  $\forall \delta$\\ \hline
$\Omega_{\rm{de}}\ge0$  & $\forall \delta$  &  $\forall \delta$\\ \hline
$r$  & $\infty$ & $0$ \\ \hline
$w^{\rm{eff}}_{\rm{dm}}$  & $0$ & $-\delta$ \\ \hline
$w^{\rm{eff}}_{\rm{de}}$ & $w+\delta$ & $w$\\ \hline
$\zeta$  & $-3 (w+\delta)$ & $-3 (w+\delta)$ \\ \hline
$w^{\rm{eff}}_{\rm{tot}}$  & $0$ & $w$\\ \hline
$q$ & $\frac{1}{2}$ & $\frac{1}{2}(1+3 w)$ \\ \hline
\end{tabular}
\caption{behavior of the model at critical points for $Q=3H \delta \left(\frac{\rho_{\rm{dm}}\rho_{\rm{de}}}{\rho_{\rm{dm}}+\rho_{\rm{de}}} \right)$.}
\label{tab:CP_B_QNL_dmde}
\end{table}

In Table~\ref{tab:CP_B_QNL_dmde} we see that for both critical points $P_{\text{m}}$ and $P_{\text{de}}$ the system behaves almost identically to the matter- and DE-dominated critical points in a non-interacting model. This indicates that the effect of the interaction effectively vanishes, $Q\rightarrow0$, in both the asymptotic past and future, as the fractional densities of DE and DM approach zero, respectively.  
The only difference introduced by the interaction is that it modifies the effective DE and DM equations of state in the past and future, respectively. This has the consequence of \textit{alleviating the coincidence problem in both the past and the future for the iDEDM regime ($\delta>0$), while worsening the coincidence problem for the iDMDE regime ($\delta<0$)}. The coincidence problem would also be solved in the special case where $\delta=-w$, but this leads to divisions by zero in the analytical solutions for $\rho_{\rm{dm}}$ and $\rho_{\rm{de}}$ in~\ref{NLID1_dm_de_BG}.  
We may now consider the stability of this system by examining the sign of the doom factor for this interaction:
\begin{gather} \label{DSA.QNLdmde.doom}
\begin{split}
\textbf{d}=  \frac{Q}{3H\rho_{\rm{de}}(1+w)}=\frac{3H \delta \left(\frac{\rho_{\rm{dm}}\rho_{\rm{de}}}{\rho_{\rm{dm}}+\rho_{\rm{de}}} \right)}{3H\rho_{\rm{de}}(1+w)}=\frac{\delta}{(1+w)} \left(\frac{\rho_{\rm{dm}}}{\rho_{\rm{dm}}+\rho_{\rm{de}}} \right),
\end{split}
\end{gather}
where we also impose the conditions $\rho_{\rm{dm}}>0$ and $\rho_{\rm{de}}>0$, so the terms in brackets remain positive. Since we require $\textbf{d} < 0$ to ensure a stable universe, it follows from \eqref{DSA.QNLdmde.doom} that this condition is satisfied only if $\delta$ and $(1+w)$ have opposite signs. This implies that \textit{DE may lie in either the quintessence or the phantom regime}. Different parameter combinations and their effects on energy densities and stability are summarized in Table~\ref{tab:QNLdmde_stability_criteria}.

\begin{table}[h]
    \centering
    \renewcommand{\arraystretch}{1.4} 
    \begin{tabular}{|c|c|c|c|c|c|c|c|c|}
        \hline 
        $\delta$ & Energy flow & $w$ & Dark energy & $d$ & a priori stable & $\rho_{\text{dm}} > 0$ & $\rho_{\text{de}} > 0$ & \textbf{Viable} \\
        \hline \hline
        $+$ & DE $\to$ DM & $< -1$ & Phantom & - & $\checkmark$ & $\checkmark$ & $\checkmark$ & $\checkmark$ \\
        \hline
        $+$ & DE $\to$ DM & $> -1$ & Quintessence & + & X & $\checkmark$ & $\checkmark$ & \textbf{X} \\
        \hline
        $-$ & DM $\to$ DE & $< -1$ & Phantom & + & X & $\checkmark$  & $\checkmark$ & \textbf{X}  \\
        \hline
        $-$ & DM $\to$ DE & $> -1$ & Quintessence & - & $\checkmark$ &  $\checkmark$  &  $\checkmark$ & $\checkmark$ \\
        \hline
    \end{tabular}
    \caption{Stability and positive energy criteria for $Q=3H \delta \left(\frac{\rho_{\rm{dm}}\rho_{\rm{de}}}{\rho_{\rm{dm}}+\rho_{\rm{de}}} \right)$.}
    \label{tab:QNLdmde_stability_criteria}
\end{table}

\subsection{Dynamical system analysis: Interaction function $Q_2=3H \delta \left(\frac{\rho^2_{\rm{dm}}}{\rho_{\rm{dm}}+\rho_{\rm{de}}} \right)$} \label{DSA.Q.dmdm}

\begin{figure}[htbp]
    \centering
    \begin{subfigure}[b]{0.495\linewidth}
        \centering
        \includegraphics[width=\linewidth]{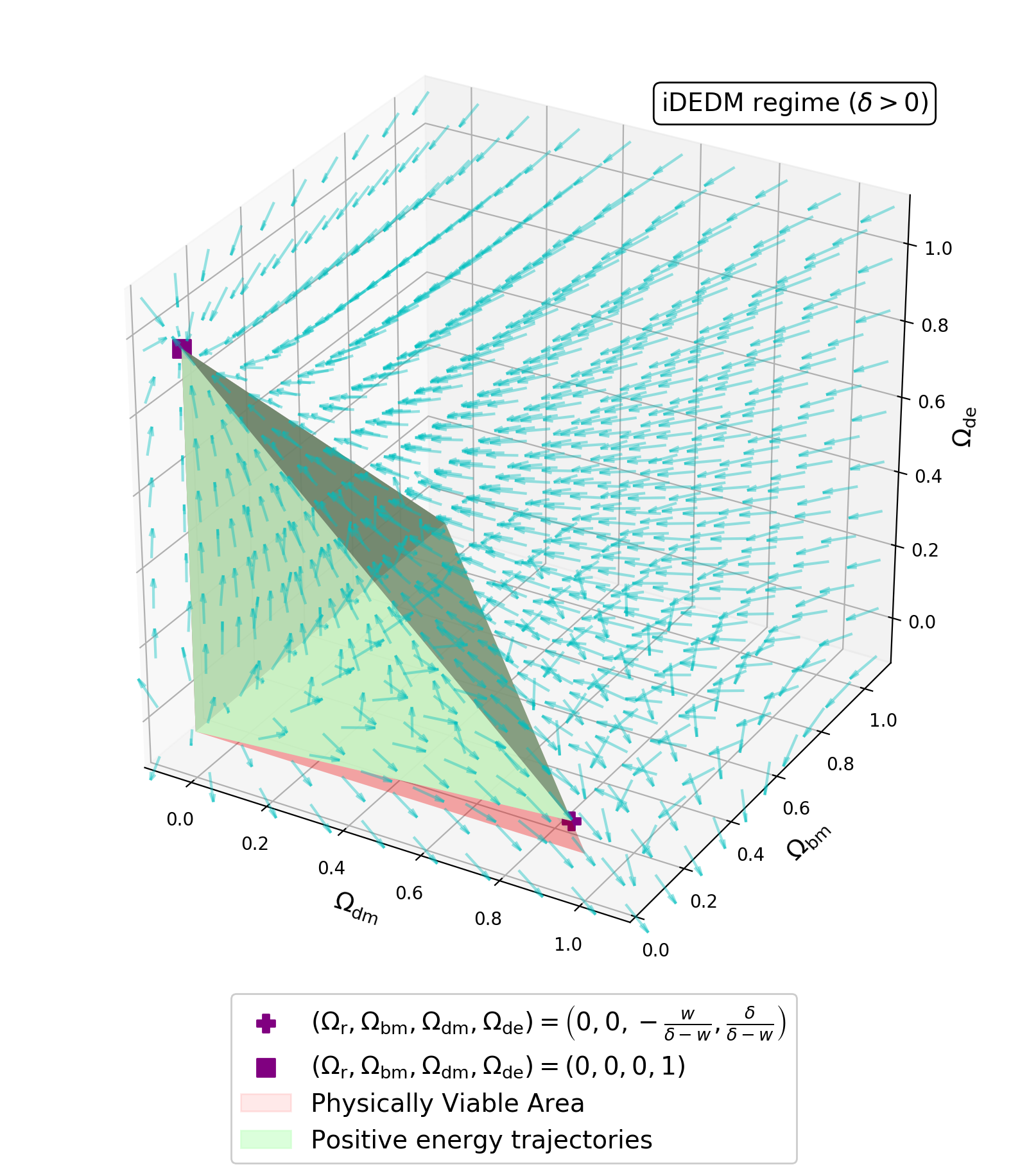}
    \end{subfigure}%
    \hspace{0pt} 
    \begin{subfigure}[b]{0.495\linewidth}
        \centering
        \includegraphics[width=\linewidth]{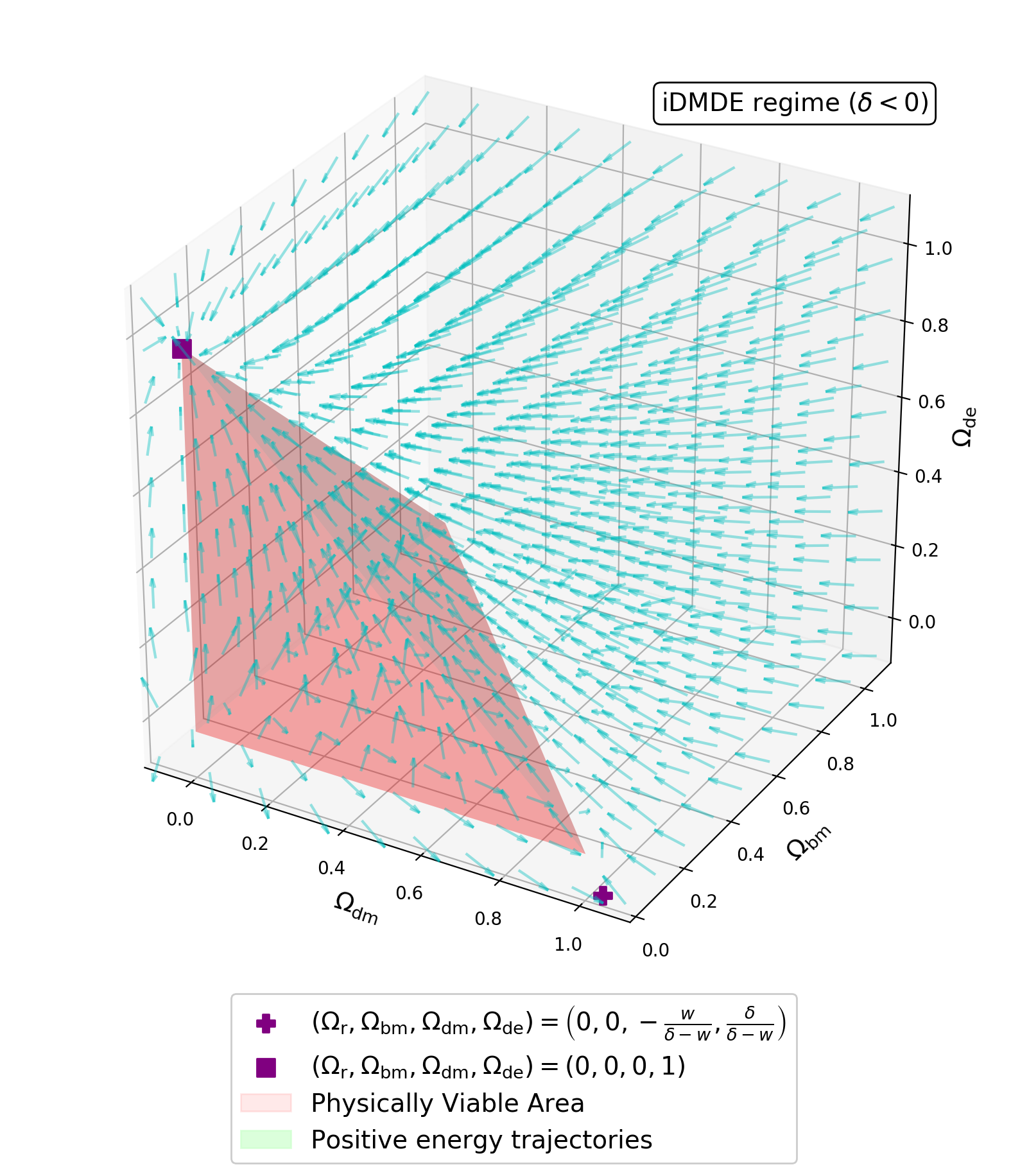}
    \end{subfigure}   
    \caption{3D phase portraits for $Q=3H \delta \left(\frac{\rho^2_{\rm{dm}}}{\rho_{\rm{dm}}+\rho_{\rm{de}}} \right)$, showing positive-energy trajectories in the iDEDM regime ($\delta=+0.1$, left panel) and negative DE trajectories in the iDMDE regime ($\delta=-0.1$, right panel).}
    \label{fig:3D_QNLdm_phase_portraits}
\end{figure}

\begin{figure}[htbp]
    \centering
    \begin{subfigure}[b]{0.49\linewidth}
        \centering
        \includegraphics[width=\linewidth]{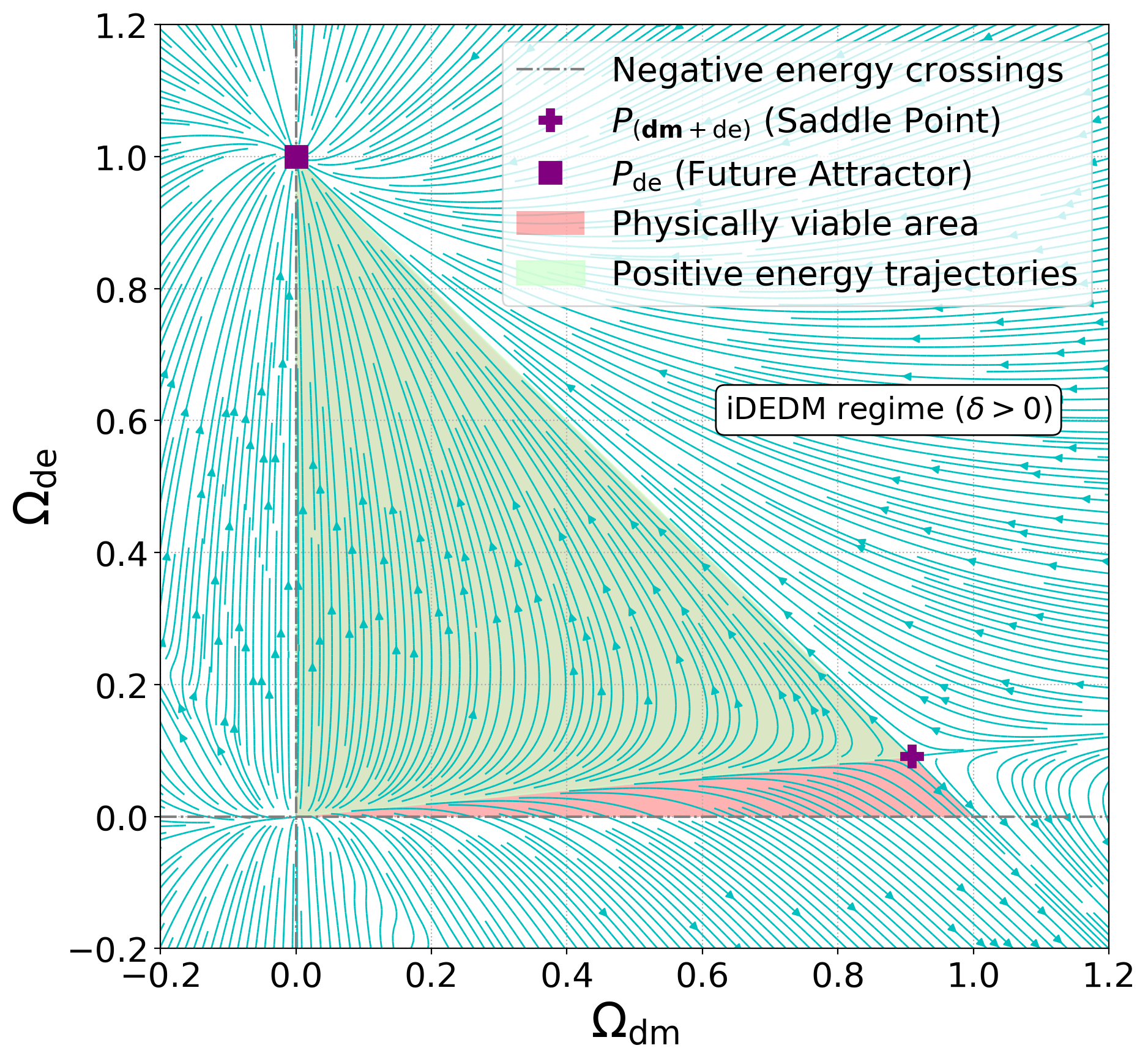}
    \end{subfigure}%
    \hspace{0pt} 
    \begin{subfigure}[b]{0.49\linewidth}
        \centering
        \includegraphics[width=\linewidth]{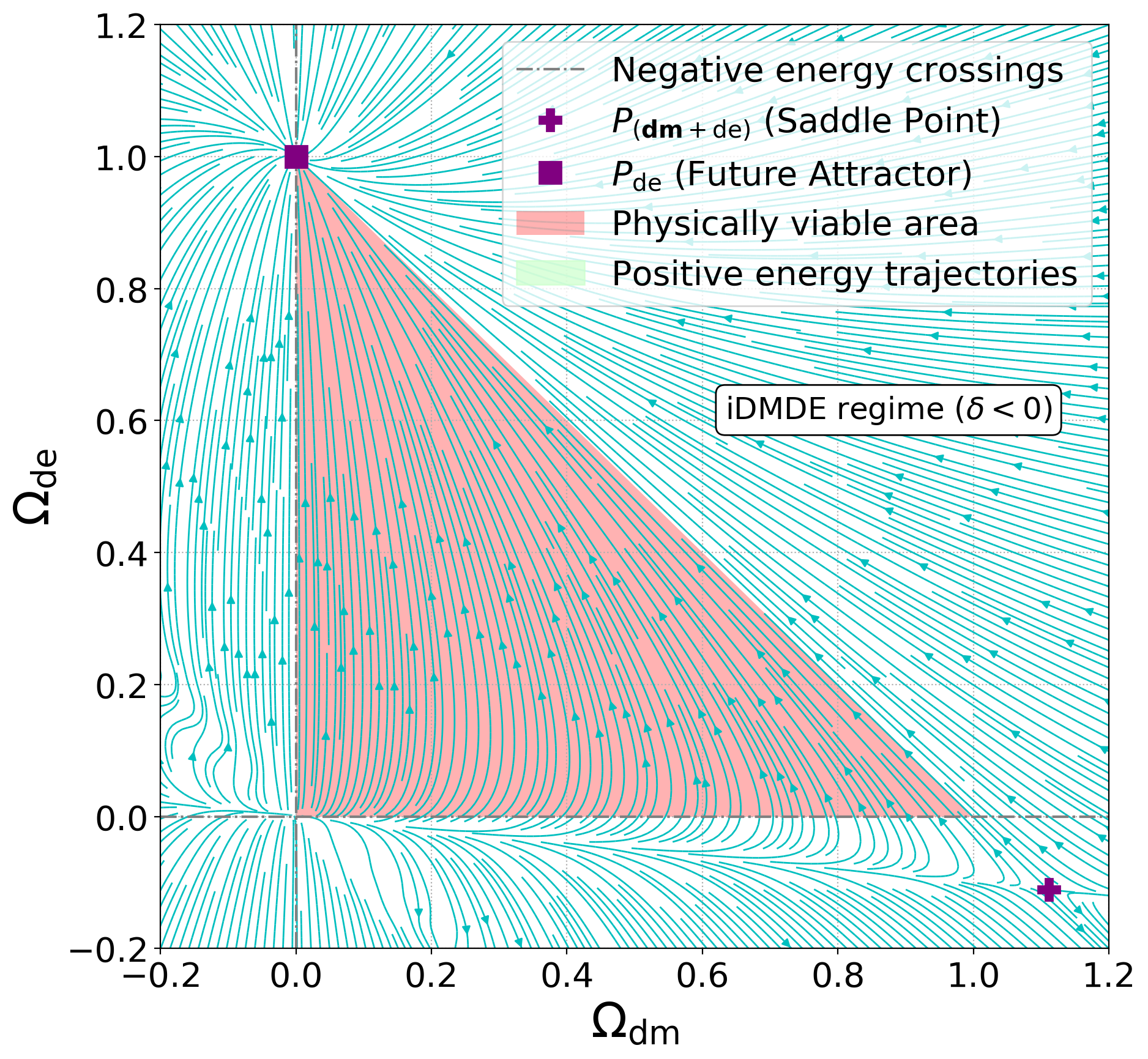}
    \end{subfigure}
    \caption{2D projection of the phase portraits for $Q=3H \delta \left(\frac{\rho^2_{\rm{dm}}}{\rho_{\rm{dm}}+\rho_{\rm{de}}} \right)$, showing positive-energy trajectories in the iDEDM regime ($\delta=+0.1$, left panel) and negative DE trajectories in the iDMDE regime ($\delta=-0.1$, right panel).}
    \label{fig:2D_QNLdm_phase_portraits}
\end{figure}

We now consider the dynamical system behavior of non-linear interacting dark energy models. In the second of these models, the interaction strength depends primarily on the DM density. For the interaction $Q=3H \delta \left(\frac{\rho^2_{\rm{dm}}}{\rho_{\rm{dm}}+\rho_{\rm{de}}} \right)$, the dynamical system (\ref{DSA.9}) becomes:
\begin{gather} \label{DSA.QNLdm.1}
\begin{split}
{\Omega}'_{\rm{de}} &= \Omega_{\rm{de}} \left[ 1-  \Omega_{\rm{bm}} -\Omega_{\rm{dm}} - \Omega_{\rm{de}}  \left(1 -3 w \right) - 3w \right]  - 3 \delta \left(\frac{\Omega^2_{\rm{dm}}}{\Omega_{\rm{dm}} + \Omega_{\rm{de}}} \right), \\
{\Omega}'_{\rm{dm}} &= \Omega_{\rm{dm}} \left[ 1-  \Omega_{\rm{bm}} -\Omega_{\rm{dm}} - \Omega_{\rm{de}}  \left(1 -3 w \right)  \right]  + 3 \delta \left(\frac{\Omega^2_{\rm{dm}}}{\Omega_{\rm{dm}} + \Omega_{\rm{de}}} \right), \\
{\Omega}'_{\rm{bm}} &= \Omega_{\rm{bm}} \left[ 1-  \Omega_{\rm{bm}} -\Omega_{\rm{dm}} - \Omega_{\rm{de}}  \left(1 -3 w \right)  \right]. \\
\end{split}
\end{gather}
From the dynamical system \eqref{DSA.QNLdm.1}, we find two critical points:  

\underline{Critical Point $P_{\text{\textbf{dm}+de}}$: dark matter–dark energy hybrid dominated phase.}
\begin{gather} \label{DSA.QNLdm.2}
\Omega_{\rm{bm}} = 0, \quad \Omega_{\rm{dm}} = - \frac{w}{\delta-w}, \quad \Omega_{\rm{de}} = \frac{\delta}{\delta-w},  \quad \rightarrow \quad \Omega_{\rm{r}} = 0  \quad ; \quad  \lambda = \begin{bmatrix}
-3w \\
\dfrac{\delta(3w - 1) + w}{\delta - w} \\
\dfrac{3\delta w}{\delta - w}
\end{bmatrix}.
\end{gather}
From the coordinates we see that $\Omega_{\rm{dm}}+\Omega_{\rm{de}}=1$ at this critical point, corresponding to a hybrid dominance of DM and DE. DM will typically be more dominant here, since $(\Omega_{\rm{r}},  \Omega_{\rm{bm}}, \Omega_{\rm{dm}},\Omega_{\rm{de}} )=(0,0,1,0)$ when the interaction is switched off ($\delta=0$) in \eqref{DSA.QNLdm.2}.  
As all eigenvalues are real and there is one additional critical point, we expect this point to behave as a saddle point. Immediately from the coordinates of the critical point, together with the assumptions $\delta<-w$ and $w<0$, we obtain $\delta-w>0$. We note that DE becomes negative if:
\begin{gather} \label{DSA.QNLdm.4}
\begin{split}
\underline{\text{Conditions for}} :  \begin{cases} 
\Omega_{\rm{de}} <0  \quad \text{if} \quad w<0 \; \text{and } \; w<\delta<0 \; \text{(iDMDE regime)}, \\ 
\Omega_{\rm{de}} >0 \quad \text{if} \quad w<0 \; \text{and } \; \delta>0 \; \text{(iDEDM regime)}.
\end{cases}
\end{split}
\end{gather} \\ 
Since $w<0$, the first eigenvalue is positive, $-3 w>0$. Together with $\delta-w>0$, this implies that the third eigenvalue is negative, $\dfrac{3\delta w}{\delta - w}<0$, when $\delta>0$, and positive when $\delta<0$ (which corresponds to unphysical negative energies and may therefore be discarded). Thus, this critical point has real eigenvalues with different signs, and is classified as a saddle point, as shown in Figures~\ref{fig:3D_QNLdm_phase_portraits} and~\ref{fig:2D_QNLdm_phase_portraits}.
\begin{gather} \label{DSA.QNLdm.6}
\begin{split}
\underline{\text{Conditions for}} : \text{ Positive dark matter–dark energy hybrid-dominated saddle point} \begin{cases} 
w<0,  \\
\delta>0,
\end{cases}
\end{split}
\end{gather} \\ 

\underline{Critical Point $P_{\text{de}}$: dark energy-dominated phase.}
\begin{gather} \label{DSA.QNLdmde.7}
\Omega_{\rm{bm}} = 0, \quad \Omega_{\rm{dm}} = 0, \quad \Omega_{\rm{de}} = 1,  \quad \rightarrow \quad \Omega_{\rm{r}} = 0  \quad ; \quad  \lambda = \begin{bmatrix}
3 w -1 \\
3 w \\
3 w
\end{bmatrix}.
\end{gather}
Since we assume $w<0$, all the eigenvalues are negative. Therefore, this critical point is a stable node (sink), as shown in Figures~\ref{fig:3D_QNLdm_phase_portraits} and~\ref{fig:2D_QNLdm_phase_portraits}, if:
\begin{gather} \label{DSA.QNLdm.9}
\begin{split}
\underline{\text{Conditions for}} : \text{ Dark energy-dominated stable node (sink)} \begin{cases} 
 w<0 
\end{cases}
\end{split}
\end{gather}

In Figures~\ref{fig:3D_QNLdm_phase_portraits} and~\ref{fig:2D_QNLdm_phase_portraits}, we again see the absence of an explicit radiation-dominated past attractor and a baryonic matter-dominated saddle point, due to the presence of the $(\rho_{\rm{dm}}+\rho_{\rm{de}})$ term in the denominator, as discussed above.  
This model shows behavior similar to the linear interaction model $Q=3H\rho_{\rm{dm}}$, where the sign of the interaction constant $\delta$ determines whether the saddle point exhibits negative energies. \textbf{For the iDEDM regime ($\delta>0$), all energies remain positive, while in the iDMDE regime ($\delta<0$), DE becomes negative in the past ($\Omega_{\rm{de}}<0$).} In contrast, DM remains positive at all times, regardless of the choice of parameters. This is a direct consequence of the fact that as $\rho_{\rm{dm}}\rightarrow 0$, the interaction $Q=3H \delta \left(\frac{\rho^2_{\rm{dm}}}{\rho_{\rm{dm}}+\rho_{\rm{de}}} \right)$ also tends to zero, so the energy flow stops before negative energies can be reached.  
This is confirmed by the boundary analysis for this model: at the boundary $\Omega_{\rm{dm}}=0$ there is an invariant submanifold. Substituting $\Omega_{\rm{dm}}=0$ into $\Omega'_{\rm{dm}}$ in the dynamical system \eqref{DSA.QNLdm.1} gives $\Omega'_{\rm{dm}}=0$, which implies that the flow lines cannot cross into negative ${\Omega}_{\rm{dm}}$, thereby ensuring positive DM densities. \\ 
For DE to remain positive, we note from Figure~\ref{fig:2D_QNLdm_phase_portraits} that there is an invariant submanifold between the origin and the saddle point, along which the trajectories flow tangent to the line. Trajectories above this line (within the indicated "positive energy trajectories") remain positive, while those below cross the zero-energy boundary and become negative.  
This straight line in the $({\Omega}_{\rm{dm}},{\Omega}_{\rm{de}})$ plane, connecting the origin $({\Omega}_{\rm{dm}},{\Omega}_{\rm{de}})=(0,0)$ and $P_{\text{\textbf{dm}+de}}$ $({\Omega}_{\rm{dm}},{\Omega}_{\rm{de}})=\left(-\tfrac{w}{\delta -w},\tfrac{\delta}{\delta -w}\right)$, may be described by the equation:  
\begin{gather} \label{DSA.QNLdm.10}
\begin{split}
\Omega_{\rm{de}} &= m \Omega_{\rm{dm}} +c,  \quad 
m=\frac{\Delta \Omega_{\rm{de}}}{\Delta \Omega_{\rm{dm}}}
= \frac{\tfrac{\delta}{\delta -w} -0}{-\tfrac{w}{\delta -w}-0}
= -\frac{\delta}{w},  
\quad c=0, \\
\Omega_{\rm{de}} &= -\frac{\delta}{w}\,\Omega_{\rm{dm}}   
\quad \rightarrow \quad \Omega_{\rm{de}}+\frac{\delta}{w}\,\Omega_{\rm{dm}} =0.
\end{split}
\end{gather}
It should be noted that the slope is positive, $m>0$, provided $\delta>0$, as required by the constraint previously derived in \eqref{DSA.QNLdm.6}. A positive slope ensures that the line lies in the positive DE domain ($\Omega_{\rm{de}}>0$).  
To demonstrate that the line described by \eqref{DSA.QNLdm.10} is an invariant submanifold, we substitute the expressions for $\Omega'_{\rm{de}}$ and $\Omega'_{\rm{dm}}$ from \eqref{DSA.QNLdm.1} into \eqref{DSA.QNLdm.10}, showing that $\Omega'_{\rm{de}}+\tfrac{\delta}{w}\Omega'_{\rm{dm}}=0$ along the line where $\Omega_{\rm{de}}=-\tfrac{\delta}{w}\Omega_{\rm{dm}}$.

To determine constraints on the initial coordinates $({\Omega}_{\rm{(dm,0)}},{\Omega}_{\rm{(de,0)}})$ that ensure positive energy, these coordinates must lie above the invariant line described in \eqref{DSA.QNLdm.10}, such that:
\begin{gather} \label{DSA.QNLdm.12}
\begin{split}
\Omega_{\rm{(de,0)}} &\ge -\frac{\delta}{w} \Omega_{\rm{(dm,0)}}  
\quad \rightarrow \quad 
\delta \le -\frac{w}{r_0}. \\
\end{split}
\end{gather}
The condition \eqref{DSA.QNLdm.12} may now be combined with the previously derived constraint for positive critical points in \eqref{DSA.QNLdm.6}, yielding the final set of conditions that ensure all components have positive energies at all points in the cosmological evolution:
\begin{gather} \label{DSA.QNLdm.13}
\begin{split}
\boxed{\underline{\text{Conditions for} \; \Omega_{\rm{dm}}\ge0 \; ; \; \Omega_{\rm{de}} \ge0 \; \text{throughout cosmological evolution:}} \quad 
\text{iDEDM with } 0 \le \delta \le -\frac{w}{r_0}}.
\end{split}
\end{gather}
The impact of different parameter choices on the DM and DE densities in both the past and future expansion is summarized in Table~\ref{tab:QNLdm_energy_conditions}. The behavior of the model at each critical point, with respect to various cosmological parameters, is given in Table~\ref{tab:CP_B_QNL_dm}. \\

\begin{table}[h]
    \centering
    \renewcommand{\arraystretch}{1.4} 
    \begin{tabular}{|c|c|c|c|c|c|c|}
        \hline
        \textbf{Conditions} & Energy flow & $\rho_{\text{dm}}$ (Past) & $\rho_{\text{dm}}$ (Future) & $\rho_{\text{de}}$ (Past) & $\rho_{\text{de}}$ (Future) & \textbf{Physical} \\
        \hline         \hline
        $0 < \delta <-\frac{w}{r_0} $ & DE $\to$ DM & $+$ & $+$ & $+$ & $+$ & $\checkmark$ \\
        \hline
        $\delta < 0$ & DM $\to$ DE & $+$ & $+$ & $-$ & $+$  & \textbf{X} \\
        \hline
         $\delta > -\frac{w}{r_0} $ & DE $\to$ DM & $+$ & $+$ & $+$ & $-$ & \textbf{X} \\
        \hline        
    \end{tabular}
    \caption{Conditions for positive energy densities throughout cosmic evolution with $w<0$, for $Q=3H \delta \left(\frac{\rho^2_{\rm{dm}}}{\rho_{\rm{dm}}+\rho_{\rm{de}}} \right)$.}
    \label{tab:QNLdm_energy_conditions}
\end{table}

\begin{table}[H]
\centering
\renewcommand{\arraystretch}{1.5} 
\setlength{\tabcolsep}{10pt} 
\begin{tabular}{|c|c|c|}
\hline
\textbf{} & $P_{\text{\textbf{dm}+de}}$ & $P_{\text{de}}$ \\ \hline
\textbf{Class} &  Saddle Point & Stable node (sink) \\ \hline
$\Omega_{\rm{r}}$ & $0$ & $0$ \\ \hline
$\Omega_{\rm{bm}}$& $0$  & $0$\\ \hline
$\Omega_{\rm{dm}}$  & $- \frac{w}{\delta-w}$  & $0$\\ \hline
$\Omega_{\rm{de}}$ & $\frac{\delta}{\delta-w}$ & $1$ \\ \hline
$\Omega_{\rm{dm}}\ge0$  & $\forall \delta$  &  $\forall \delta$\\ \hline
$\Omega_{\rm{de}}\ge0$  & $\delta\ge0$  &  $\forall \delta$\\ \hline
$r$  & $-\frac{w}{\delta}$ & $0$ \\ \hline
$w^{\rm{eff}}_{\rm{dm}}$  & $\frac{\delta w}{\delta-w}$ & $0$ \\ \hline
$w^{\rm{eff}}_{\rm{de}}$ & $\frac{\delta w}{\delta-w}$ & $w$\\ \hline
$\zeta$  & $0$ & $-3 w$ \\ \hline
$w^{\rm{eff}}_{\rm{tot}}$  & $\frac{\delta w}{\delta-w}$ & $w$\\ \hline
$q$ & $\frac{1}{2}(1+3 \frac{\delta w}{\delta-w})$ & $\frac{1}{2}(1+3 w)$ \\ \hline
\end{tabular}
\caption{behavior of the model at critical points for $Q=3H \delta \left(\frac{\rho^2_{\rm{dm}}}{\rho_{\rm{dm}}+\rho_{\rm{de}}} \right)$.}
\label{tab:CP_B_QNL_dm}
\end{table}

In Table~\ref{tab:CP_B_QNL_dm} we see that at $P_{\text{\textbf{dm}+de}}$ the system behaves identically to the corresponding dark energy-dominated critical point in the non-interacting $w$CDM model. This indicates that the effect of the interaction completely vanishes in the future, as the DM density approaches zero.  
Furthermore, at $P_{\text{\textbf{dm}+de}}$ in the past, DM and DE redshift at the same rate, $w^{\rm{eff}}_{\rm{dm}}=w^{\rm{eff}}_{\rm{de}}$, which implies a fixed ratio $r=-\tfrac{w}{\delta}$ and $\zeta=0$. This solves the coincidence problem in the past, while leaving it unchanged in the future.  
As in the $w$CDM model and in the case of the interaction $Q=3H\rho_{\rm{dm}}$, this model exhibits accelerated expansion in the late-time DE-dominated era as long as $w<-1/3$, and it will experience a big rip singularity if $w^{\rm{eff}}_{\rm{tot}}=w<-1$. 
We may now consider the stability of this system by examining the sign of the doom factor for this interaction:
\begin{gather} \label{DSA.QNLdm.doom}
\begin{split}
\textbf{d}=  \frac{Q}{3H\rho_{\rm{de}}(1+w)}=\frac{3H \delta \left(\frac{\rho^2_{\rm{dm}}}{\rho_{\rm{dm}}+\rho_{\rm{de}}} \right)}{3H\rho_{\rm{de}}(1+w)}=\frac{\delta}{(1+w)} \left(\frac{\rho^2_{\rm{dm}}}{\rho_{\rm{de}}(\rho_{\rm{dm}}+\rho_{\rm{de}})} \right),
\end{split}
\end{gather}
where we also impose the conditions $\rho_{\rm{dm}}>0$ and $\rho_{\rm{de}}>0$, ensuring that the terms in brackets remain positive. Since we require $\textbf{d} < 0$ to ensure a stable universe, it follows from \eqref{DSA.QNLdm.doom} that this condition is satisfied only if $\delta$ and $(1+w)$ have opposite signs. Because positive energies demand $\delta>0$, this implies that $(1+w)>0$ is needed for a priori stability. This corresponds to $w<-1$, meaning that \textit{DE must lie in the phantom regime}. Different parameter combinations and their effects on energy density and stability are summarized in Table~\ref{tab:QNLdm_stability_criteria}.

\begin{table}[h]
    \centering
    \renewcommand{\arraystretch}{1.4} 
    \begin{tabular}{|c|c|c|c|c|c|c|c|c|}
        \hline 
        $\delta$ & Energy flow & $w$ & Dark energy & $d$ & a priori stable & $\rho_{\text{dm}} > 0$ & $\rho_{\text{de}} > 0$ & \textbf{Viable} \\
        \hline \hline
        $+$ & DE $\to$ DM & $< -1$ & Phantom & - & $\checkmark$ & $\checkmark$ & $\checkmark$ & $\checkmark$ \\
        \hline
        $+$ & DE $\to$ DM & $> -1$ & Quintessence & + & \textbf{X} & $\checkmark$ & $\checkmark$ & \textbf{X} \\
        \hline
        $-$ & DM $\to$ DE & $< -1$ & Phantom & + & \textbf{X} & $\checkmark$ & X & \textbf{X} \\
        \hline
        $-$ & DM $\to$ DE & $> -1$ & Quintessence & - & $\checkmark$ & $\checkmark$ & X & \textbf{X} \\
        \hline
    \end{tabular}
    \caption{Stability and positive energy criteria for $Q=3H \delta \left(\frac{\rho^2_{\rm{dm}}}{\rho_{\rm{dm}}+\rho_{\rm{de}}} \right)$.}
    \label{tab:QNLdm_stability_criteria}
\end{table}

\subsection{Dynamical system analysis: Interaction function $Q_3=3H \delta \left(\frac{\rho^2_{\rm{de}}}{\rho_{\rm{dm}}+\rho_{\rm{de}}} \right)$} \label{DSA.Q.dede}

For the interaction $Q=3H \delta \left(\frac{\rho^2_{\rm{de}}}{\rho_{\rm{dm}}+\rho_{\rm{de}}} \right)$, the dynamical system (\ref{DSA.9}) becomes:
\begin{gather} \label{DSA.QNLde.1}
\begin{split}
{\Omega}'_{\rm{de}} &= \Omega_{\rm{de}} \left[ 1 - \Omega_{\rm{bm}} - \Omega_{\rm{dm}} - \Omega_{\rm{de}} \left(1 - 3w \right) - 3w \right]  
- 3 \delta \left(\frac{\Omega^2_{\rm{de}}}{\Omega_{\rm{dm}} + \Omega_{\rm{de}}} \right), \\
{\Omega}'_{\rm{dm}} &= \Omega_{\rm{dm}} \left[ 1 - \Omega_{\rm{bm}} - \Omega_{\rm{dm}} - \Omega_{\rm{de}} \left(1 - 3w \right) \right]  
+ 3 \delta \left(\frac{\Omega^2_{\rm{de}}}{\Omega_{\rm{dm}} + \Omega_{\rm{de}}} \right), \\
{\Omega}'_{\rm{bm}} &= \Omega_{\rm{bm}} \left[ 1 - \Omega_{\rm{bm}} - \Omega_{\rm{dm}} - \Omega_{\rm{de}} \left(1 - 3w \right) \right]. \\
\end{split}
\end{gather}
From the dynamical system \eqref{DSA.QNLde.1} we find: \\ 

\begin{figure}[htbp]
    \centering
    \begin{subfigure}[b]{0.495\linewidth}
        \centering
        \includegraphics[width=\linewidth]{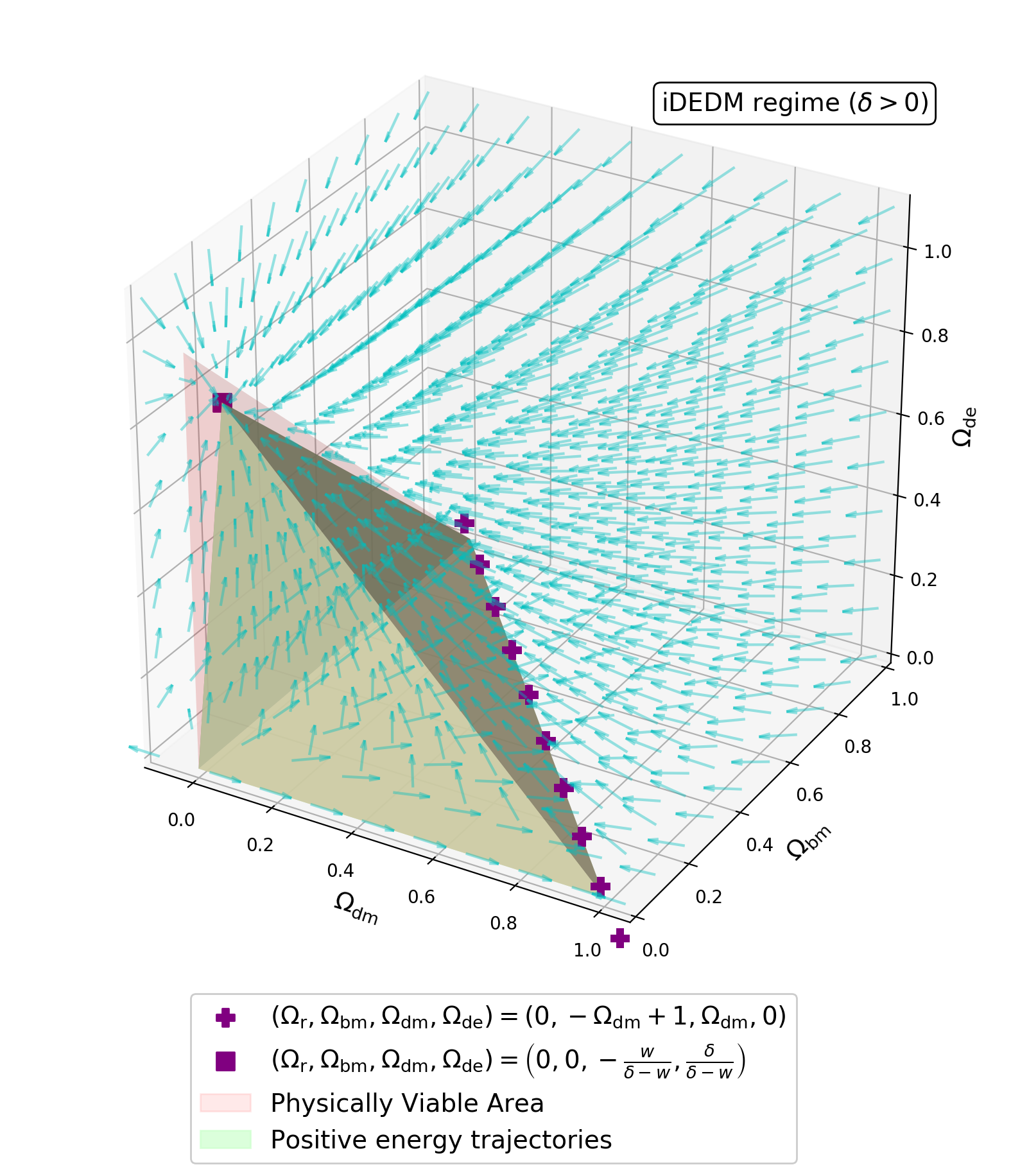}
    \end{subfigure}%
    \hspace{0pt} 
    \begin{subfigure}[b]{0.495\linewidth}
        \centering
        \includegraphics[width=\linewidth]{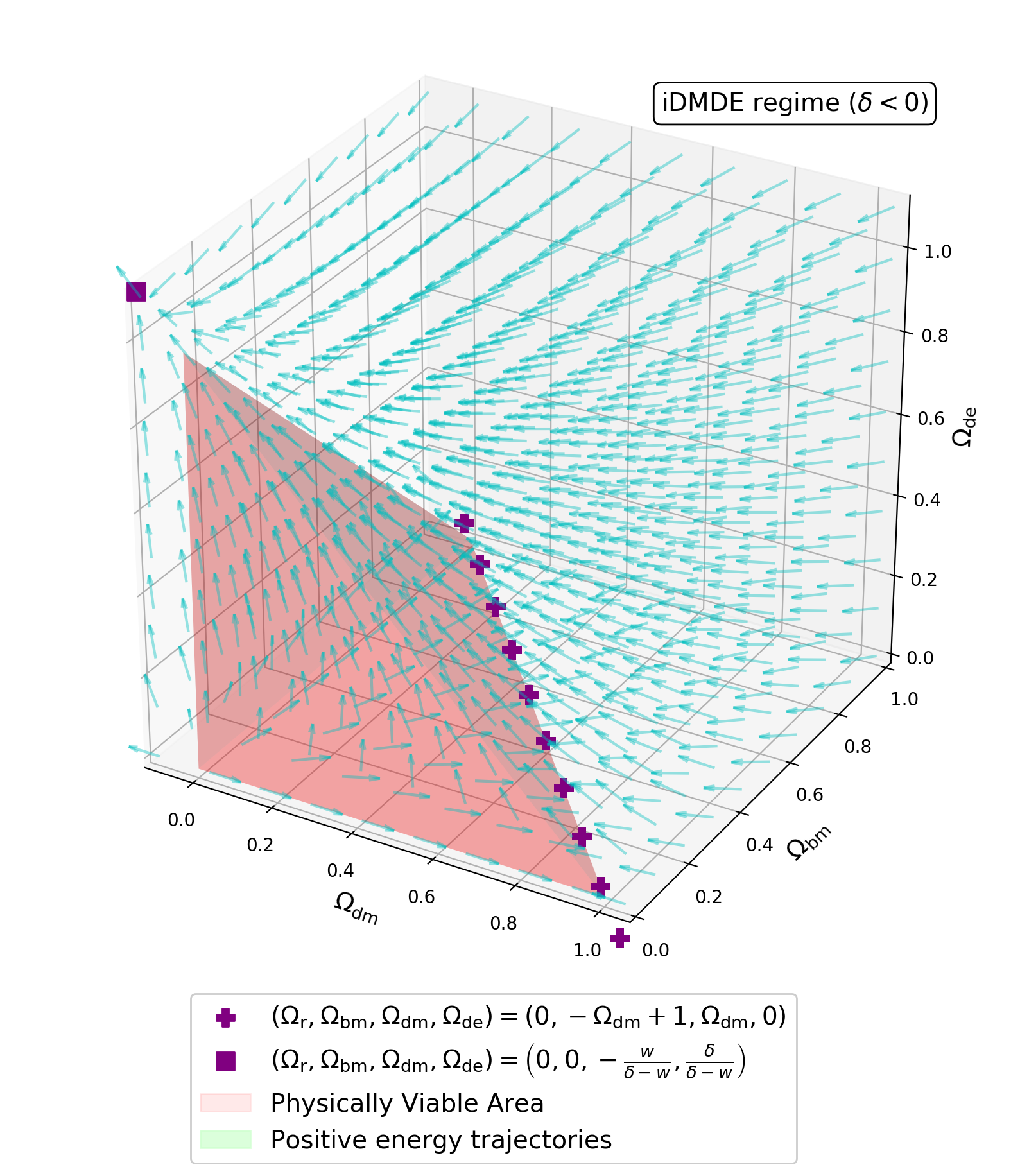}
    \end{subfigure}
    \caption{3D phase portraits for $Q=3H \delta \left(\frac{\rho^2_{\rm{de}}}{\rho_{\rm{dm}}+\rho_{\rm{de}}} \right)$, showing positive-energy trajectories in the iDEDM regime ($\delta=+0.1$, left panel) and negative DM trajectories in the iDMDE regime ($\delta=-0.1$, right panel).}
    \label{fig:3D_QNLde_phase_portraits}
\end{figure}

\underline{Critical Point $P_{\text{m}}$: matter-dominated phase.}
\begin{gather} \label{DSA.QNLde.2}
\Omega_{\rm{bm}} = -\Omega_{\rm{dm}}+1, \quad \Omega_{\rm{dm}} = \Omega_{\rm{dm}}, \quad \Omega_{\rm{de}} = 0  
\quad \rightarrow \quad \Omega_{\rm{r}} = 0   \quad ; \quad  
\lambda = \begin{bmatrix}
0 \\
-1 \\
-3w
\end{bmatrix}.
\end{gather}
Similar to \eqref{DSA.QNLdmde.2}, the coordinates and the zero eigenvalue in \eqref{DSA.QNLde.2} correspond to a line or manifold consisting of a continuous set of equilibria where the sum of baryonic and DM equals one, as shown in Figure~\ref{fig:3D_QNLde_phase_portraits}. Since the other two eigenvalues are negative, this line behaves as a saddle point.

\begin{gather} \label{DSA.QNLde.4}
\begin{split}
\underline{\text{Conditions for}} : \text{ Matter-dominated saddle manifold } \begin{cases} 
w<0 
\end{cases}
\end{split}
\end{gather} \\ 

\begin{figure}[htbp]
    \centering
    \begin{subfigure}[b]{0.49\linewidth}
        \centering
        \includegraphics[width=\linewidth]{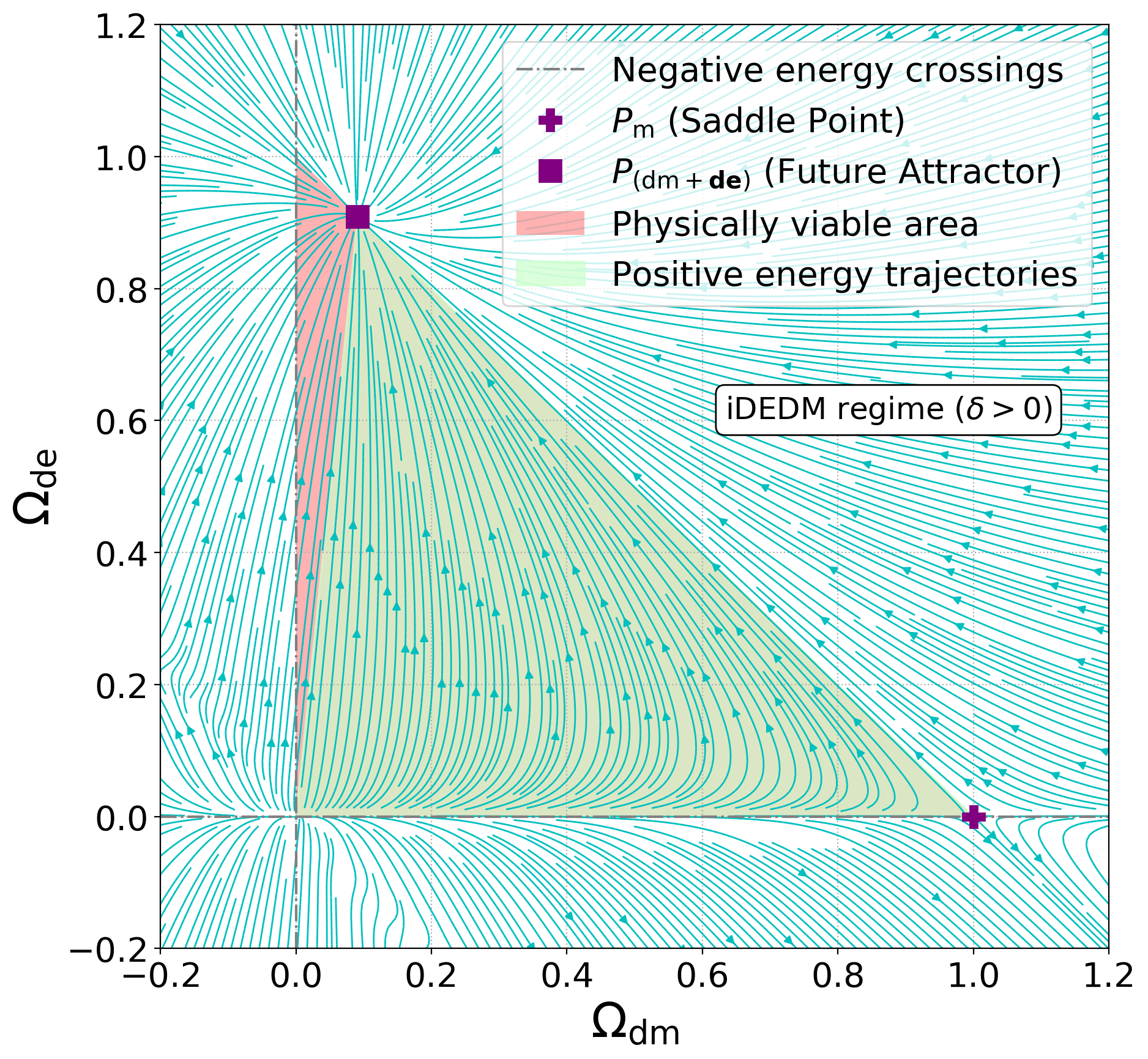}
    \end{subfigure}%
    \hspace{0pt} 
    \begin{subfigure}[b]{0.49\linewidth}
        \centering
        \includegraphics[width=\linewidth]{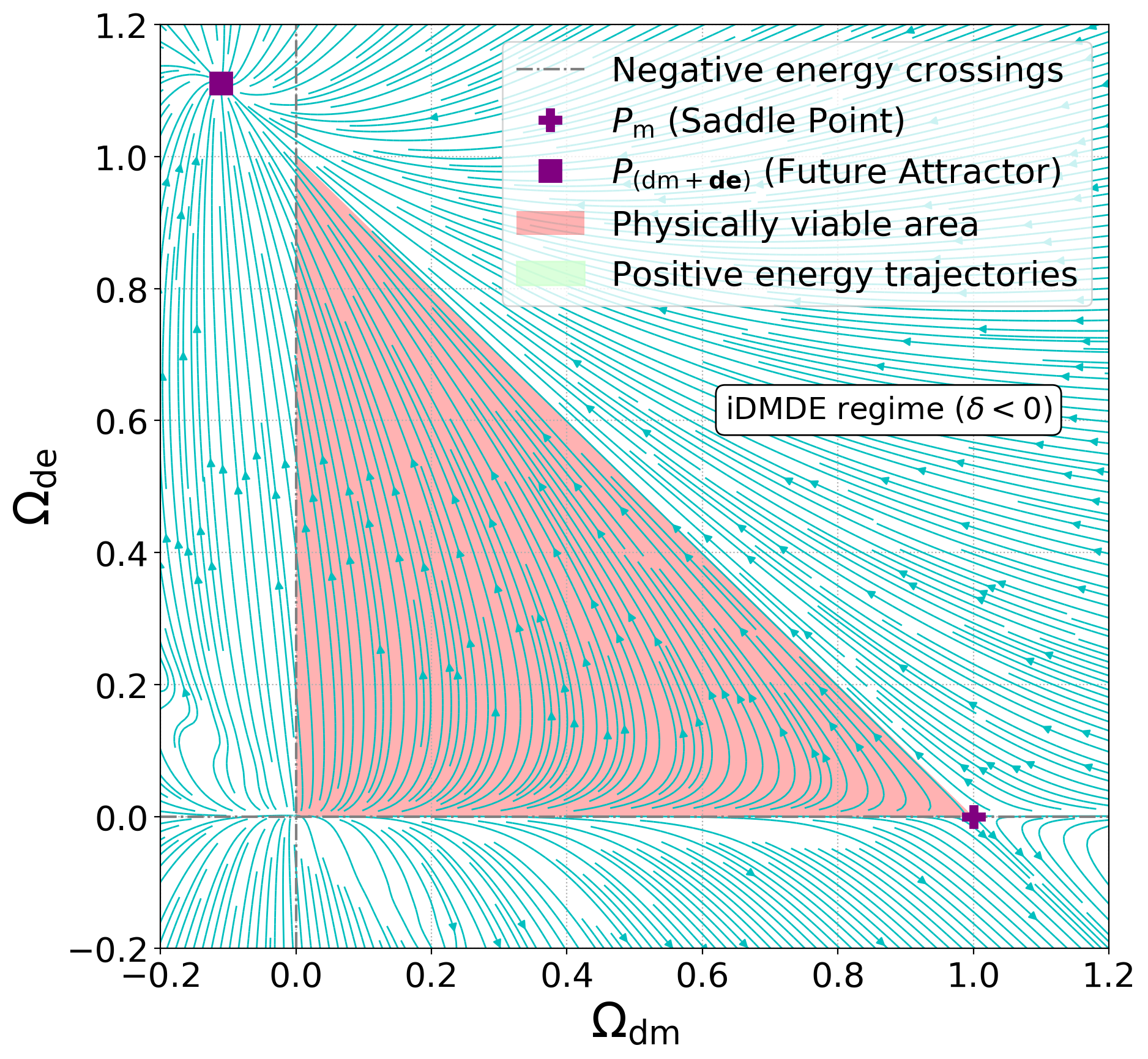}
    \end{subfigure}
    \caption{2D projection of the phase portraits for $Q=3H \delta \left(\frac{\rho^2_{\rm{de}}}{\rho_{\rm{dm}}+\rho_{\rm{de}}} \right)$, showing positive-energy trajectories in the iDEDM regime ($\delta=+0.1$, left panel) and negative DM trajectories in the iDMDE regime ($\delta=-0.1$, right panel).}
    \label{fig:2D_QNLde_phase_portraits}
\end{figure}

\underline{Critical Point $P_{\text{dm+\textbf{de}}}$: dark matter–dark energy hybrid dominated phase.}
\begin{gather} \label{DSA.QNLde.5}
\Omega_{\rm{bm}} = 0, \quad \Omega_{\rm{dm}} = \frac{\delta}{\delta-w}, \quad \Omega_{\rm{de}} = - \frac{w}{\delta-w}   
\quad \rightarrow \quad \Omega_{\rm{r}} = 0 \quad ; \quad  
\lambda = \begin{bmatrix}
3w  \\
\dfrac{-3 w^2-(\delta - w)}{\delta - w} \\
-\dfrac{3w^2}{\delta - w}
\end{bmatrix}.
\end{gather}
From the coordinates we see that $\Omega_{\rm{dm}}+\Omega_{\rm{de}}=1$ at this critical point, which corresponds to a DM–DE hybrid dominated phase. DE will generally be more dominant here, since $(\Omega_{\rm{r}}, \Omega_{\rm{bm}}, \Omega_{\rm{dm}}, \Omega_{\rm{de}})=(0,0,0,1)$ when the interaction is switched off ($\delta=0$) in \eqref{DSA.QNLde.5}.  
Because we assume $w<0$ and $\delta<-w$, we find $\Omega_{\rm{de}}>0$ at this point.
From the DM coordinate of the critical point we note that DM will be negative if:
\begin{gather} \label{DSA.QNLde.7}
\begin{split}
\underline{\text{Conditions for}} :  \begin{cases} 
\Omega_{\rm{dm}} <0 \quad \text{if} \quad w<0 \; \text{and } w<\delta<0 \; \text{(iDMDE regime)}, \\ 
\Omega_{\rm{dm}} >0 \quad \text{if} \quad w<0 \; \text{and } \delta>0 \; \text{(iDEDM regime)} 
\end{cases}.
\end{split}
\end{gather} \\ 
Since $w<0$, the first eigenvalue is negative, $3w<0$. For the third eigenvalue, we note that $w^2>0$ and $\delta-w>0$, which implies $-\tfrac{3w^2}{\delta - w}<0$, so it is also negative. For the second eigenvalue, the numerator $(-3w^2-(\delta-w))$ is negative, while the denominator $(\delta-w)$ is positive, hence the eigenvalue is negative as well.  
Since all eigenvalues are negative, this critical point is a stable node (sink), as shown in Figures~\ref{fig:3D_QNLde_phase_portraits} and~\ref{fig:2D_QNLde_phase_portraits}.  
\begin{gather} \label{DSA.QNLde.8}
\begin{split}
\underline{\text{Conditions for}} : \text{ Positive dark matter–dark energy hybrid dominated stable node} \begin{cases} 
w<0, \\
\delta>0
\end{cases}
\end{split}
\end{gather} \\ 


In both Figures~\ref{fig:3D_QNLde_phase_portraits} and~\ref{fig:2D_QNLde_phase_portraits}, we see that this model shows similar behavior to the linear interaction model $Q=3H\rho_{\rm{de}}$, as the interaction also shifts the stable node from complete DE dominance to a dark matter–dark energy hybrid dominance. Similarly, the sign of the interaction constant $\delta$ determines whether the future attractor has negative energies in the DM component. \textbf{For the iDEDM regime ($\delta>0$), all energies remain positive, while in the iDMDE regime ($\delta<0$), the DM density is negative in the future ($\Omega_{\rm{dm}}<0$).} On the contrary, DE remains positive at all times, regardless of the choice of parameters. This is a direct consequence of the fact that as $\rho_{\rm{de}}\rightarrow 0$, then 
$Q=3H \delta \left(\frac{\rho^2_{\rm{de}}}{\rho_{\rm{dm}}+\rho_{\rm{de}}} \right)\rightarrow 0$, 
and the energy flow stops before negative energies can be reached.  
The boundary analysis for this model shows that there is an invariant line at the boundary $\Omega_{\rm{de}}=0$. When we substitute the condition $\Omega_{\rm{de}}=0$ into $\Omega'_{\rm{de}}$ found in the dynamical system \eqref{DSA.QNLde.1}, we obtain $\Omega'_{\rm{de}}=0$, which implies that the flow lines cannot cross into negative ${\Omega}_{\rm{de}}$, ensuring positive DE densities.
For DM to remain positive, we may notice in Figure~\ref{fig:2D_QNLde_phase_portraits} that there is an invariant submanifold between the origin and the future attractor. Trajectories below this line (within the indicated "Positive energy trajectories") remain positive, while those above this line cross the zero-energy border and become negative. The straight line in the $({\Omega}_{\rm{dm}},{\Omega}_{\rm{de}})$ plane connecting the origin $({\Omega}_{\rm{dm}},{\Omega}_{\rm{de}})=(0,0)$ and $P_{\text{dm+\textbf{de}}}$ $({\Omega}_{\rm{dm}},{\Omega}_{\rm{de}})=(\tfrac{\delta}{\delta -w},-\tfrac{w}{\delta -w})$ can be described by the straight line equation: 
\begin{gather} \label{DSA.QNLde.9}
\begin{split}
\Omega_{\rm{de}}&=m \Omega_{\rm{dm}} +c  \quad ; \quad m=\frac{\Delta \Omega_{\rm{de}}}{\Delta \Omega_{\rm{dm}}}=\frac{-\frac{w}{\delta -w}-0}{\frac{\delta}{\delta -w}-0}= -\frac{w}{\delta}, \quad c=0, \\
\Omega_{\rm{de}}&=-\frac{w}{\delta} \, \Omega_{\rm{dm}}   \quad \rightarrow \quad \Omega_{\rm{de}}-\frac{w}{\delta} \, \Omega_{\rm{dm}} =0.
\end{split}
\end{gather} 
The slope of the line \eqref{DSA.QNLde.9} will be positive $m>0$ as long as $\delta>0$, consistent with the constraint previously derived in \eqref{DSA.QNLde.8}. The positive slope ensures that the line lies in the positive DE domain ($\Omega_{\rm{de}}>0$).  
To show that the line described by \eqref{DSA.QNLde.9} is an invariant submanifold, we can substitute the expressions for $\Omega'_{\rm{de}}$ and $\Omega'_{\rm{dm}}$ from \eqref{DSA.QNLde.1} into \eqref{DSA.QNLde.9}, and verify that $\Omega'_{\rm{de}}+\tfrac{w}{\delta}\Omega'_{\rm{dm}}=0$ on the line where $\Omega_{\rm{de}}=-\tfrac{w}{\delta} \Omega_{\rm{dm}}$.

To find constraints on the initial coordinates $({\Omega}_{\rm{(dm,0)}},{\Omega}_{\rm{(de,0)}})$ that ensure positive energy, we require these coordinates to lie below the invariant line described in \eqref{DSA.QNLde.9}, such that:
\begin{gather} \label{DSA.QNLde.10}
\begin{split}
\Omega_{\rm{(de,0)}}&\le -\frac{w}{\delta} \Omega_{\rm{(dm,0)}} \quad \rightarrow \quad
\delta \le -w r_0.
\end{split}
\end{gather}

The condition \eqref{DSA.QNLde.10} may now be combined with the previously derived constraint for positive critical points in \eqref{DSA.QNLde.8}, from which we obtain the final set of conditions to ensure that all components have positive energies at all points in the cosmological evolution:
\begin{gather} \label{DSA.QNLde.11}
\begin{split}
\boxed{\underline{\text{Conditions for $\Omega_{\rm{dm}}\ge0 \; ; \; \Omega_{\rm{de}} \ge0 $ at all points in cosmological evolution:}} \; \text{iDEDM with }  
0<\delta<-w r_0}
\end{split}
\end{gather} 
The impact of different parameter choices on the DM and DE densities in both the past and future expansion is summarized in Table~\ref{tab:QNLde_stability_criteria}. The behavior of the system at each critical point is given in Table~\ref{tab:CP_B_QNL_dm}. 

\begin{table}[h]
    \centering
    \renewcommand{\arraystretch}{1.4} 
    \begin{tabular}{|c|c|c|c|c|c|c|}
        \hline
        \textbf{Conditions} & Energy flow & $\rho_{\text{dm}}$ (Past) & $\rho_{\text{dm}}$ (Future) & $\rho_{\text{de}}$ (Past) & $\rho_{\text{de}}$ (Future) & \textbf{Physical} \\
        \hline         \hline
        $ 0\le\delta \le - w r_0  $ & DE $\to$ DM & $+$ & $+$ & $+$ & $+$ & $\checkmark$ \\
        \hline
        $\delta < 0$ & DM $\to$ DE & $+$ & $-$ & $+$ & $+$  & \textbf{X} \\
        \hline
         $\delta >  - w r_0 $ & DE $\to$ DM & $-$ & $+$ & $+$ & $+$ & \textbf{X} \\
        \hline
    \end{tabular}
    \caption{{Conditions for positive energy densities throughout cosmic evolution, with $w<0$} ($Q=3H \delta \left(\tfrac{\rho^2_{\rm{de}}}{\rho_{\rm{dm}}+\rho_{\rm{de}}} \right)$).}
    \label{tab:QNLde_energy_conditions}
\end{table}

\begin{table}[H]
\centering
\renewcommand{\arraystretch}{1.5} 
\setlength{\tabcolsep}{10pt} 
\begin{tabular}{|c|c|c|}
\hline
\textbf{} & $P_{\text{m}}$ & $P_{\text{dm+\textbf{de}}}$ \\ \hline
\textbf{Class} &  Saddle Point & Stable node (sink) \\ \hline
$\Omega_{\rm{r}}$ & $0$ & $0$ \\ \hline
$\Omega_{\rm{bm}}$& $-\Omega_{\rm{dm}}+1$  & $0$\\ \hline
$\Omega_{\rm{dm}}$  & $\Omega_{\rm{dm}}$  & $\frac{\delta}{\delta-w}$\\ \hline
$\Omega_{\rm{de}}$ & $0$ & $- \frac{w}{\delta-w}$ \\ \hline
$\Omega_{\rm{dm}}\ge0$  & $\forall \delta$  &  $\delta\ge0$\\ \hline
$\Omega_{\rm{de}}\ge0$  & $\forall \delta$  &  $\forall \delta$\\ \hline
$r$  & $\infty$ & $-\frac{\delta}{w}$ \\ \hline
$w^{\rm{eff}}_{\rm{dm}}$  & $0$ & $-\frac{w^2}{\delta-w}$ \\ \hline
$w^{\rm{eff}}_{\rm{de}}$ & $w$ & $-\frac{w^2}{\delta-w}$\\ \hline
$\zeta$  & $-3 w$ & $0$ \\ \hline
$w^{\rm{eff}}_{\rm{tot}}$  & $0$ & $-\frac{w^2}{\delta-w}$\\ \hline
$q$ & $\frac{1}{2}$ & $\frac{1}{2}(1-3 \frac{w^2}{\delta-w})$ \\ \hline
\end{tabular}
\caption{behavior of model at critical points – $Q=3H \delta \left(\tfrac{\rho^2_{\rm{de}}}{\rho_{\rm{dm}}+\rho_{\rm{de}}} \right)$.}
\label{tab:CP_B_QNL_de}
\end{table}

In Table~\ref{tab:CP_B_QNL_de}, we see that for $P_{\text{m}}$ the system behaves identically to the corresponding matter-dominated critical point in the non-interacting $w$CDM model, indicating that the effect of the interaction vanishes in the past as the DE fractional density approaches zero. 
In the future, during the dark matter–dark energy hybrid dominance ($P_{\text{dm+\textbf{de}}}$), DE and DM redshift at the same rate $w^{\rm{eff}}_{\rm{dm}}=w^{\rm{eff}}_{\rm{de}}$, which implies a fixed ratio $r=-\left(\tfrac{\delta}{w}\right)$ and $\zeta=0$, \textit{solving the coincidence problem in the future}. This model will also exhibit accelerated expansion in the late-time attractor, provided that $\delta<w(3w+1)$.  
Finally, if the DE equation of state lies in the phantom regime $w<-1$, \textit{a future big rip may still be avoided in the iDEDM regime if $\delta$ is sufficiently positive}, such that $w^{\rm{eff}}_{\rm{tot}}\ge-1$, which is obtained if $\delta\ge w(1+w)$. We may now consider the stability of this system by looking at the sign of the doom factor for this interaction:
\begin{gather} \label{DSA.QNLde.doom}
\begin{split}
\textbf{d}=  \frac{Q}{3H\rho_{\rm{de}}(1+w)}=\frac{3H \delta \left(\frac{\rho^2_{\rm{de}}}{\rho_{\rm{dm}}+\rho_{\rm{de}}} \right)}{3H\rho_{\rm{de}}(1+w)}=\frac{\delta}{(1+w)} \left(\frac{\rho_{\rm{de}}}{\rho_{\rm{dm}}+\rho_{\rm{de}}} \right),
\end{split}
\end{gather}
where we also apply the conditions $\rho_{\rm{dm}}>0$ and $\rho_{\rm{de}}>0$, so the terms in brackets remain positive.
Since we need $\textbf{d} < 0$ to ensure a stable universe, we see from \eqref{DSA.QNLde.doom} that this will only occur if $\delta$ and $(1+w)$ have opposite signs. Since we require $\delta>0$ for positive energy \eqref{DSA.QNLde.11}, this implies that $(1+w)>0$ is needed for a priori stability, corresponding to $w<-1$, which means \textit{DE must be in the phantom regime}. Different combinations of parameters and their effects on energy density and stability are summarized in Table~\ref{tab:QNLde_stability_criteria}.

\begin{table}[h]
    \centering
    \renewcommand{\arraystretch}{1.4} 
    \begin{tabular}{|c|c|c|c|c|c|c|c|c|}
        \hline 
        $\delta$ & Energy flow & $w$ & Dark energy & $d$ & a priori stable & $\rho_{\text{dm}} > 0$ & $\rho_{\text{de}} > 0$ & \textbf{Viable} \\
        \hline \hline
        $+$ & DE $\to$ DM & $< -1$ & Phantom & - & $\checkmark$ & $\checkmark$ & $\checkmark$ & $\checkmark$ \\
        \hline
        $+$ & DE $\to$ DM & $> -1$ & Quintessence & + & X & $\checkmark$ & $\checkmark$ & \textbf{X} \\
        \hline
        $-$ & DM $\to$ DE & $< -1$ & Phantom & + & X & X  & $\checkmark$ & \textbf{X}  \\
        \hline
        $-$ & DM $\to$ DE & $> -1$ & Quintessence & - & $\checkmark$ & X  &  $\checkmark$ & \textbf{X} \\
        \hline
    \end{tabular}
    \caption{{Stability and positive energy criteria – $Q=3H \delta \left(\tfrac{\rho^2_{\rm{de}}}{\rho_{\rm{dm}}+\rho_{\rm{de}}} \right)$.}}
    \label{tab:QNLde_stability_criteria}
\end{table}

\section{Background cosmology for each interaction kernel} \label{Background_cosmology}

All the results in this section are obtained using the same methods applied in Section 4 of our companion paper \cite{vanderWesthuizen:2025I}, now using the new analytical solutions for the energy densities $\rho_{\text{dm}}$ and $\rho_{\text{de}}$. 
The results from the analytical solutions are consistent with those from the dynamical system analysis in Section~\ref{Sec.DSA}, and in both cases, when $\delta=0$ and $w=-1$, the relevant expressions for the $\Lambda$CDM model are recovered, thus validating both approaches.

\subsection{Non-linear IDE model 1: $Q_1=3\delta H  
\left(\frac{\rho_{\text{dm}}\rho_{\text{de}}}{\rho_{\text{dm}}+\rho_{\text{de}}} \right)$} \label{analytical.dmde}

\begin{figure}[H]
    \centering
    \includegraphics[width=0.95\linewidth]{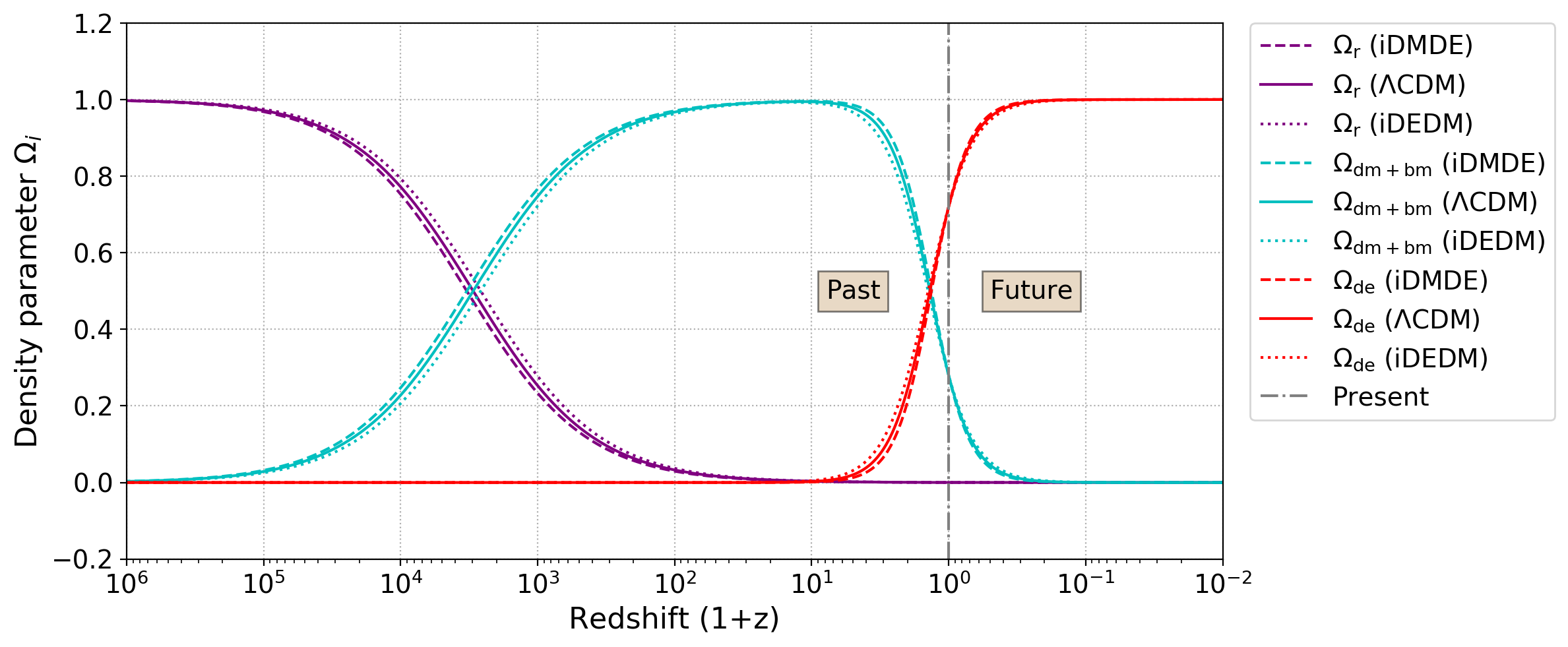}
    \caption{Density parameters vs. redshift – $Q=3\delta H \left(\tfrac{\rho_{\text{dm}}\rho_{\text{de}}}{\rho_{\text{dm}}+\rho_{\text{de}}} \right)$, with positive DM and DE densities always guaranteed in both the iDEDM regime ($\delta=+0.1$) and the iDMDE regime ($\delta=-0.1$).}   
    \label{fig:Omega_NLID1}
\end{figure}

This interaction changes the dynamics in both the past and future during DM and DE domination, respectively. Although the model approaches the non-interacting case $Q=0$ in both the asymptotic past (when $\rho_{\text{de}}\rightarrow0$) and future (when $\rho_{\text{dm}}\rightarrow0$), as shown in Figure~\ref{fig:Q_nonLinear}, the interaction has a distinct effect in intermediate epochs. Furthermore, since $Q=0$ whenever either $\rho_{\text{dm}}=0$ or $\rho_{\text{de}}=0$, both the DM and DE densities remain positive for any choice of parameters. These two key features of the model are illustrated in Figure~\ref{fig:Omega_NLID1}. 
The DM and DE densities for this model were derived in~\cite{Arevalo:2011hh} and are given by:  
\begin{equation}  
\begin{split} 
\rho_{\text{dm}} &= \rho_{\text{(dm,0)}}  a^{-3(1-\delta)}     
\left[\frac{1+\left( \frac{  \rho_{\text{(dm,0)}}}{\rho_{\text{(de,0)}}} \right) a^{3\left[ w + \delta\right] }}
{1+\left( \frac{  \rho_{\text{(dm,0)}}}{\rho_{\text{(de,0)}}} \right)} \right]^{-\frac{\delta}{(w +\delta) }}   , \\
\rho_{\text{de}} &=  \rho_{\text{(de,0)}}  a^{-3(1+w)}     
\left[\frac{1+\left( \frac{  \rho_{\text{(dm,0)}}}{\rho_{\text{(de,0)}}} \right) a^{3\left[ w + \delta\right] }}
{1+\left( \frac{  \rho_{\text{(dm,0)}}}{\rho_{\text{(de,0)}}} \right)} \right]^{-\frac{\delta}{(w +\delta) }},
\label{NLID1_dm_de_BG}
\end{split} 
\end{equation}
where $\delta\neq -w$ to avoid division by zero in the exponent.

\begin{gather} \label{Positive_energy.Q.dmde}
\begin{split}
\boxed{\underline{\text{Conditions for} \; \rho_{\rm{dm}}\ge0 \; ; \; \rho_{\rm{de}} \ge0 \; \text{at all points in cosmological evolution:}} \;  
\forall\delta \ }.
\end{split}
\end{gather} 
The fractional densities of both DM and DE converge asymptotically in the past and future (noting that in the presence of baryons there will be hybrid DM–baryon domination) to the following expressions:
\begin{equation}
\begin{split}
\Omega_{\text{(dm,past)}} =1 \quad &; \quad \Omega_{\text{(de,past)}}=0, \\
\Omega_{\text{(dm,future)}} =0 \quad &; \quad \Omega_{\text{(de,future)}}=1.
\end{split}\label{eq:frac_den_NLID_1}
\end{equation}
The dark matter–dark energy equality occurs at the redshift where $\rho_{\text{dm}}=\rho_{\text{de}}$:
\begin{equation}
\begin{split} 
z _{\text{(dm=de)}} &= \left[ \frac{\Omega_{\text{(de,0)}}}{\Omega_{\text{(dm,0)}}}  \right]^{-\frac{1}{3(\delta+w)}} -1.
\label{NLID1_dm=de_BG}
\end{split} 
\end{equation}
\begin{figure}[htbp]
    \centering
    \begin{subfigure}[b]{0.502\linewidth}
        \centering
        \includegraphics[width=\linewidth]{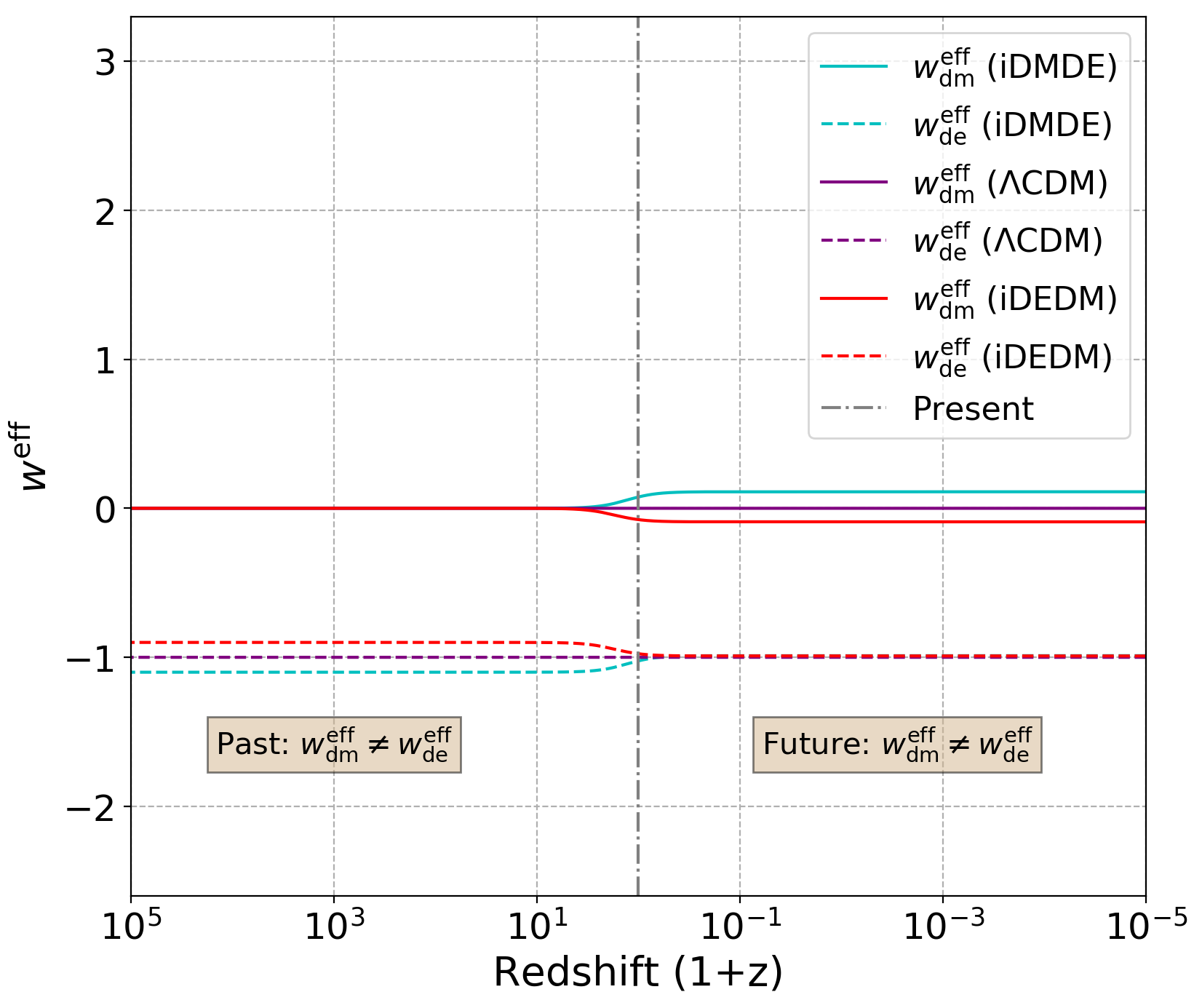}
        \label{fig:omega_dmde_NLID1}
    \end{subfigure}    
    \hspace{0pt} 
    \begin{subfigure}[b]{0.483\linewidth}
        \centering
        \includegraphics[width=\linewidth]{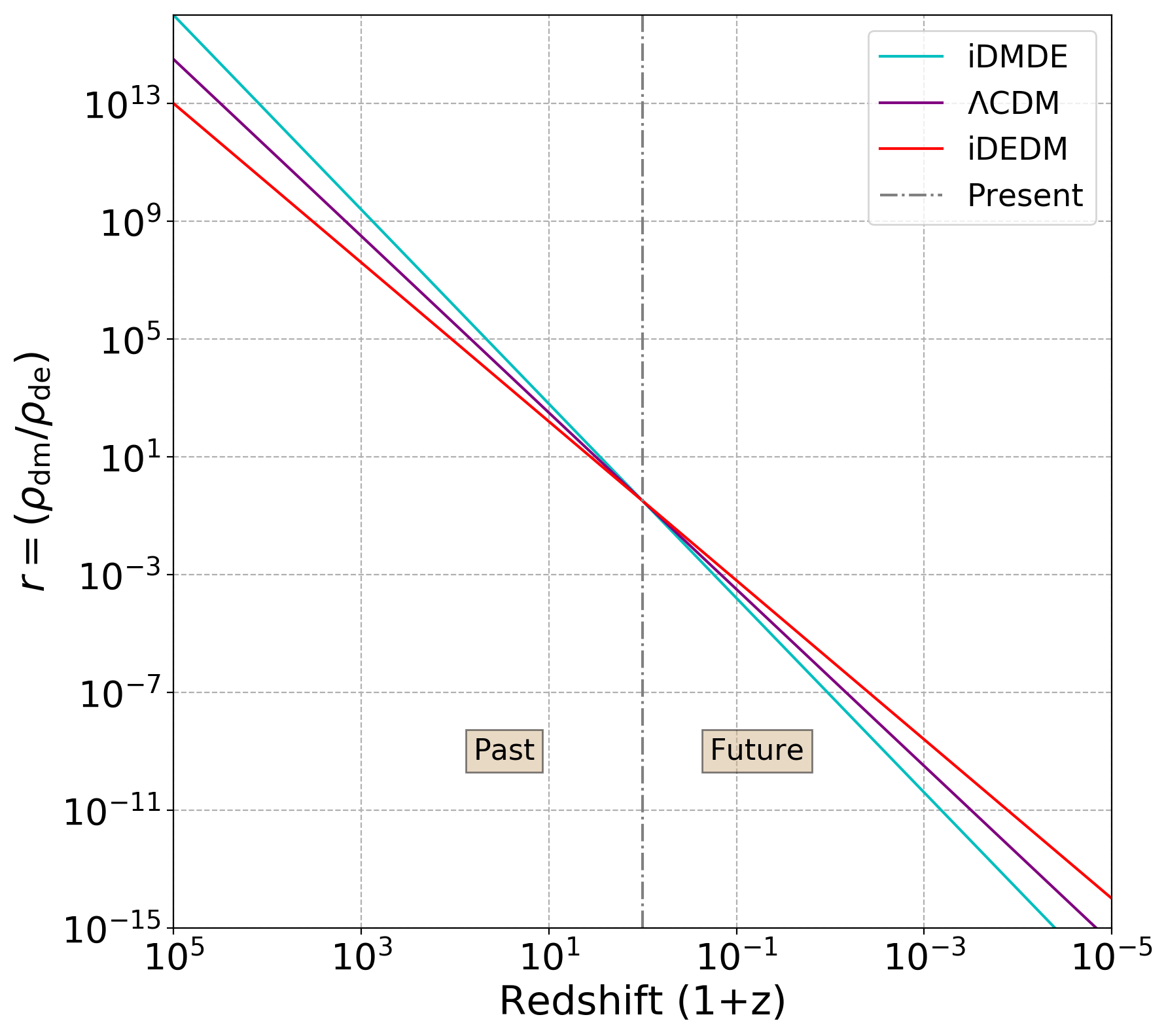}
        \label{fig:CP_NLID1}
    \end{subfigure}%
    \caption{Effective equations of state and Coincidence Problem (CP) vs. redshift – $Q=3\delta H \left(\tfrac{\rho_{\text{dm}}\rho_{\text{de}}}{\rho_{\text{dm}}+\rho_{\text{de}}} \right)$. In the iDEDM regime ($\delta=+0.1$), the CP is alleviated, but not solved (since $w^{\rm{eff}}_{\rm{dm}} \neq w^{\rm{eff}}_{\rm{de}}$), in both the past and the future, while it is worsened in the iDMDE regime ($\delta=-0.1$).}
    \label{fig:CP+omega_dmde_NLID1}
\end{figure}
The DM-to-DE ratio $r$ converges to the following expressions in the past and future:
\begin{equation}
\begin{split}
r &=  r_0 a^{3 \left( w + \delta\right)} \quad ; \quad
r_{\text{past}} (a\rightarrow0)=\frac{\rho_{\text{dm}}}{\rho_{\text{de}}} \approx  \infty \quad ; \quad 
r_{\text{future}} (a\rightarrow\infty)=\frac{\rho_{\text{dm}}}{\rho_{\text{de}}} \approx 0.
\end{split}%
\label{NLID1_r_PF}
\end{equation}
The DM and DE effective equations of state \eqref{DSA.H2} for this interaction are:
\begin{gather} \label{NLID1_omega_eff_dm_de_BG}
\begin{split}
w^{\rm{eff}}_{\rm{dm}} = -\delta \left(\frac{1}{r+1}\right), 
\quad 
w^{\rm{eff}}_{\rm{de}} = w + \delta \left(\frac{1}{1+\tfrac{1}{r}}\right).
\end{split}
\end{gather}
Substituting \eqref{NLID1_r_PF} into \eqref{NLID1_omega_eff_dm_de_BG} gives $w^{\rm{eff}}_{\rm{dm}}$ and $w^{\rm{eff}}_{\rm{de}}$ in the asymptotic past and future:
\begin{equation}
\begin{split}
w^{\rm{eff}}_{\rm{(dm,past)}}  &= 0, 
\; \;w^{\rm{eff}}_{\rm{(de,past)}}   = w+\delta,  \quad \rightarrow \quad
\zeta_{\text{past}} \approx -3(w + \delta) \; \text{(alleviates coincidence problem)}. \\
w^{\rm{eff}}_{\rm{(dm,future)}}  &= \delta, 
\; \;w^{\rm{eff}}_{\rm{(de,future)}}   = w, \; \; \quad \rightarrow \quad \zeta_{\text{future}} \approx -3(w + \delta) \; \text{(alleviates coincidence problem)}. \\
\end{split}%
\label{eq:omega_eff_dm_de_NLID1_subbed_past_future}
\end{equation}
This interaction model therefore alleviates the coincidence problem in both the past and the future as long as $\delta>0$ (iDEDM regime), while it worsens the coincidence problem if $\delta<0$ (iDMDE). This behavior is also shown in Figure~\ref{fig:CP+omega_dmde_NLID1}.

\begin{figure}[htbp]    \centering
\begin{subfigure}[b]{0.515\linewidth}
        \centering
        \includegraphics[width=\linewidth]{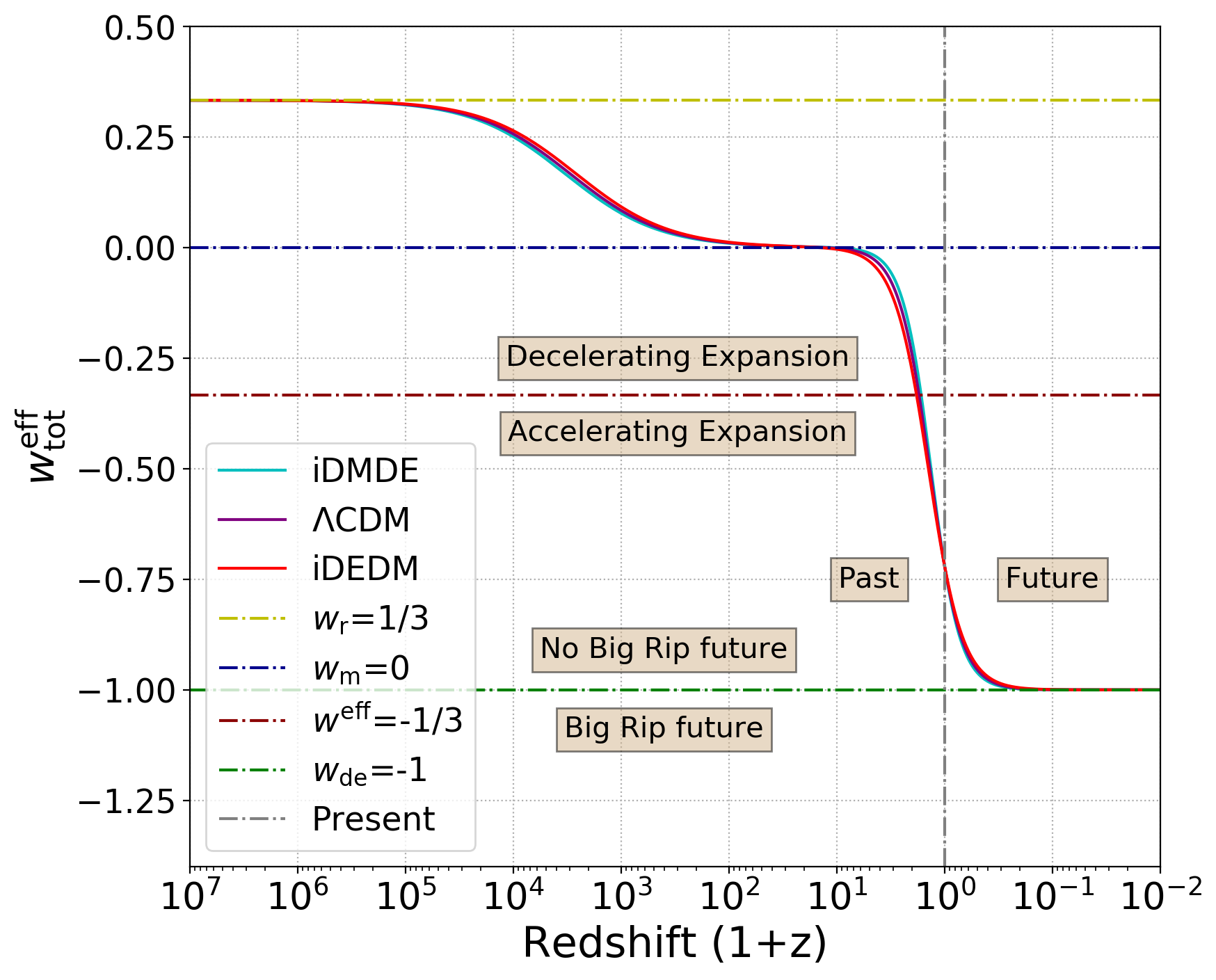}
        \label{q+eos_tot_NLID1}
    \end{subfigure}    
    \begin{subfigure}[b]{0.475\linewidth}
        \centering
        \includegraphics[width=\linewidth]{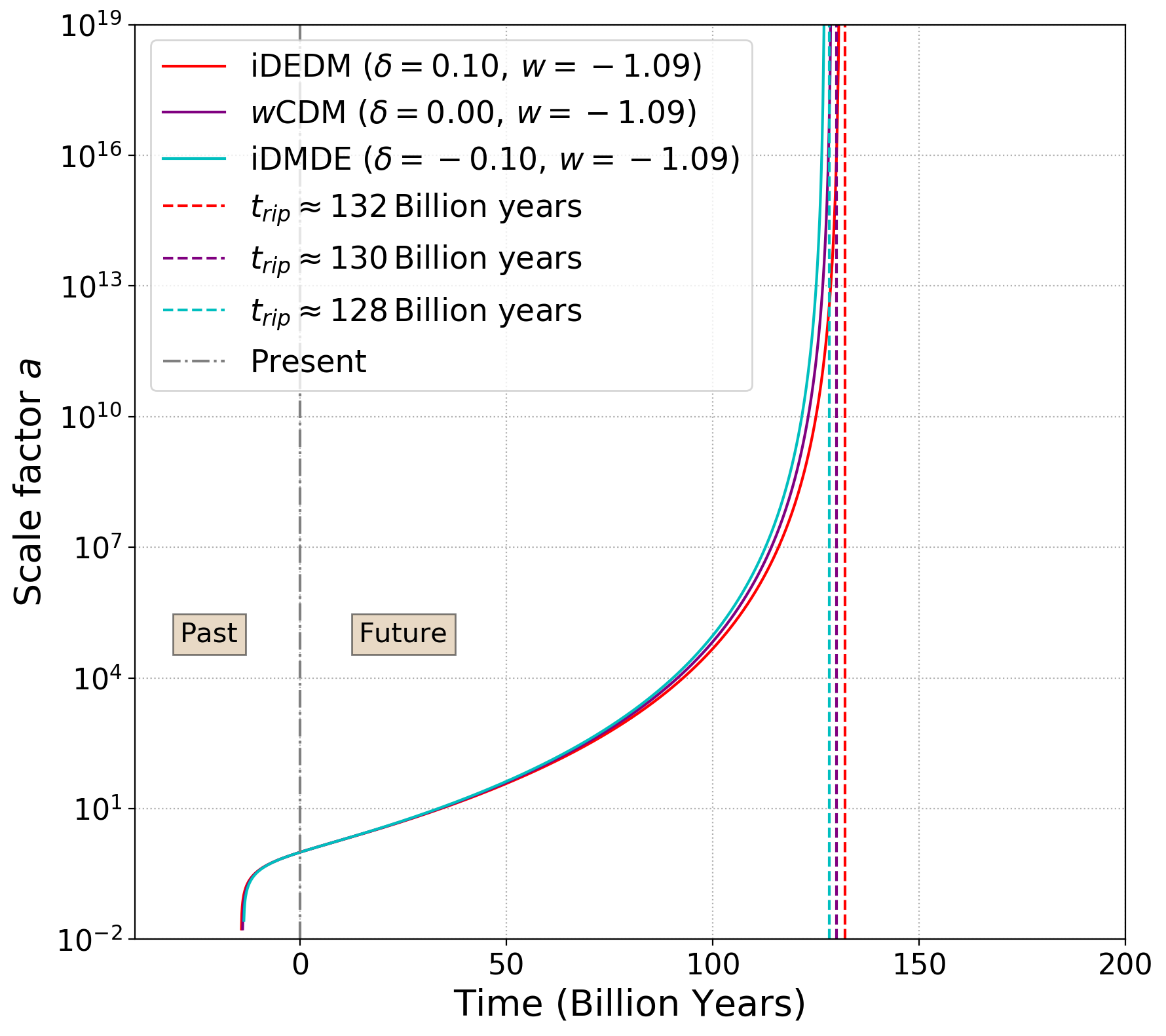}
        \label{fig:Big_rip_NLID1}
    \end{subfigure}    
    \hspace{0pt} 
    \caption{Total effective equation of state $w^{\rm{eff}}_{\rm{tot}}$ and big rip future singularities — $Q=3\delta H \left(\frac{\rho_{\text{dm}}\rho_{\text{de}}}{\rho_{\text{dm}}+\rho_{\text{de}}} \right)$, with $w=-1$ (left panel) and $w=-1.09$ (right panel). In both the iDEDM regime ($\delta=+0.1$) and the iDMDE regime ($\delta=-0.1$), the effect of the interaction diminishes in the asymptotic future, leading to $w^{\rm{eff}}_{\rm{tot}}=w$. This implies that $w^{\rm{eff}}_{\rm{tot}}<-1$ whenever $w<-1$, thereby guaranteeing a future big rip singularity in both cases.}
    \label{fig:eos_tot_BR_NLID1}
\end{figure}

The DE phantom crossing (pc) for this interaction model can only occur if $(1+w)$ and $\delta$ have opposite signs. The predicted redshift at which this happens is obtained by setting $w^{\text{eff}}_{\text{de}}=-1$ in \eqref{NLID1_omega_eff_dm_de_BG}:
\begin{gather} \label{NLID1_phantom_crossing_BG}
\begin{split}
z_{\text{pc}} &=\left[-\frac{\Omega_{\text{(dm,0)}}}{\Omega_{\text{(de,0)}}} \left(\frac{\delta}{1 +w}+1\right)  \right]^{\tfrac{1}{3 \left( w + \delta\right)}} -1.
\end{split}
\end{gather}
As seen from \eqref{eq:omega_eff_dm_de_NLID1_subbed_past_future} and Figure~\ref{fig:CP+omega_dmde_NLID1}, for given values of $w$ and $\delta$, there are two possible directions for the phantom crossing (pc), with both cases maintaining positive energies:
\begin{equation}
\begin{split}
\text{pc direction} \begin{cases}
\text{iDMDE: } \text{Phantom }   (w^{\rm{eff}}_{\rm{(de,past)}}<-1) \;\;\rightarrow\;\; \text{Quintessence }   (w^{\rm{eff}}_{\rm{(de,future)}}>-1), \;\; \rho_{\rm{dm/de}}>0.  \\
\text{iDEDM: } \text{Quintessence } (w^{\rm{eff}}_{\rm{(de,past)}}>-1) \;\;\rightarrow\;\; \text{Phantom }   (w^{\rm{eff}}_{\rm{(de,future)}}<-1), \;\; \rho_{\rm{dm/de}}>0.
\end{cases}
\end{split}\label{eq:z_pc_dmde_direction}
\end{equation}
We also have $q=\tfrac{1}{2} \left(1+3w^{\rm{eff}}_{\rm{tot}} \right)$ and $w^{\rm{eff}}_{\rm{tot}}=w^{\rm{eff}}_{\rm{de}}=w$ in the distant future, as seen in Figure~\ref{fig:eos_tot_BR_NLID1}, which leads to the following condition for a big rip to occur:
\begin{gather} \label{omega_eff_tot_NLID1_BG}
\begin{split}
\text{Big rip condition: \quad } w^{\rm{eff}}_{\rm{(tot,future)}} &\approx w <-1.
\end{split}
\end{gather}
This implies that both the deceleration parameter and the effective equations of state will asymptotically match those of any uncoupled model in the distant future, as seen in Figure~\ref{fig:eos_tot_BR_NLID1}. In the case where $w^{\rm{eff}}_{\rm{tot}}=w<-1$, the universe will experience a future big rip singularity at time $t_{\text{rip}}$:
\begin{equation}
\begin{split}
t_{\text{rip}}-t_0 &\approx - \frac{2}{3H_0(1+w) \sqrt{\Omega_{\text{(de,0)}} \left(1+\frac{\Omega_{\text{(dm,0)}}}{\Omega_{\text{(de,0)}}}\right)^{\tfrac{\delta}{(w +\delta) }}}}   
\end{split}\label{eq:Big_Rip_general_NLID1_BG}
\end{equation}
The effect of the coupling on $w^{\rm{eff}}_{\rm{tot}}$ and how the interaction can either cause or prevent a big rip can be seen in Figure~\ref{fig:eos_tot_BR_NLID1}. The time of the big rip predicted by \eqref{eq:Big_Rip_general_NLID1_BG} is indicated by the dashed lines in Figure~\ref{fig:eos_tot_BR_NLID1}, which agrees with the point where the scale factor diverges $a \rightarrow \infty$ within a finite time. We note that for phantom DE ($w<-1$), a big rip is inevitable.

\subsection{Non-linear IDE model 2: $Q_2=3\delta H  
\left(\frac{\rho_{\text{dm}}^2}{\rho_{\text{dm}}+\rho_{\text{de}}} \right)$} \label{analytical.dmdm}

Similar to the kernel $Q=3H\delta \rho_{\text{dm}}$, this interaction primarily changes the dynamics in the distant past during DM domination, while having a smaller effect at late times and during DE domination. For this interaction, $Q=0$ when $\rho_{\text{dm}}=0$, which guarantees that $\rho_{\text{dm}}\ge 0$ at all times. In contrast, this interaction leads to past negative DE densities in the iDMDE regime, but this can be avoided in the iDEDM regime if the interaction is sufficiently small, as given by condition~\ref{Positive_energy.Q.NLID2} and illustrated in Figure~\ref{fig:Omega_NLID2}. The DM and DE densities for this model are derived in Appendix~\ref{Appendix_A} and given by:
\begin{equation}
\begin{split} 
 \rho_{\text{dm}} &=   \rho_{\text{(dm,0)}}  a^{-3\left(1- \frac{w\delta }{w -\delta} \right)} \left[\frac{\left[w+\delta \left( \frac{  \rho_{\text{(dm,0)}}}{\rho_{\text{(de,0)}}} \right) \right]a^{-3w } + \left( \frac{  \rho_{\text{(dm,0)}}}{\rho_{\text{(de,0)}}} \right)(w -\delta) }{ w\left[1+\left( \frac{  \rho_{\text{(dm,0)}}}{\rho_{\text{(de,0)}}} \right) \right]}    \right]^{\frac{\delta }{w -\delta}}, \\
 \rho_{\text{de}} &=   \rho_{\text{(de,0)}}   a^{-3\left(1- \frac{w\delta }{w -\delta}  \right)} \left(\frac{\left[w+\delta \left( \frac{  \rho_{\text{(dm,0)}}}{\rho_{\text{(de,0)}}} \right) \right]a^{-3w} -\delta \left( \frac{  \rho_{\text{(dm,0)}}}{\rho_{\text{(de,0)}}} \right)} {w } \right) \left[\frac{\left[w+\delta \left( \frac{  \rho_{\text{(dm,0)}}}{\rho_{\text{(de,0)}}} \right) \right]a^{-3w } + \left( \frac{  \rho_{\text{(dm,0)}}}{\rho_{\text{(de,0)}}} \right)(w -\delta) }{ w\left[1+\left( \frac{  \rho_{\text{(dm,0)}}}{\rho_{\text{(de,0)}}} \right) \right]}    \right]^{\frac{\delta }{w -\delta}},
\label{NLID2_dm_de_BG}
\end{split} 
\end{equation}
where $\delta\neq w$ to avoid divisions by zero. If the power $\left(\frac{\delta }{w -\delta}\right)<0$, we require the additional conditions $w<0$ and $w<\delta\le -\frac{w}{r_0}$ to ensure that no divisions by zero occur for any scale factor $a$.

\begin{figure}[H]
    \centering
    \includegraphics[width=0.95\linewidth]{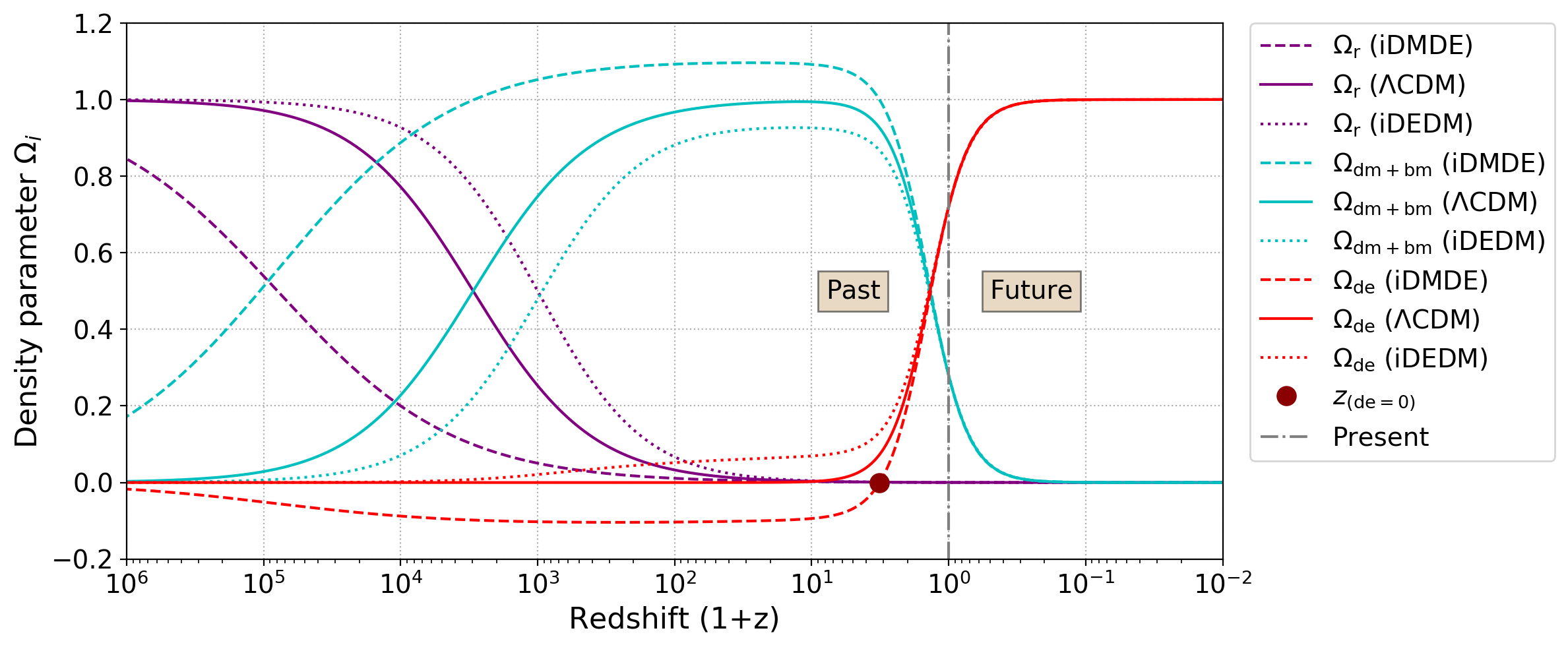}
    \caption{Density parameters vs redshift – $Q=3\delta H  \left(\frac{\rho_{\text{dm}}^2}{\rho_{\text{dm}}+\rho_{\text{de}}} \right)$, with positive energy densities found only in the iDEDM regime ($\delta=+0.1$), while negative DE densities (in the past) are always present in the iDMDE regime ($\delta=-0.1$).}
    \label{fig:Omega_NLID2}
\end{figure}

\begin{gather} \label{Positive_energy.Q.NLID2}
\begin{split}
\boxed{\underline{\text{Conditions for}  \text{ $\rho_{\rm{dm}}\ge0 \; ; \; \rho_{\rm{de}} \ge0 $ at all points in cosmological evolution:}} \; \text{ iDEDM with }  
0\le\delta \le -\frac{w}{r_0} }.
\end{split}
\end{gather} 
The fractional densities of DM and DE converge asymptotically in both the past and the future (noting that in the presence of baryons, a hybrid DM–baryon dominated phase arises) to the following expressions:
\begin{equation}
\begin{split}
\Omega_{\text{(dm,past)}} =\left( \frac{w}{w-\delta} \right) \quad &; \quad \Omega_{\text{(de,past)}}=-\left( \frac{\delta}{w-\delta} \right)\\
\Omega_{\text{(dm,future)}} =0 \quad &; \quad \Omega_{\text{(de,future)}}=1.
\end{split}\label{eq:frac_den_NLID2}
\end{equation}
For this model, the DM density will always remain positive, but DE may become negative at the redshift:
\begin{equation}
\begin{split} \label{NLID2_de=0_z_BG}
z _{\text{(de=0)}}= \left[\frac{\delta }{w \left( \frac{  \Omega_{\text{(de,0)}}}{\Omega_{\text{(dm,0)}}} \right) + \delta } \right]^{\frac{1}{{3w}}} -1.
\end{split} 
\end{equation} 
The dark matter–dark energy equality occurs at the redshift where $\rho_{\text{dm}}=\rho_{\text{de}}$:
\begin{equation}
\begin{split} 
z _{\text{(dm=de)}} &= \left[ \frac{ w+ \delta   }{  w\left( \frac{  \Omega_{\text{(de,0)}}}{\Omega_{\text{(dm,0)}}} \right) + \delta  } \right]^{\frac{1}{3w}} -1. 
\label{NLID2_dm=de_BG}
\end{split} 
\end{equation}

\begin{figure}[htbp]
    \centering
    \begin{subfigure}[b]{0.502\linewidth}
        \centering
        \includegraphics[width=\linewidth]{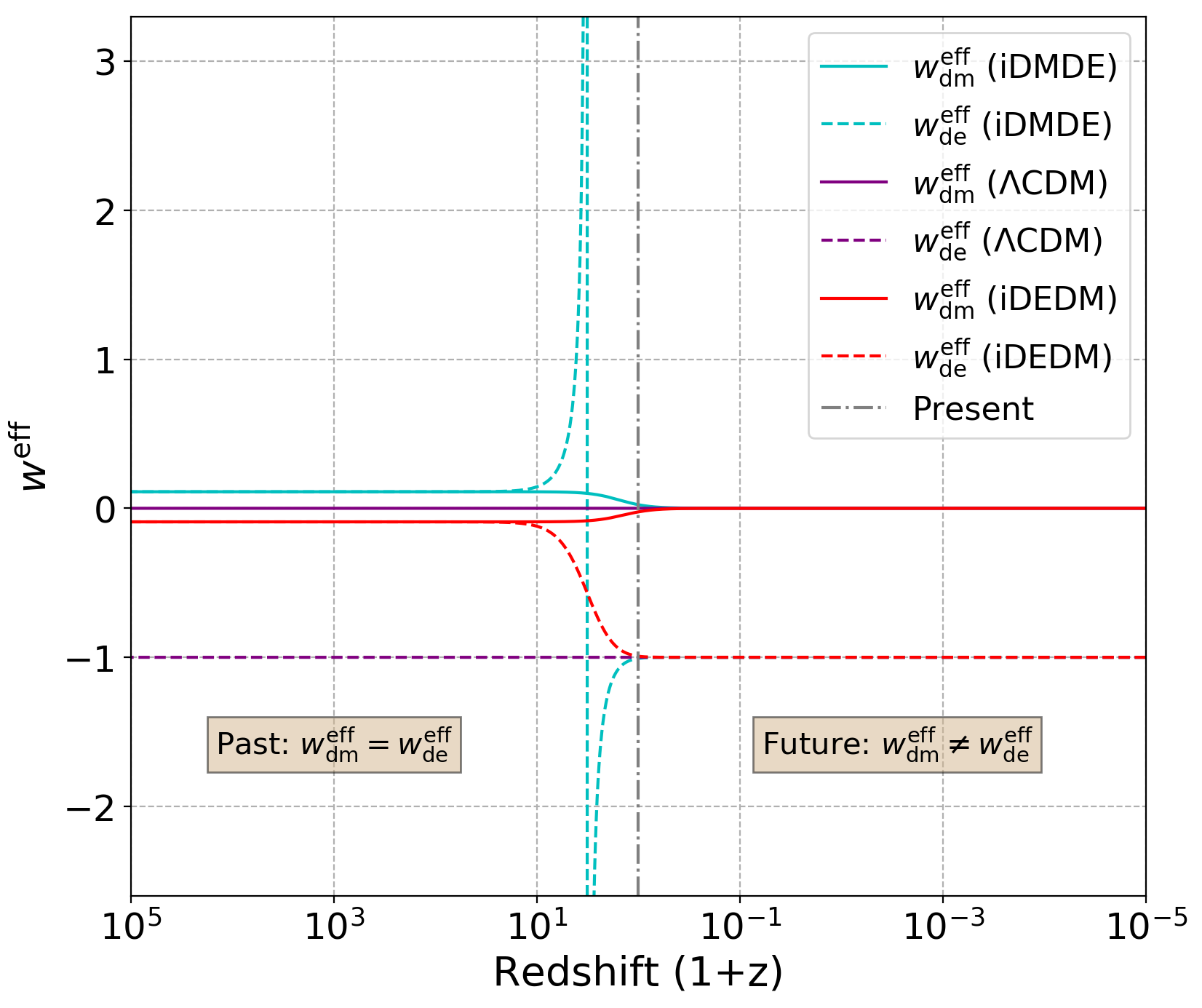}
        \label{fig:omega_dmde_NLID2}
    \end{subfigure}    
    \hspace{0pt} 
    \begin{subfigure}[b]{0.483\linewidth}
        \centering
        \includegraphics[width=\linewidth]{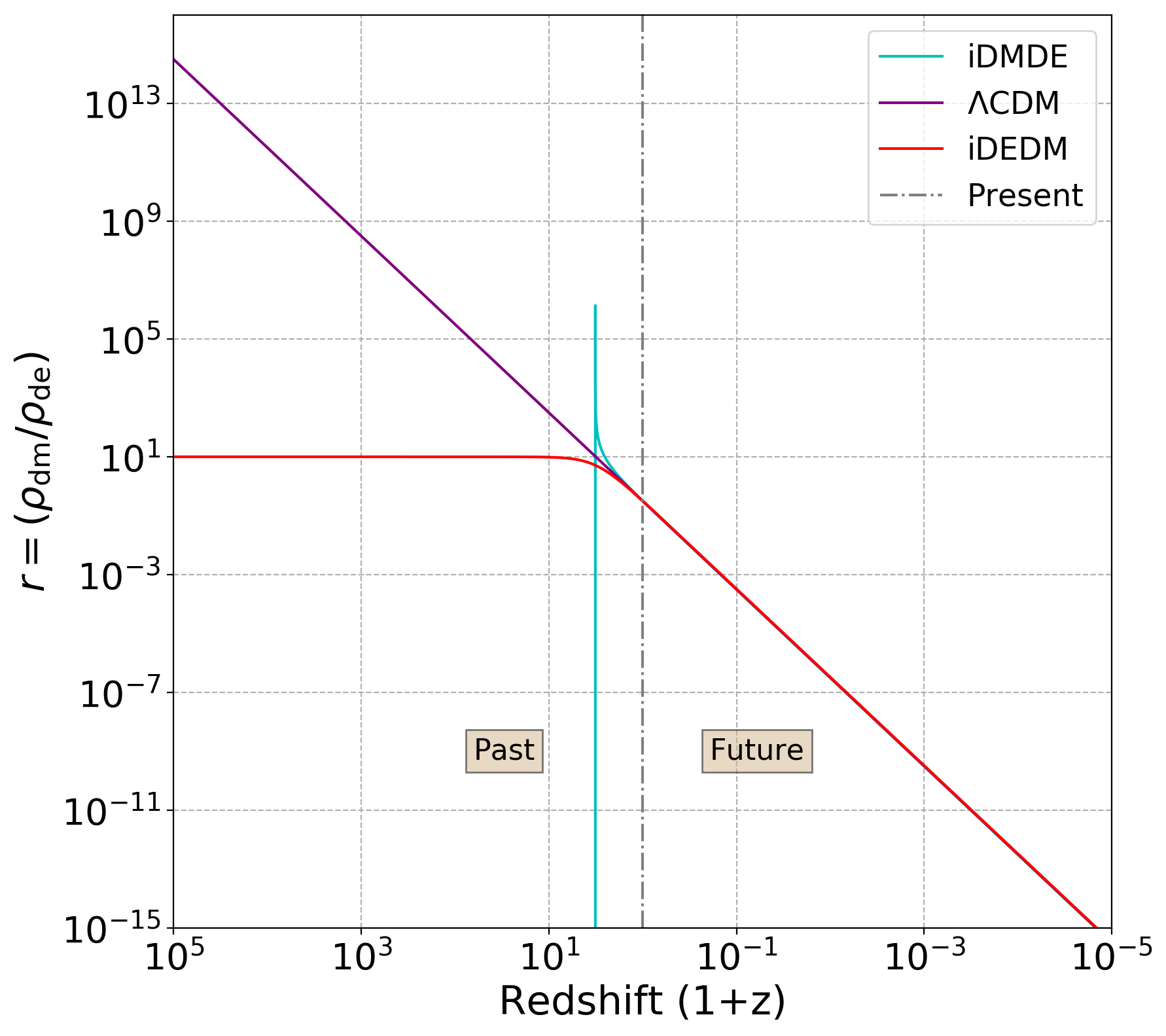}
        \label{fig:CP_NLID2}
    \end{subfigure}%
    \caption{Effective equations of state and Coincidence Problem (CP) vs redshift — $Q=3\delta H  \left(\frac{\rho_{\text{dm}}^2}{\rho_{\text{dm}}+\rho_{\text{de}}} \right)$. In the iDEDM regime ($\delta=+0.1$), $w^{\rm{eff}}_{\rm{dm}}=w^{\rm{eff}}_{\rm{de}}$ in the past, thereby solving the CP ($r=\text{constant}$). In the iDMDE regime ($\delta=-0.1$), negative DE densities and divergent $w^{\rm{eff}}_{\rm{de}}$ appear in the past. As the interaction diminishes in the future, there is no change to the CP at late times.}
    \label{fig:CP+omega_dmde_NLID2}
\end{figure}

The DM to DE ratio $r$ converges to the following expressions in the past and future:
\begin{equation}
\begin{split}
r &= \frac{wr_0 }{\left( w+ \delta r_0\right) a^{-3w} -\delta r_0} 
\quad ; \quad
r_{\text{past}} (a\rightarrow0)=\frac{\rho_{\text{dm}}}{\rho_{\text{de}}} \approx  -\frac{w}{\delta} 
\quad ; \quad 
r_{\text{future}} (a\rightarrow\infty)=\frac{\rho_{\text{dm}}}{\rho_{\text{de}}} \approx 0 .
\end{split}%
\label{NLID2_r_PF}
\end{equation}
The DM and DE effective equations of state from \eqref{DSA.H2} for this interaction are:
\begin{gather} \label{NLID2_omega_eff_dm_de_BG}
\begin{split}
w^{\rm{eff}}_{\rm{dm}} &= -\delta \left(\frac{1}{1+\frac{1}{r}}\right), 
\quad 
w^{\rm{eff}}_{\rm{de}} = w + \delta \left(\frac{r}{1+\frac{1}{r}}\right).
\end{split}
\end{gather}
Substituting \eqref{NLID2_r_PF} into \eqref{NLID2_omega_eff_dm_de_BG} gives $w^{\rm{eff}}_{\rm{dm}}$ and $w^{\rm{eff}}_{\rm{de}}$ in the asymptotic past and future:
\begin{equation}
\begin{split}
w^{\rm{eff}}_{\rm{(dm,past)}} &=w^{\rm{eff}}_{\rm{(de,past)}}  =  \frac{\delta w}{\delta-w}, \quad \rightarrow \quad  \zeta_{\text{past}} = 0 \; \text{(solves the coincidence problem)}.\\
w^{\rm{eff}}_{\rm{(dm,future)}}  &= 0, 
\; \;w^{\rm{eff}}_{\rm{(de,future)}}   = w,  \quad \rightarrow \quad
\zeta_{\text{future}} = -3w \; \text{(no change)}.
\end{split}%
\label{eq:omega_eff_dm_de_NLID2_subbed_past_future}
\end{equation}
This interaction model therefore solves the coincidence problem in the past, while leaving the problem unchanged in the future, as illustrated in Figure~\ref{fig:CP+omega_dmde_NLID2}.

The predicted redshift at which the DE phantom crossing occurs can be obtained by setting $w^{\text{eff}}_{\text{de}}=-1$ in \eqref{NLID2_omega_eff_dm_de_BG}, from which we calculate:
\begin{gather} \label{NLID2_phantom_crossing_BG}
\begin{split}
z_{\text{pc}}  &= \left[ \frac{ \delta }{{w \left( \frac{  \Omega_{\text{(de,0)}}}{\Omega_{\text{(dm,0)}}} \right)+ \delta }}\left(\frac{2 w}{-(1+w) \pm \sqrt{(1+ w)^2-4 \delta(1+w)}}+1  \right) \right]^{\frac{1}{3w}}-1.
\end{split}
\end{gather}

For suitable choices of $w$ and $\delta$, one of the $\pm$ branches will yield a solution with $z>-1$. As seen from \eqref{eq:omega_eff_dm_de_NLID2_subbed_past_future} and Figure~\ref{fig:CP+omega_dmde_NLID2}, the direction of the phantom crossing depends on the regime, with the iDMDE case suffering from divergent behavior and negative energies:
\begin{equation}
\begin{split}
 \text{pc direction} \begin{cases}
   \text{iDMDE: } \text{Divergent pc for  $w^{\rm{eff}}_{\rm{de}} $ at $z_{\text{(de=0)}}$ \eqref{NLID2_de=0_z_BG}, with $\rho_{\rm{de}}<0$.} \\
   \text{iDEDM: } \text{Quintessence } (w^{\rm{eff}}_{\rm{(de,past)}}>-1)  \;\;\rightarrow\;\;   \text{Phantom }   (w^{\rm{eff}}_{\rm{(de,future)}}<-1), \text{ with $\rho_{\rm{dm/de}}>0$.}
\end{cases}
\end{split} \label{eq:z_pc_dmdm_direction}
\end{equation}

\begin{figure}[htbp]    \centering
\begin{subfigure}[b]{0.515\linewidth}
        \centering
        \includegraphics[width=\linewidth]{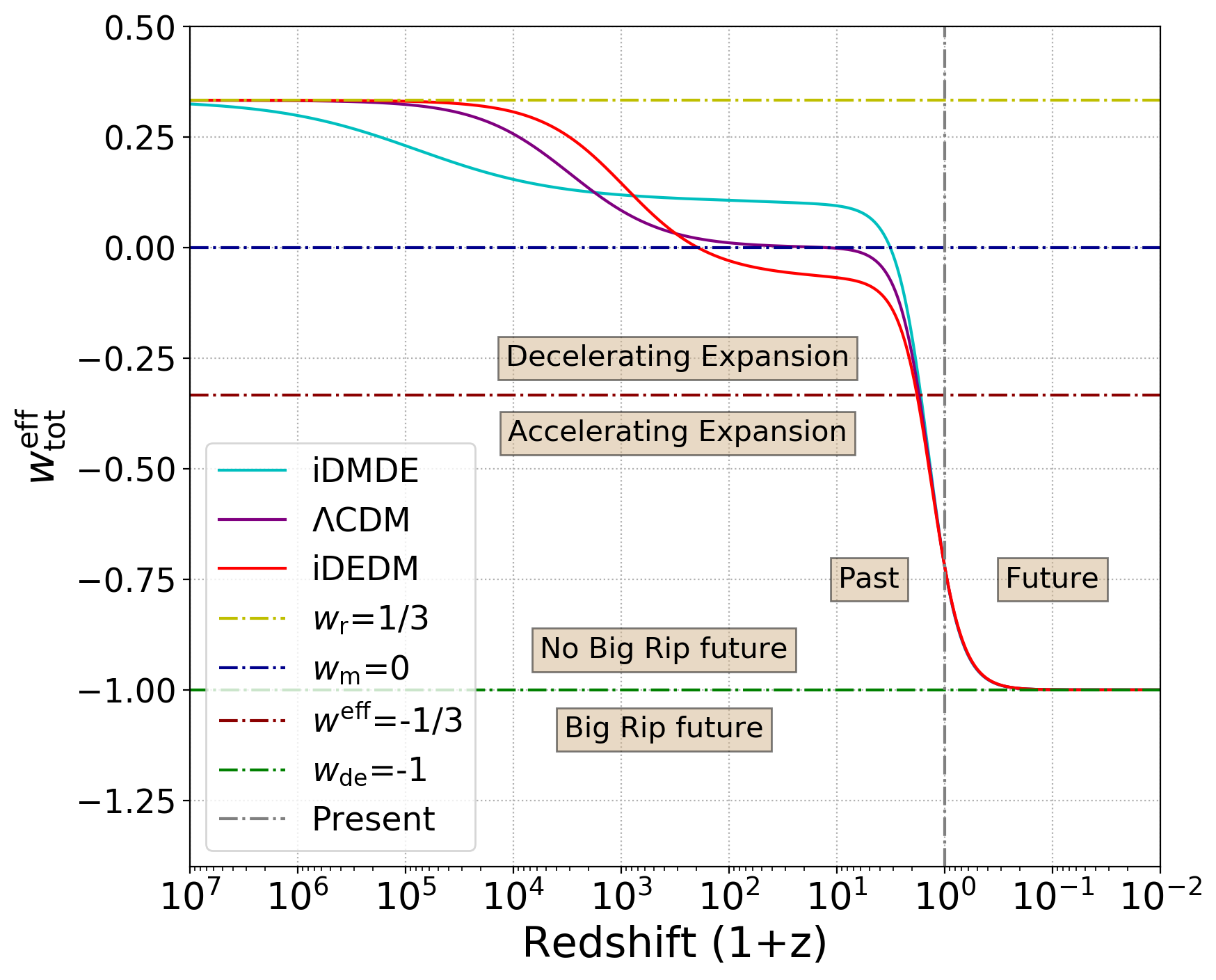}
        \label{q+eos_tot_NLID2}
    \end{subfigure}    
    \begin{subfigure}[b]{0.475\linewidth}
        \centering
        \includegraphics[width=\linewidth]{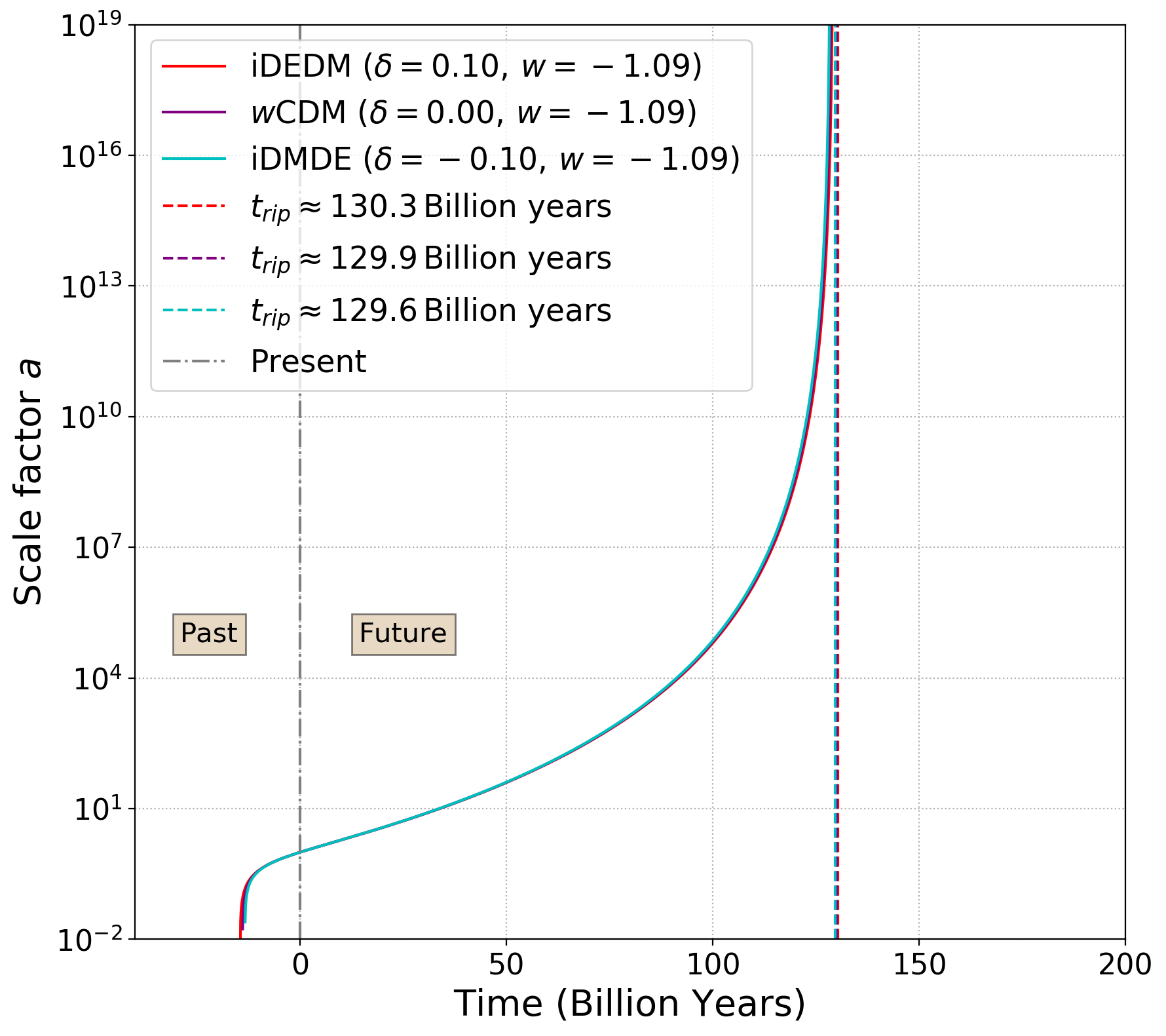}
        \label{fig:Big_rip_NLID2}
    \end{subfigure}    
    \hspace{0pt} 
    \caption{Total effective equation of state $w^{\rm{eff}}_{\rm{tot}}$ and big rip future singularities — $Q=3\delta H \left(\frac{\rho_{\text{dm}}^2}{\rho_{\text{dm}}+\rho_{\text{de}}} \right)$, with $w=-1$ (left panel) and $w=-1.09$ (right panel). In both the iDEDM regime ($\delta=+0.1$) and iDMDE regime ($\delta=-0.1$), the effect of the interaction diminishes in the asymptotic future and $w^{\rm{eff}}_{\rm{tot}}=w$. This implies that whenever $w<-1$, we necessarily have $w^{\rm{eff}}_{\rm{tot}}<-1$, thus guaranteeing a future big rip singularity in both cases.}
    \label{fig:eos_tot_BR_NLID2}
\end{figure}

We also have $q=\frac{1}{2} \left(1+3w^{\rm{eff}}_{\rm{tot}} \right)$ and $w^{\rm{eff}}_{\rm{tot}}=w^{\rm{eff}}_{\rm{de}}=w$ in the distant future, as seen in Figure~\ref{fig:eos_tot_BR_NLID2}, which leads to the following condition for a big rip to occur:
\begin{gather} \label{omega_eff_tot_NLID2_BG}
\begin{split}
\text{Big rip condition: \quad} w^{\rm{eff}}_{\rm{(tot,future)}} &\approx w < -1.
\end{split}
\end{gather}
In the case where $w<-1$, the universe will encounter a big rip future singularity at time $t_{\text{rip}}$:
\begin{equation}
\begin{split}
t_{\text{rip}}-t_0 &\approx -\frac{2}{3H_0(1+w)\sqrt{\Omega_{\text{(de,0)}}\left[1+\frac{\delta}{w}\left(\frac{\Omega_{\text{(dm,0)}}}{\Omega_{\text{(de,0)}}}\right)\right]\left[\frac{1+\frac{\delta}{w}\left(\frac{\Omega_{\text{(dm,0)}}}{\Omega_{\text{(de,0)}}}\right)}{1+\left(\frac{\Omega_{\text{(dm,0)}}}{\Omega_{\text{(de,0)}}}\right)}\right]^{\frac{\delta}{w-\delta}}}}.
\end{split}\label{eq:Big_Rip_general_NLID2_BG}
\end{equation}

The effect of the coupling on $w^{\rm{eff}}_{\rm{tot}}$ and how the interaction can either cause or avoid a big rip can be seen in Figure~\ref{fig:eos_tot_BR_NLID2}. The time of the big rip predicted using \eqref{eq:Big_Rip_general_NLID2_BG} is shown by the dashed lines in Figure~\ref{fig:eos_tot_BR_NLID2}, which is consistent with the point where the scale factor diverges $a\rightarrow\infty$ within a finite time. We note that for phantom DE, a big rip remains inevitable.

\subsection{Non-linear IDE model 3: $Q_3=3\delta H  
\left(\frac{\rho_{\text{de}}^2}{\rho_{\text{dm}}+\rho_{\text{de}}} \right)$} \label{analytical.dede}

Similar to the kernel $Q=3H\delta \rho_{\text{de}}$, this interaction mostly affects the dynamics of the late-time expansion during DE domination, while having a smaller impact on past dynamics during DM domination. For this interaction $Q=0$ if $\rho_{\text{de}}=0$, which guarantees that $\rho_{\text{de}}\ge 0$ at all times. In contrast, this interaction leads to future negative DM densities in the iDMDE regime, but this can be avoided with a sufficiently small interaction in the iDEDM regime, as required by the conditions in \eqref{Positive_energy.Q.NLID2} and illustrated in Figure~\ref{fig:Omega_NLID3}. The DM and DE densities for this model are derived in Appendix~\ref{Appendix_B} and are given by: 
\begin{equation}
\begin{split} 
\rho_{\text{dm}} &=  \rho_{\text{(dm,0)}}a^{-3\left(1+ \frac{w^2 }{w -\delta} \right)}  \left(\frac{\left[w \left( \frac{\rho_{\text{(dm,0)}}}{\rho_{\text{(de,0)}}}\right)+\delta \right]a^{3w }   -\delta }{w \left( \frac{\rho_{\text{(dm,0)}}}{\rho_{\text{(de,0)}}}\right) }  \right) \left[\frac{\left[w \left( \frac{\rho_{\text{(dm,0)}}}{\rho_{\text{(de,0)}}}\right)+\delta \right]a^{3w } + w -\delta }{w \left[1+ \frac{\rho_{\text{(dm,0)}}}{\rho_{\text{(de,0)}}} \right]}\right]^{\frac{\delta}{w -\delta} }, \\
\rho_{\text{de}} &=  \rho_{\text{(de,0)}}   a^{-3\left(1+ \frac{w^2 }{w -\delta} \right)}   \left[\frac{\left[w \left( \frac{\rho_{\text{(dm,0)}}}{\rho_{\text{(de,0)}}}\right)+\delta \right]a^{3w } + w -\delta }{w \left[1+ \frac{\rho_{\text{(dm,0)}}}{\rho_{\text{(de,0)}}} \right]}\right]^{\frac{\delta}{w -\delta} },
\label{NLID3_dm_de_BG}
\end{split} 
\end{equation}
where $\delta\neq w$ to avoid divisions by zero. If the power $\left(\frac{\delta }{w -\delta}\right)<0$, we require the additional conditions $w<0$ and $w<\delta\le - w r_0$ to avoid singularities for all $a$.

\begin{figure}[H]
    \centering
    \includegraphics[width=0.95\linewidth]{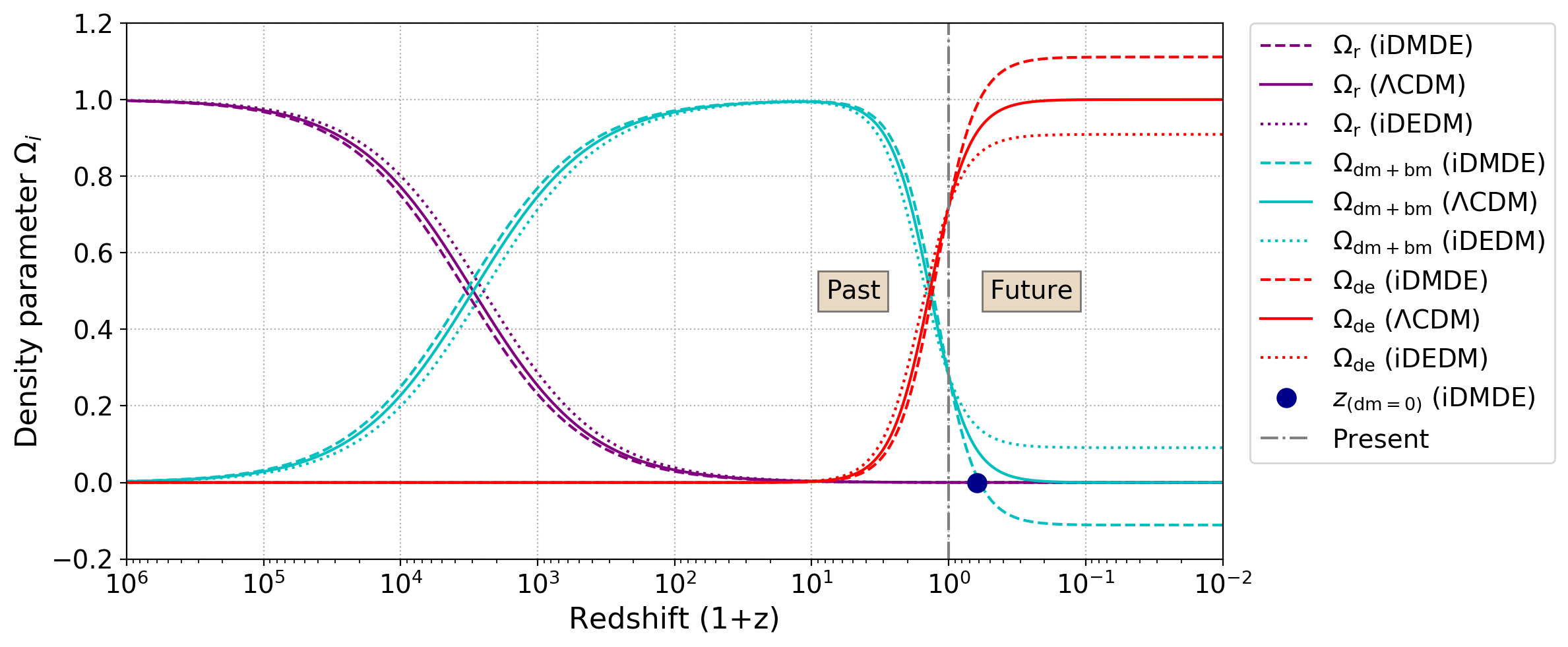}
    \caption{Density parameters vs redshift - $Q=3\delta H  \left(\frac{\rho_{\text{de}}^2}{\rho_{\text{dm}}+\rho_{\text{de}}} \right)$, with positive energy densities only found in the iDEDM regime ($\delta=+0.1$), while negative DM densities (in the future) are always present in the iDMDE regime ($\delta=-0.1$).}
    \label{fig:Omega_NLID3}
\end{figure}

\begin{gather} \label{Positive_energy.Q.NLID3}
\begin{split}
\boxed{\underline{\text{Conditions for}  \text{ $\rho_{\rm{dm}}\ge0 \; ; \; \rho_{\rm{de}} \ge0 $ at all points in cosmological evolution:}} \; \text{ iDEDM with }  
0<\delta< -w r_0}
\end{split}
\end{gather}
The fractional densities of both DM and DE will converge asymptotically in the future and past (note that in the presence of baryons there will be hybrid DM and baryon domination) to the following expressions:
\begin{equation}
\begin{split}
\Omega_{\text{(dm,past)}} =1 \quad &; \quad \Omega_{\text{(de,past)}}=0 \\
\Omega_{\text{(dm,future)}} =\left( \frac{\delta }{\delta-w}\right) \quad &; \quad \Omega_{\text{(de,future)}}=-\left( \frac{w }{\delta-w}\right)
\end{split}\label{eq:frac_den_NLID3}
\end{equation}
For this model, the DE density will always remain positive, but the DM density may become negative at the redshift:
\begin{equation}
\begin{split} \label{NLID3_dm=0_z_BG}
z _{\text{(dm=0)}} &= \left[\frac{\delta}{w \left( \frac{\Omega_{\text{(dm,0)}}}{\Omega_{\text{(de,0)}}}\right)+ \delta} \right]^{-\frac{1}{3w}} -1
\end{split} 
\end{equation}
The dark matter–dark energy equality occurs at the redshift where $\rho_{\text{dm}}=\rho_{\text{de}}$:
\begin{equation}
\begin{split} 
z_{\text{(dm=de)}} &= \left[ \frac{w+\delta}{w \left( \frac{\Omega_{\text{(dm,0)}}}{\Omega_{\text{(de,0)}}} \right) + \delta} \right]^{-\tfrac{1}{3w}} - 1
\label{NLID3_dm=de_BG}
\end{split} 
\end{equation}

\begin{figure}[htbp]
    \centering
    \begin{subfigure}[b]{0.502\linewidth}
        \centering
        \includegraphics[width=\linewidth]{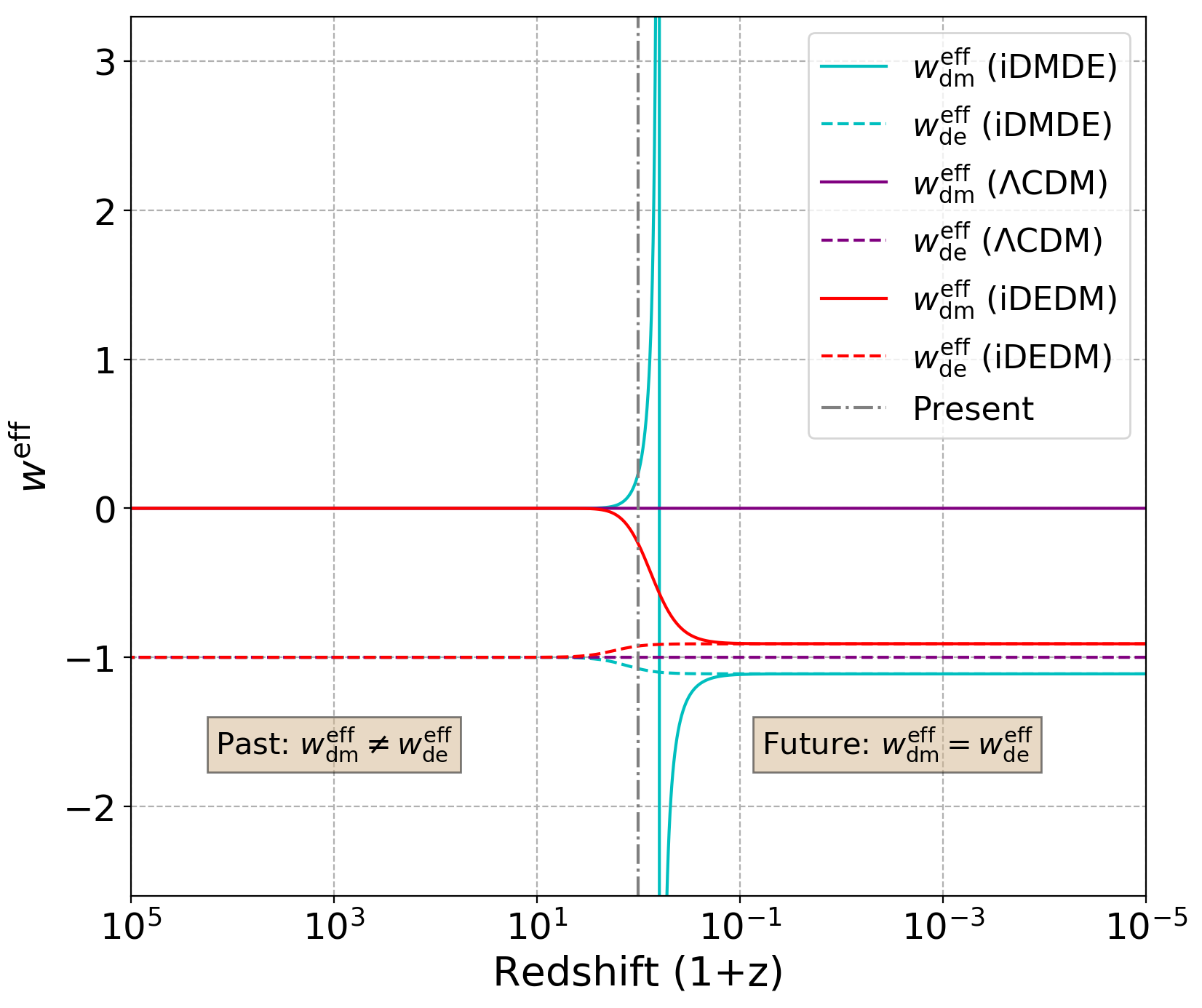}
        \label{fig:omega_dmde_NLID3}
    \end{subfigure}    
    \hspace{0pt} 
    \begin{subfigure}[b]{0.483\linewidth}
        \centering
        \includegraphics[width=\linewidth]{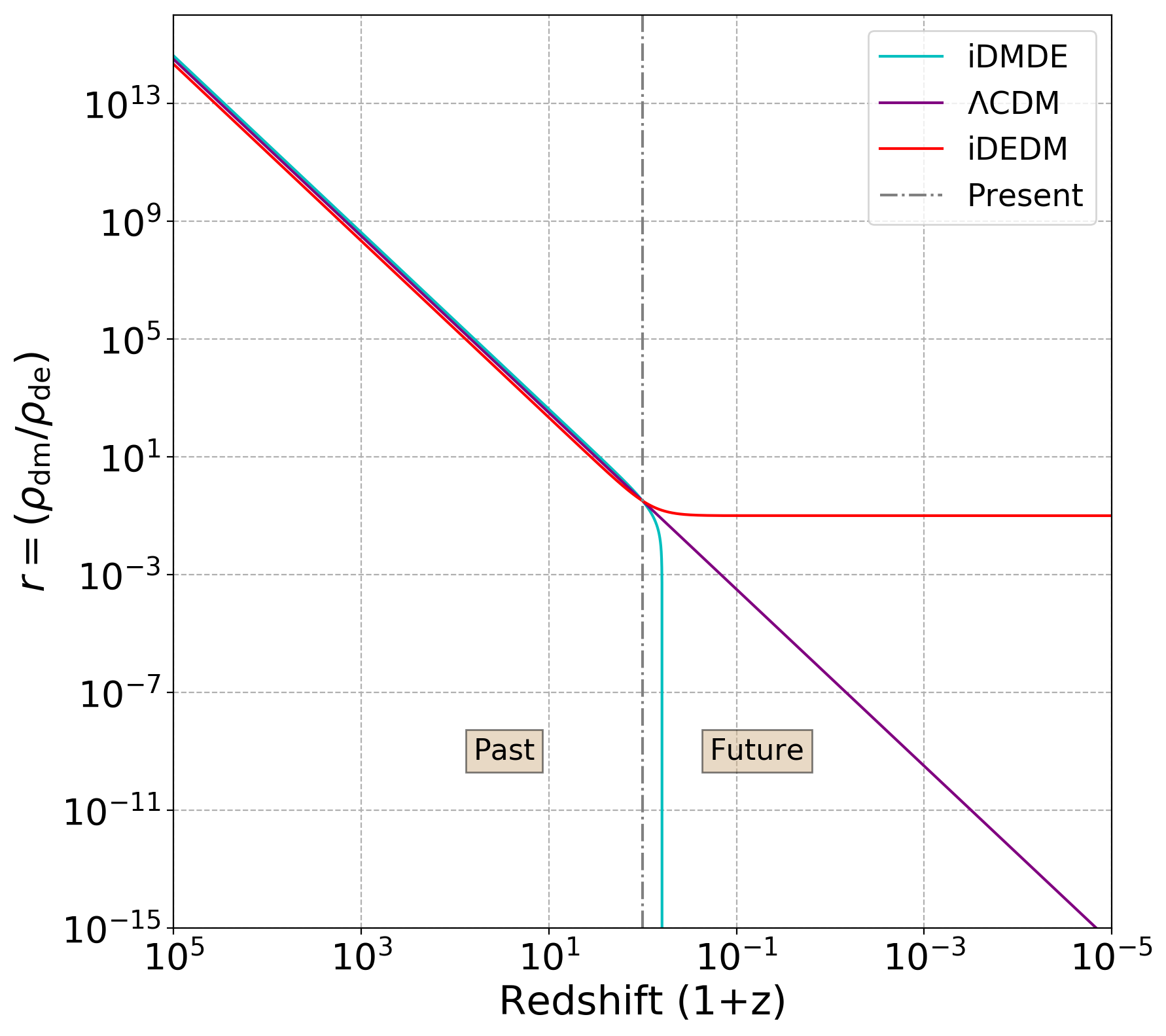}
        \label{fig:CP_NLID3}
    \end{subfigure}%
    \caption{Effective equations of state and Coincidence Problem (CP) vs redshift - $Q=3\delta H  \left(\frac{\rho_{\text{de}}^2}{\rho_{\text{dm}}+\rho_{\text{de}}} \right)$, with $w^{\rm{eff}}_{\rm{dm}}=w^{\rm{eff}}_{\rm{de}}$ in the future, thus solving the CP ($r=constant$) in the future for the iDEDM regime ($\delta=+0.1$). In the iDMDE regime ($\delta=-0.1$), negative DM densities and divergent $w^{\rm{eff}}_{\rm{dm}}$ (in the future) are always present. As the interaction diminishes in the past, there is no change to the CP for the past.}
    \label{fig:CP+omega_dmde_NLID3}
\end{figure}

The DM-to-DE ratio $r$ converges to the following expressions in the past and future:
\begin{equation}
\begin{split}
r &= \left(r_0 +\frac{\delta}{w} \right) a^{3 w} -\frac{\delta}{w} \quad ; \quad
r_{\text{past}} (a\rightarrow0)=\frac{\rho_{\text{dm}}}{\rho_{\text{de}}} \approx  \infty \quad ; \quad 
r_{\text{future}} (a\rightarrow\infty)=\frac{\rho_{\text{dm}}}{\rho_{\text{de}}} \approx -\frac{\delta}{w}
\end{split}%
\label{NLID3_r_PF}
\end{equation}
The DM and DE effective equations of state \eqref{DSA.H2} for this interaction are given by:
\begin{gather} \label{NLID3_omega_eff_dm_de_BG}
\begin{split}
w^{\rm{eff}}_{\rm{dm}} =- \delta\left(\frac{1}{r(r+1)}\right) \quad ; \quad w^{\rm{eff}}_{\rm{de}} = w + \delta \left(\frac{1}{r+1}\right).
\end{split}
\end{gather}
Substituting \eqref{NLID3_r_PF} into \eqref{NLID3_omega_eff_dm_de_BG} gives $w^{\rm{eff}}_{\rm{dm}}$ and $w^{\rm{eff}}_{\rm{de}}$ in the asymptotic past and future:
\begin{equation}
\begin{split}
w^{\rm{eff}}_{\rm{(dm,past)}}  &= 0, \; \;w^{\rm{eff}}_{\rm{(de,past)}}   = w, \;  \quad \rightarrow \quad \zeta_{\text{past}} = -3w \;\text{(no change)}. \\
w^{\rm{eff}}_{\rm{(dm,future)}} &=w^{\rm{eff}}_{\rm{(de,future)}}  =  \frac{w^2}{w-\delta}, \;\rightarrow \quad  \zeta_{\text{future}} = 0 \;\text{(solves the coincidence problem)}.
\end{split}%
\label{eq:omega_eff_dm_de_NLID3_subbed_past_future}
\end{equation}
This interaction model will therefore solve the coincidence problem in the future, but will have no effect on the problem in the past, as illustrated in Figure~\ref{fig:CP+omega_dmde_NLID3}.

The predicted redshift at which the phantom crossing occurs can be obtained by setting $w^{\text{eff}}_{\text{de}}=-1$ in \eqref{NLID3_omega_eff_dm_de_BG}:
\begin{equation}
\begin{split}
    z_{\text{pc}}  &= \left[\frac{w \left( \frac{  \Omega_{\text{(dm,0)}}}{\Omega_{\text{(de,0)}}} \right) +\delta}{ - \left(\frac{w\delta}{1+w}\right)-w  +\delta} \right]^{\frac{1}{3 w}} -1 
\label{NLID3_phantom_crossing_3} 
\end{split}
\end{equation}
As seen from \eqref{eq:omega_eff_dm_de_NLID3_subbed_past_future} and Figure~\ref{fig:CP+omega_dmde_NLID3}, given specific values of $w$ and $\delta$, we have two possibilities for the direction of the phantom crossing, with the iDMDE case plagued by negative energies:

\begin{equation}
\begin{split}
 \text{pc direction} \begin{cases}
   \text{iDMDE} : \text{Divergent pc for $w^{\rm{eff}}_{\rm{dm}}$ at $z_{\text{(dm=0)}}$ \eqref{NLID3_dm=0_z_BG}, with } \rho_{\rm{dm}}<0. \\
   \text{iDEDM} : \text{Phantom } (w^{\rm{eff}}_{\rm{(de,past)}}<-1) \rightarrow \text{ Quintessence } (w^{\rm{eff}}_{\rm{(de,future)}}>-1), \text{ with } \rho_{\rm{dm/de}}>0.
\end{cases}
\end{split}\label{eq:z_pc_dede_direction}
\end{equation}

\begin{figure}[htbp]    \centering
\begin{subfigure}[b]{0.515\linewidth}
        \centering
        \includegraphics[width=\linewidth]{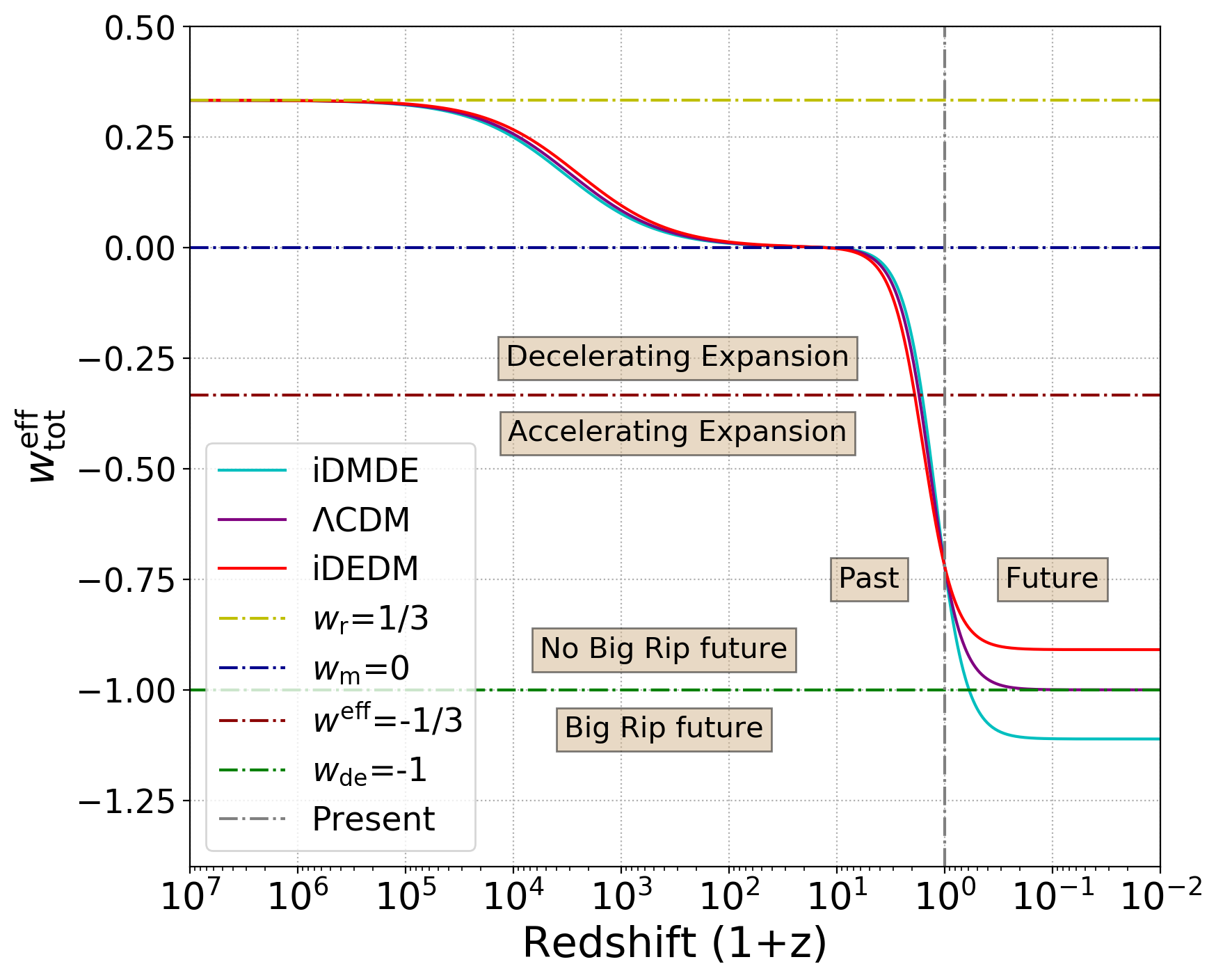}
        \label{q+eos_tot_NLID3}
    \end{subfigure}    
    \begin{subfigure}[b]{0.475\linewidth}
        \centering
        \includegraphics[width=\linewidth]{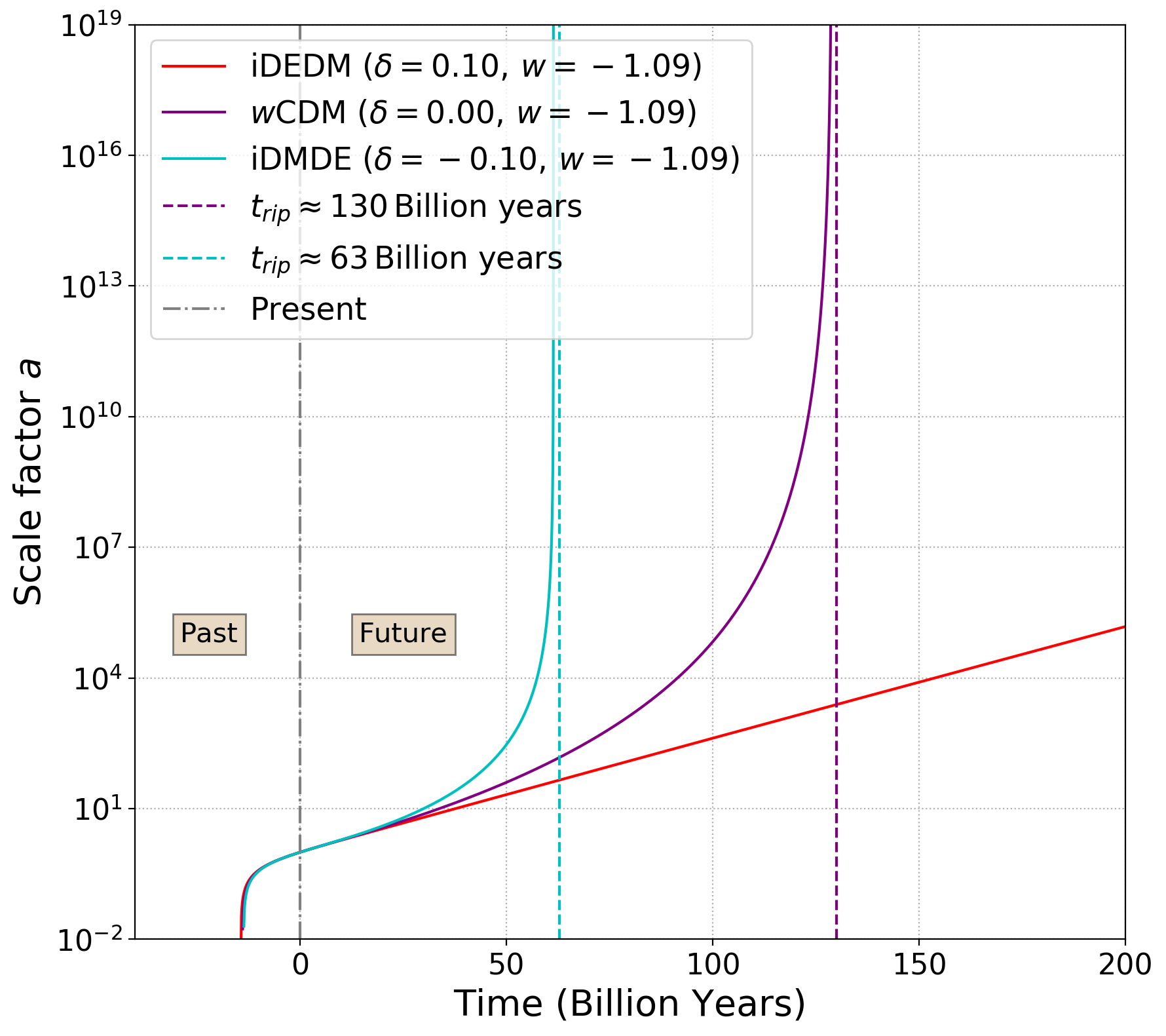}
        \label{fig:Big_rip_NLID3}
    \end{subfigure}    
    \hspace{0pt} 
    \caption{Total effective equation of state $w^{\rm{eff}}_{\rm{tot}}$ and big rip future singularities - $Q=3\delta H \left(\frac{\rho_{\text{de}}^2}{\rho_{\text{dm}}+\rho_{\text{de}}} \right)$, with $w=-1$ (left panel) and $w=-1.09$ (right panel). In the iDEDM regime ($\delta=+0.1$), in the asymptotic future we may have $w^{\rm{eff}}_{\rm{tot}}>-1$, even if $w<-1$, thus avoiding a big rip. In the iDMDE regime ($\delta=-0.1$), in the asymptotic future we will always have $w^{\rm{eff}}_{\rm{tot}}<-1$ if $w<-1$, thus guaranteeing a future big rip singularity.}
    \label{fig:eos_tot_BR_NLID3}
\end{figure}

We also have $q=\frac{1}{2} \left(1+3w^{\rm{eff}}_{\rm{tot}} \right)$ and $w^{\rm{eff}}_{\rm{tot}}=w^{\rm{eff}}_{\rm{dm}}=w^{\rm{eff}}_{\rm{de}}=\frac{w^2}{w-\delta}$ in the distant future. This implies that, even if $w<-1$, both the deceleration parameter and the effective equations of state can take larger values, such that $w^{\rm{eff}}_{\rm{tot}}>-1$ (for $\delta>0$ in the iDEDM regime), which may avoid a future big rip singularity. This leads to condition \eqref{omega_eff_tot_BLID3_BG} for a big rip to occur:
\begin{gather} \label{omega_eff_tot_BLID3_BG}
\begin{split}
\text{Big rip condition: \quad  }
w^{\rm{eff}}_{\rm{(tot,future)}} &\approx\frac{ w^2 }{ w-\delta }<-1 \quad \rightarrow \quad \delta \ge w(w+1). 
\end{split}
\end{gather}
In the case where $w^{\rm{eff}}_{\rm{tot}}<-1$, the universe will experience a big rip future singularity at time $t_{\text{rip}}$:
\begin{equation}
\begin{split}
t_{\text{rip}}-t_0 &\approx -\frac{2}{3H_0 \left(1+ \frac{w^2 }{w -\delta} \right)\sqrt{ \Omega_{\text{(de,0)}}  \left(1-\frac{\delta}{w} \right) \left[\frac{w -\delta}{w \left(1+\frac{\Omega_{\text{(dm,0)}}}{\Omega_{\text{(de,0)}}} \right)}\right]^{\frac{\delta }{w -\delta}} }}.
\end{split}\label{eq:Big_Rip_NLID3_BG}
\end{equation}

The effect of the coupling on $w^{\rm{eff}}_{\rm{tot}}$ and how the interaction may cause or avoid a big rip can be seen in Figure~\ref{fig:eos_tot_BR_NLID3}. The time of the big rip predicted using \eqref{eq:Big_Rip_NLID3_BG} is indicated by the dashed lines in Figure~\ref{fig:eos_tot_BR_NLID3}, which is in agreement with the point where the scale factor diverges $a\rightarrow\infty$ within a finite time. We note that for phantom DE, a big rip can only be avoided in the iDEDM regime.

\section{Dark interactions as a dynamical dark energy equation of state $\tilde{w}(z)$} \label{reconstructed_w}

\begin{figure}
    \centering
    \begin{subfigure}[b]{0.494\linewidth}
        \centering
        \includegraphics[width=\linewidth]{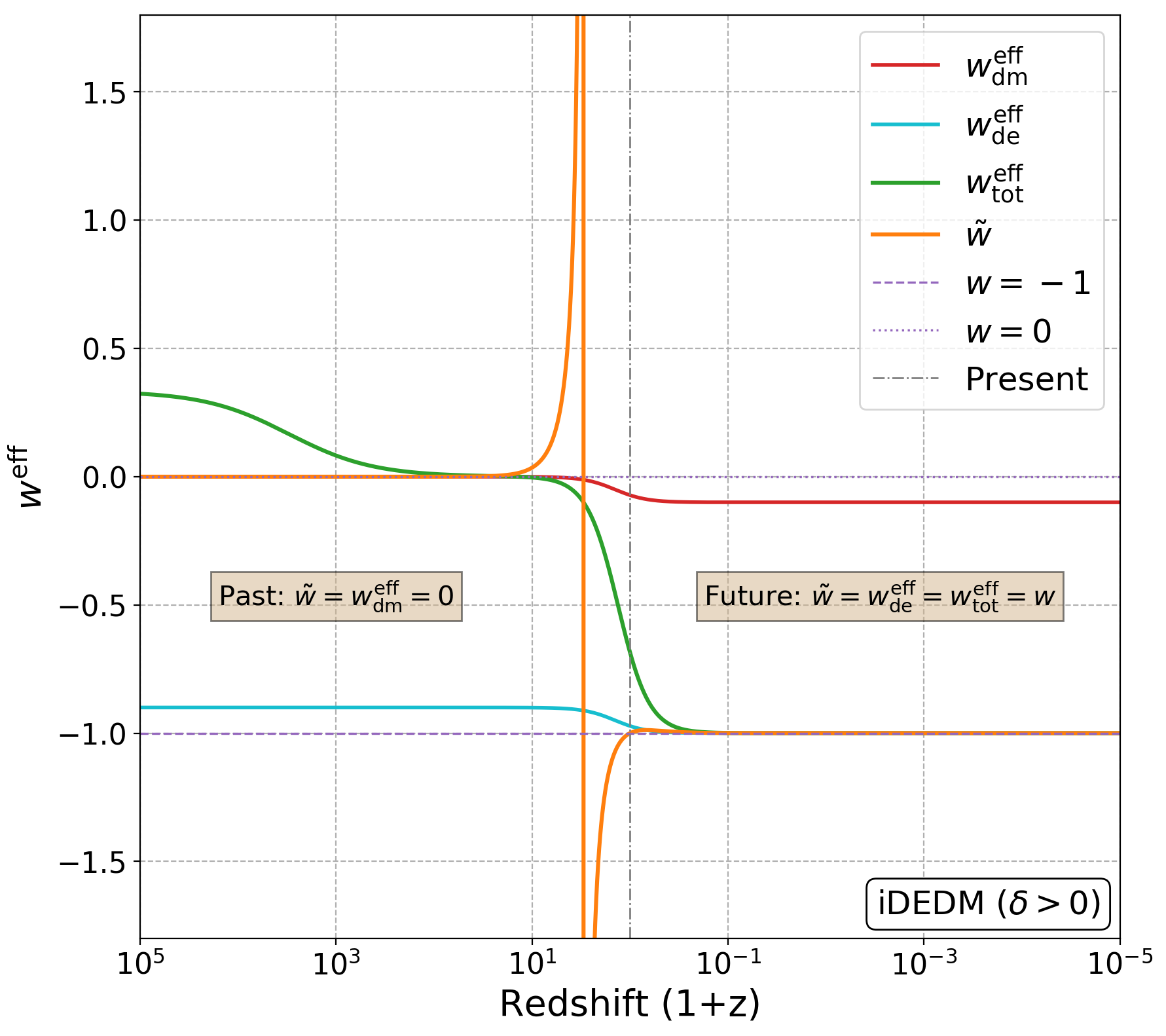}
    \end{subfigure}    
    \hspace{0pt} 
    \begin{subfigure}[b]{0.494\linewidth}
        \centering
        \includegraphics[width=\linewidth]{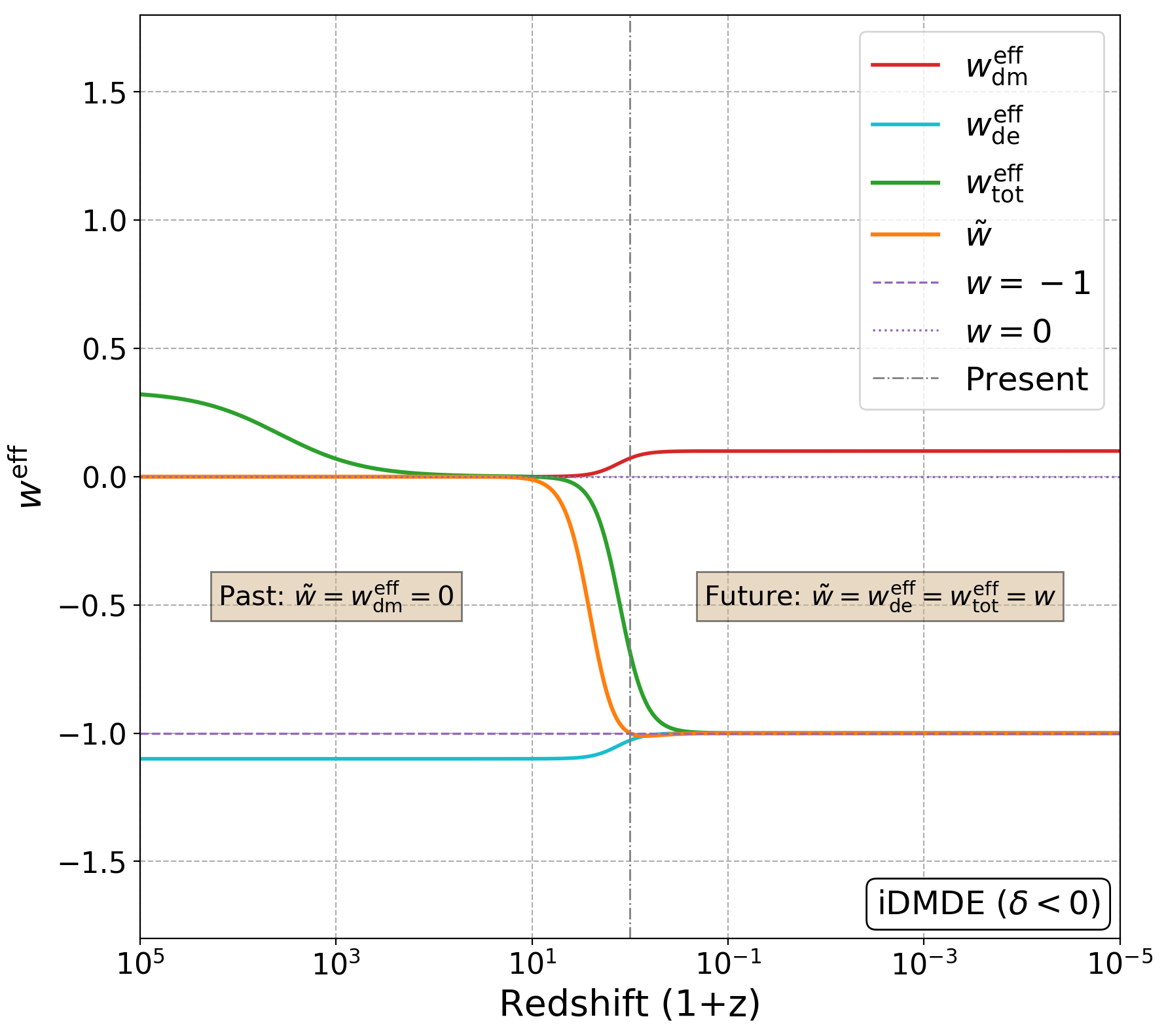}
    \end{subfigure}%
    \caption{Equations of state $\tilde{w}$, $w^{\rm eff}_{\rm de}$, $w^{\rm eff}_{\rm dm}$, $w^{\rm eff}_{\rm tot}$, and $w$ vs.\ redshift — $Q_1=3\delta H \left(\frac{\rho_{\text{dm}}\rho_{\text{de}}}{\rho_{\text{dm}}+\rho_{\text{de}}} \right)$. The left panel shows the iDEDM regime ($\delta=+0.1$), where only $\tilde{w}(z)$ exhibits a divergent phantom crossing. Conversely, the right panel shows the iDMDE regime ($\delta=-0.1$), where no divergent phantom crossings are present for any of the equations of state. Additionally, for both cases we have $\tilde{w}(0)=w$ at present, $\tilde{w}=w^{\rm eff}_{\rm de}=w^{\rm eff}_{\rm tot}=w$ in the asymptotic future, and $\tilde{w}=w^{\rm eff}_{\rm dm}=0$ in the asymptotic past.}
    \label{fig:w_all_Qdmde}
\end{figure}

\begin{figure}
    \centering
    \begin{subfigure}[b]{0.494\linewidth}
        \centering
        \includegraphics[width=\linewidth]{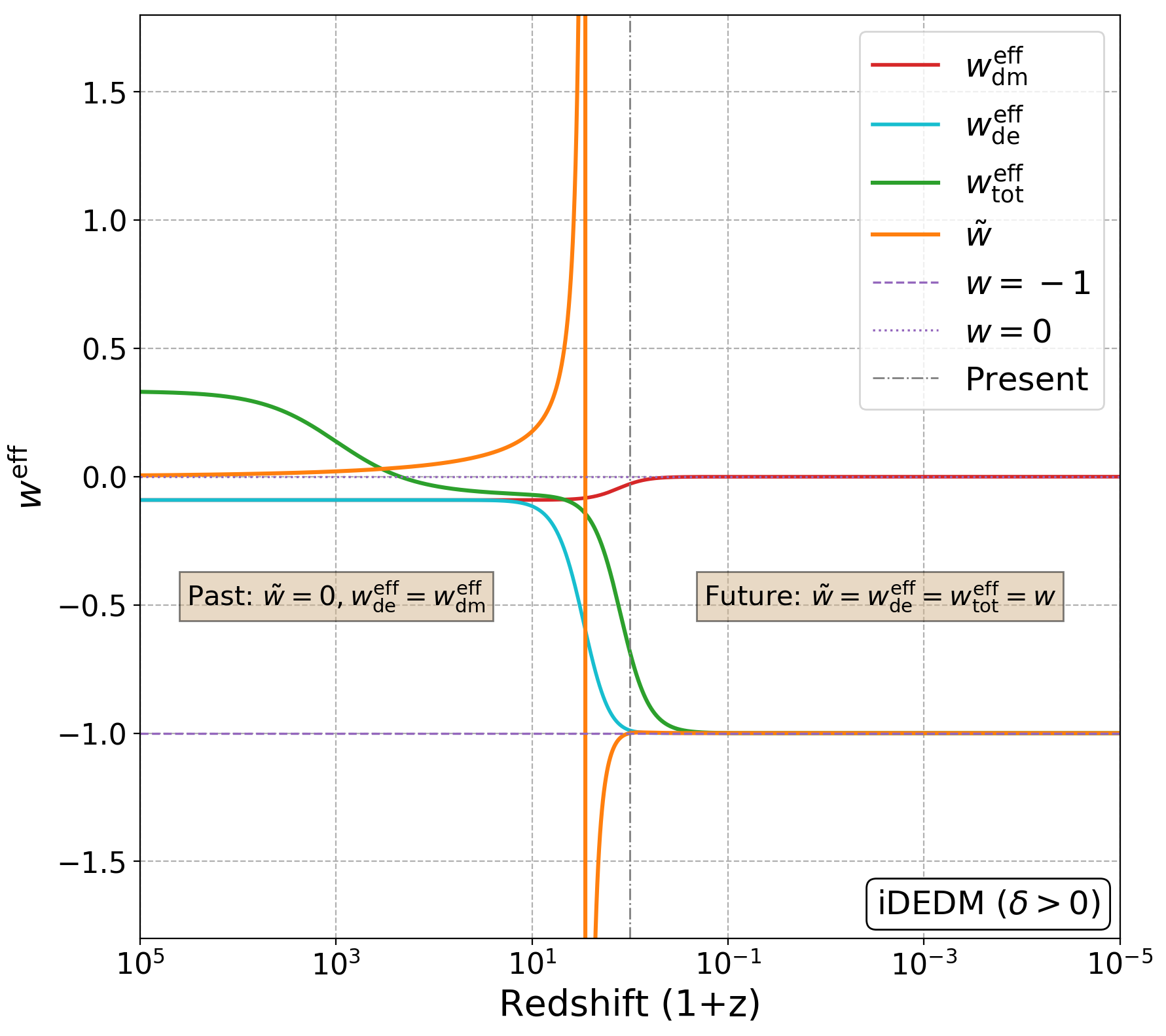}
    \end{subfigure}    
    \hspace{0pt} 
    \begin{subfigure}[b]{0.494\linewidth}
        \centering
        \includegraphics[width=\linewidth]{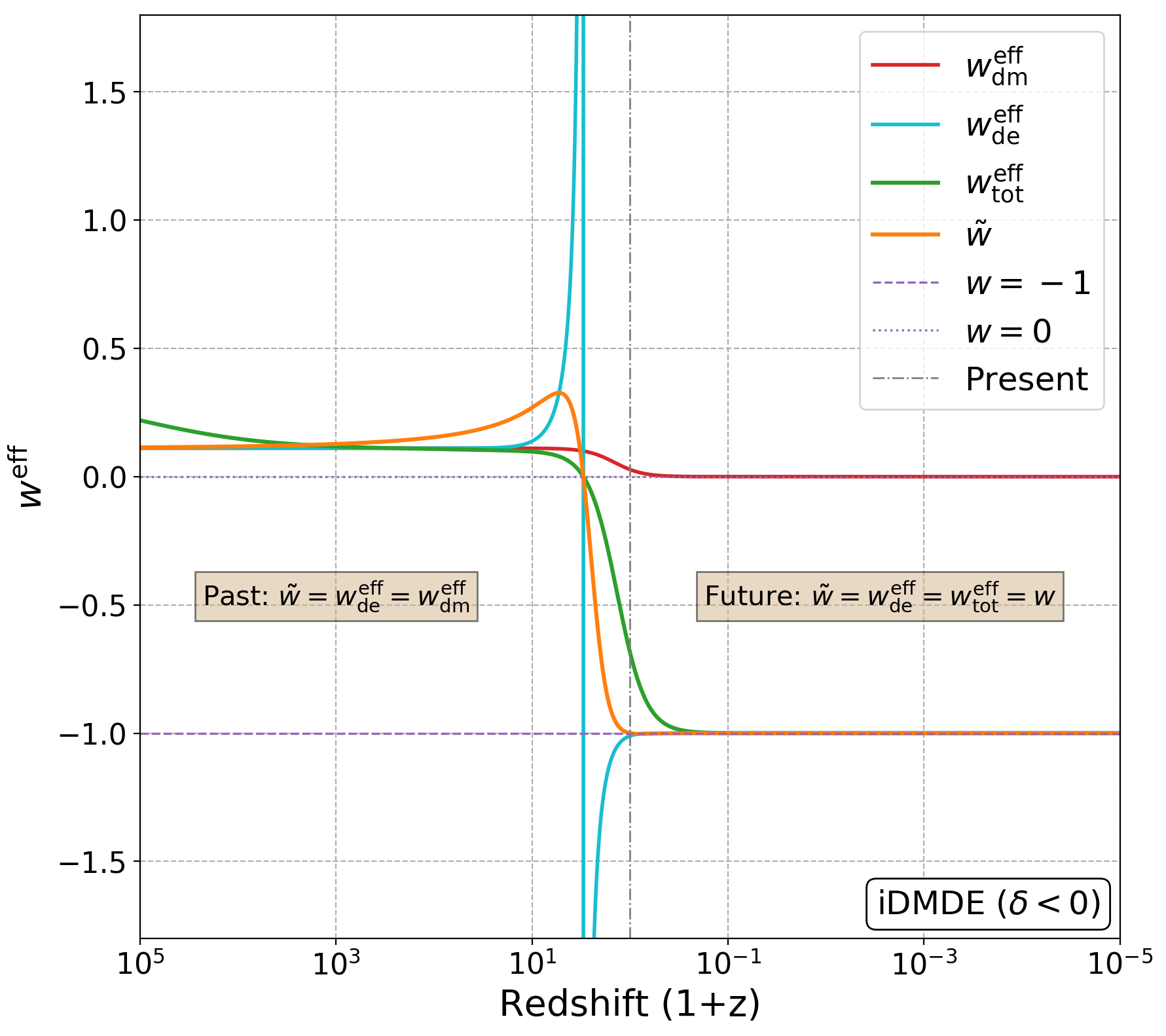}
    \end{subfigure}%
    \caption{Equations of state $\tilde{w}$, $w^{\rm eff}_{\rm de}$, $w^{\rm eff}_{\rm dm}$, $w^{\rm eff}_{\rm tot}$, and $w$ vs.\ redshift — $Q_2=3\delta H  \left(\frac{\rho_{\text{dm}}^2}{\rho_{\text{dm}}+\rho_{\text{de}}} \right)$. The left panel shows the iDEDM regime ($\delta=+0.1$), where only $\tilde{w}(z)$ exhibits a divergent phantom crossing. Conversely, the right panel shows the iDMDE regime ($\delta=-0.1$), where only $w^{\rm eff}_{\rm de}$ exhibits a divergent phantom crossing, due to $\rho_{\rm de}$ becoming negative in the effective split. Additionally, for both cases we have $\tilde{w}(0)=w$, while in the asymptotic future $\tilde{w}=w^{\rm eff}_{\rm de}=w^{\rm eff}_{\rm tot}=w$. In the asymptotic past, the iDEDM regime has $\tilde{w}=0$ and $w^{\rm eff}_{\rm de}=w^{\rm eff}_{\rm dm}$, whereas the iDMDE regime has $\tilde{w}=w^{\rm eff}_{\rm de}=w^{\rm eff}_{\rm dm}$.}
    \label{fig:w_all_Qdmdm}
\end{figure}

\begin{figure}
    \centering
    \begin{subfigure}[b]{0.494\linewidth}
        \centering
        \includegraphics[width=\linewidth]{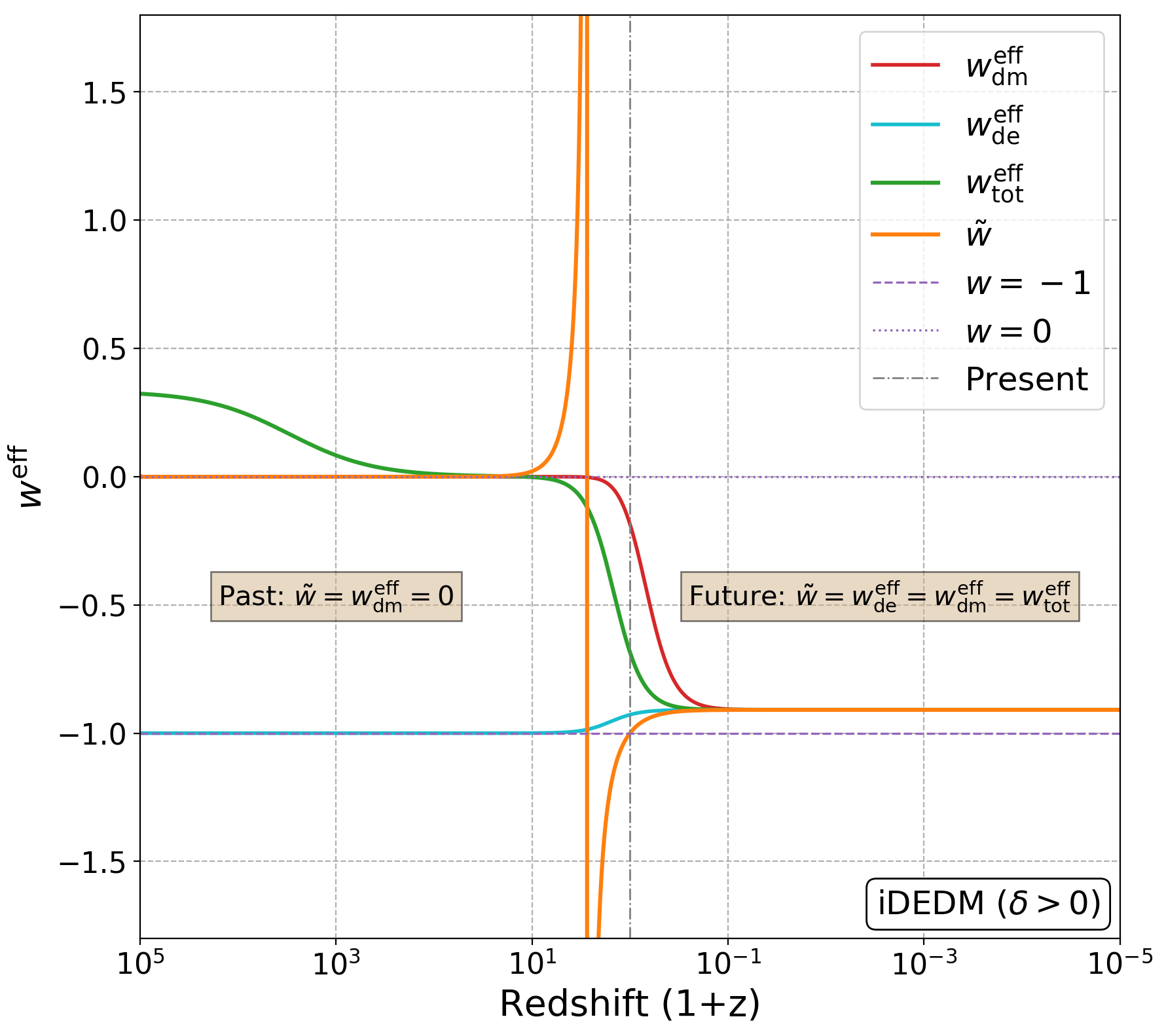}
    \end{subfigure}    
    \hspace{0pt} 
    \begin{subfigure}[b]{0.494\linewidth}
        \centering
        \includegraphics[width=\linewidth]{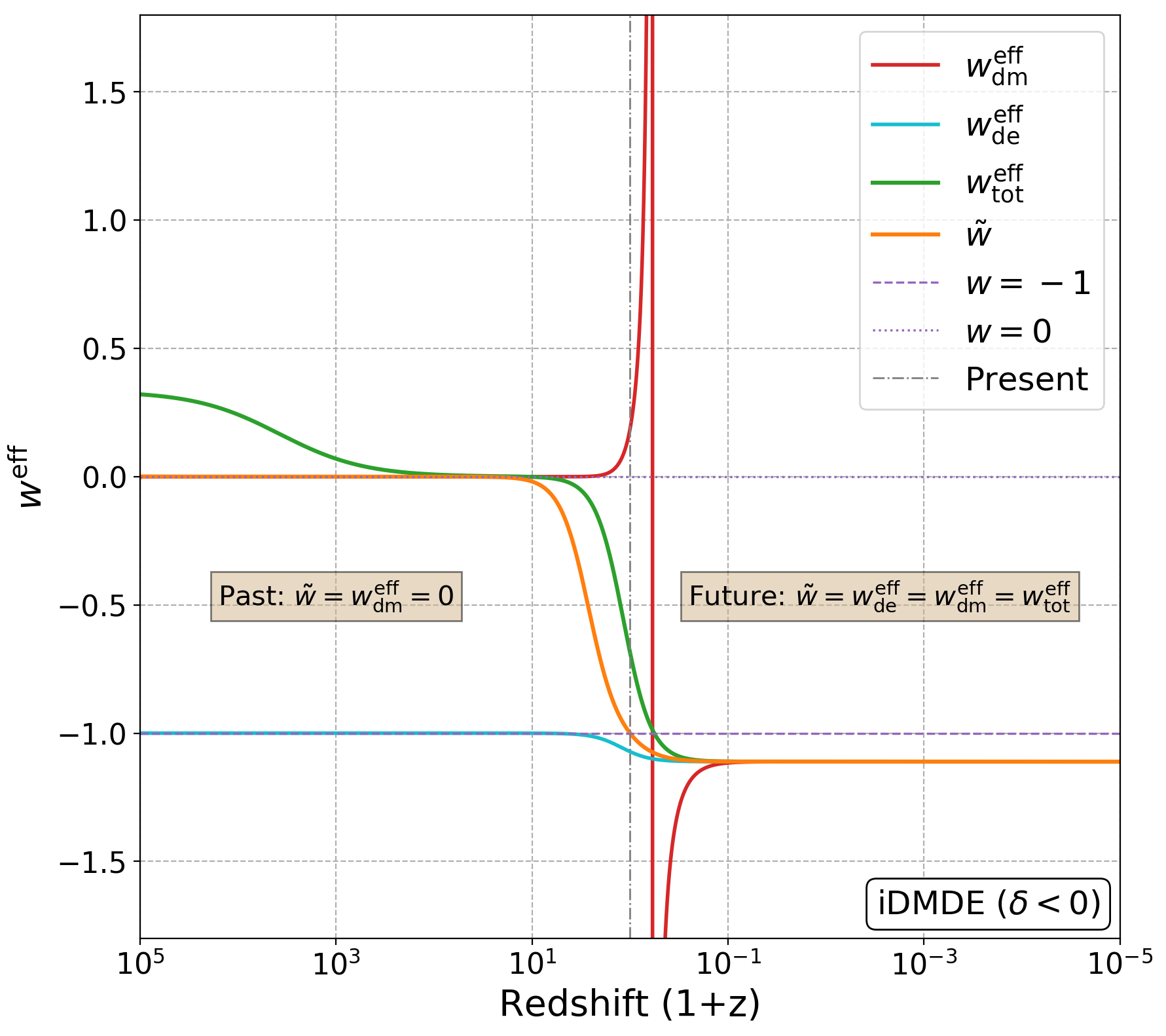}
    \end{subfigure}%
    \caption{Equations of state $\tilde{w}$, $w^{\rm eff}_{\rm de}$, $w^{\rm eff}_{\rm dm}$, $w^{\rm eff}_{\rm tot}$, and $w$ vs.\ redshift — $Q_3=3\delta H  \left(\frac{\rho_{\text{de}}^2}{\rho_{\text{dm}}+\rho_{\text{de}}} \right)$. The left panel shows the iDEDM regime ($\delta=+0.1$) where only $\tilde{w}(z)$ exhibits a divergent phantom crossing. Conversely, the right panel shows the iDMDE regime ($\delta=-0.1$) where $w^{\rm eff}_{\rm dm}$ has a divergent phantom crossing, due to $\rho_{\rm dm}$ becoming negative in the effective split. Additionally, for both cases we have $\tilde{w}(0)=w$ at present, $\tilde{w}=w^{\rm eff}_{\rm dm}=0$ in the asymptotic past, and $\tilde{w}=w^{\rm eff}_{\rm de}=w^{\rm eff}_{\rm dm}=w^{\rm eff}_{\rm tot}$ in the asymptotic future.}
    \label{fig:w_all_Qdede}
\end{figure}

In the companion paper \cite{vanderWesthuizen:2025I}, we discuss how a dynamical dark energy reconstruction of the equation of state $\tilde{w}(z)$ can be useful when comparing IDE models to other DE and modified gravity models. This reconstruction is obtained by equating $h(z)$ in any model to $h(z)$ in \eqref{wz_h} and solving for $\tilde{w}(z)$:
\begin{gather} \label{wz_h}
\begin{split}
h^2(z)&= \Omega_{\text{(r,0)}}(1+z)^{4}+ \Omega_{\text{(bm,0)}}(1+z)^{3}+ \Omega_{\text{(dm,0)}}(1+z)^{3}+  \Omega_{\text{(de,0)}} \exp\left[ 3 \int_0^z dz' \frac{1+\tilde{w}(z')}{1 + z'} \right].
\end{split}
\end{gather}
Here we want to add that a peculiar property of all IDE models with a Hubble function given by \eqref{DSA.H} and conservation equations \eqref{eq:conservation} is that they will always yield a $\tilde{w}(z)$ of the following form:
\begin{gather} \label{wz_general}
\begin{split}
\tilde{w}(z)&=\frac{w \rho_{\rm{de}}}{\rho_{\rm{dm}}+\rho_{\rm{de}}-\rho_{\rm{(dm,0)}}(1+z)^3} 
=\frac{w}{1+r-\dfrac{\rho_{\rm{(dm,0)}}(1+z)^3}{\rho_{\rm{de}}}}.
\end{split}
\end{gather}
Equation \eqref{wz_general} is especially useful for IDE models with a simple ratio $r$, such as the ones studied in this paper. A full derivation of \eqref{wz_general} is given in Appendix~\ref{Appendix_C}, where we also include a discussion of why the iDEDM regime $\tilde{w}(z)$ exhibits divergent behavior, while the iDMDE regime does not, as seen in Figures~\ref{fig:w_all_Qdmde},~\ref{fig:w_all_Qdmdm} and~\ref{fig:w_all_Qdede}.

\subsection{Non-linear IDE model 1: $Q_1=3\delta H  
\left(\frac{\rho_{\text{dm}}\rho_{\text{de}}}{\rho_{\text{dm}}+\rho_{\text{de}}} \right)$}

Substituting \eqref{NLID1_dm_de_BG} into \eqref{wz_general}, the reconstructed dynamical DE equation of state for this interaction is given by:
\begin{gather} \label{wz_Q_dmde_1}
\begin{split}
\tilde{w}(z)&=   \frac{ w  }{  1+ r_0  (1+z)^{-3(\delta +w)}      - r_0  (1+z)^{-3w} \left[\frac{1+r_0  (1+z)^{-3\left( w + \delta\right) }}{1+r_0 } \right]^{\tfrac{\delta}{(w +\delta) }}   } .  
\end{split}
\end{gather}
At present we have $\tilde{w}(0)=w$. For the asymptotic future $(z\rightarrow-1)$, after simplification we obtain:
\begin{gather} \label{wz_Q_ddm+dde_6}
\begin{split}
\tilde{w}(z\rightarrow-1)&=w= w_{\rm{(de},future)}^{\rm{eff}}=w_{\rm{(dm},future)}^{\rm{eff}}=w_{\rm{(tot},future)}^{\rm{eff}}.
\end{split}
\end{gather}
For the asymptotic past, we have:
\begin{gather} \label{wz_Q_ddm+dde_7}
\begin{split}
\tilde{w}(z\rightarrow\infty)&=0 = w_{\rm{(dm},past)}^{\rm{eff}}.
\end{split}
\end{gather}

\subsection{Non-linear IDE model 2: $Q_2=3\delta H  
\left(\frac{\rho_{\text{dm}}^2}{\rho_{\text{dm}}+\rho_{\text{de}}} \right)$}

Substituting \eqref{NLID2_dm_de_BG} into \eqref{wz_general} gives the reconstructed dynamical DE equation of state for this interaction:
\begin{gather}\label{wz_Q_dmdm_1}
\begin{split}
\tilde{w}(z)
&= \frac{w}{1+ \dfrac{wr_0}{\big(w+\delta r_0\big)(1+z)^{3w}-\delta r_0}
\left[
1
- (1+z)^{\tfrac{3w\delta}{w-\delta}}
\left(
\dfrac{(w+\delta r_0)(1+z)^{3w}+r_0(w-\delta)}{w(1+r_0)}
\right)^{-\tfrac{\delta}{w-\delta}}
\right]}.
\end{split}
\end{gather}

We have at present $\tilde{w}(0)=w$. Given our initial assumptions, \eqref{wz_Q_dmdm_1} converges in the asymptotic future $(z\rightarrow-1)$ to \eqref{wz_Q_dm_6}.
\begin{gather} \label{wz_Q_dm_6}
\begin{split}
\tilde{w}(z\rightarrow-1)&= w= w_{\rm{(de},future)}^{\rm{eff}}=w_{\rm{(tot},future)}^{\rm{eff}}.
\end{split}
\end{gather}
For the asymptotic past, there are two possible outcomes $(z\rightarrow\infty)$, depending on which power dominates in \eqref{DSA.H}. We therefore have two possibilities for the past:
\begin{gather} \label{wz_Q_dm_7}
\begin{split}
\text{if } \delta<0 \; \text{(iDMDE regime)}: &\quad \tilde{w}(z\rightarrow\infty)=-\frac{\delta w}{\delta-w}= w_{\rm{(de},past)}^{\rm{eff}}=w_{\rm{(dm},past)}^{\rm{eff}}, \\
\text{if } \delta\ge0 \; \text{(iDEDM regime)}: &\quad \tilde{w}(z\rightarrow\infty)=0.
\end{split}
\end{gather}

\subsection{Non-linear IDE model 3: $Q_3=3\delta H  
\left(\frac{\rho_{\text{de}}^2}{\rho_{\text{dm}}+\rho_{\text{de}}} \right)$}

Substituting \eqref{NLID3_dm_de_BG} into \eqref{wz_general} gives the reconstructed dynamical DE equation of state for this interaction:  
\begin{gather}\label{wz_Q_dede_1}
\begin{split}
\tilde{w}(z)
&=\frac{w}{1
+ \left(r_0+\frac{\delta}{w}\right)(1+z)^{-3w}
- \frac{\delta}{w}-\, r_0\,
(1+z)^{-\frac{3w^{2}}{\,w-\delta\,}}
\left(
\frac{\big(w r_0+\delta\big)(1+z)^{-3w}+w-\delta}{\,w(1+r_0)\,}
\right)^{-\frac{\delta}{\,w-\delta\,}}}
\end{split}
\end{gather}
We have $\tilde{w}(0)=w$. We may note that $(\delta+w)<0$ in the most general case, given our initial assumptions. Therefore, for the asymptotic future $(z\rightarrow-1)$ we obtain:
\begin{gather} \label{wz_Q_de_6}
\begin{split}
\tilde{w}(z\rightarrow-1)&= \frac{w^2}{w-\delta}= w_{\rm{(de},future)}^{\rm{eff}}=w_{\rm{(dm},future)}^{\rm{eff}}=w_{\rm{(tot},future)}^{\rm{eff}}.
\end{split}
\end{gather}
For the asymptotic past, we have:
\begin{gather} \label{wz_Q_de_7}
\begin{split}
\tilde{w}(z\rightarrow\infty)&=0=w_{\rm{(dm},past)}^{\rm{eff}}.
\end{split}
\end{gather}

\section{Statefinder diagnostics} \label{statefinder}

\begin{figure}[htbp]
    \centering
    \begin{subfigure}[b]{0.49\linewidth}
        \centering
        \includegraphics[width=\linewidth]{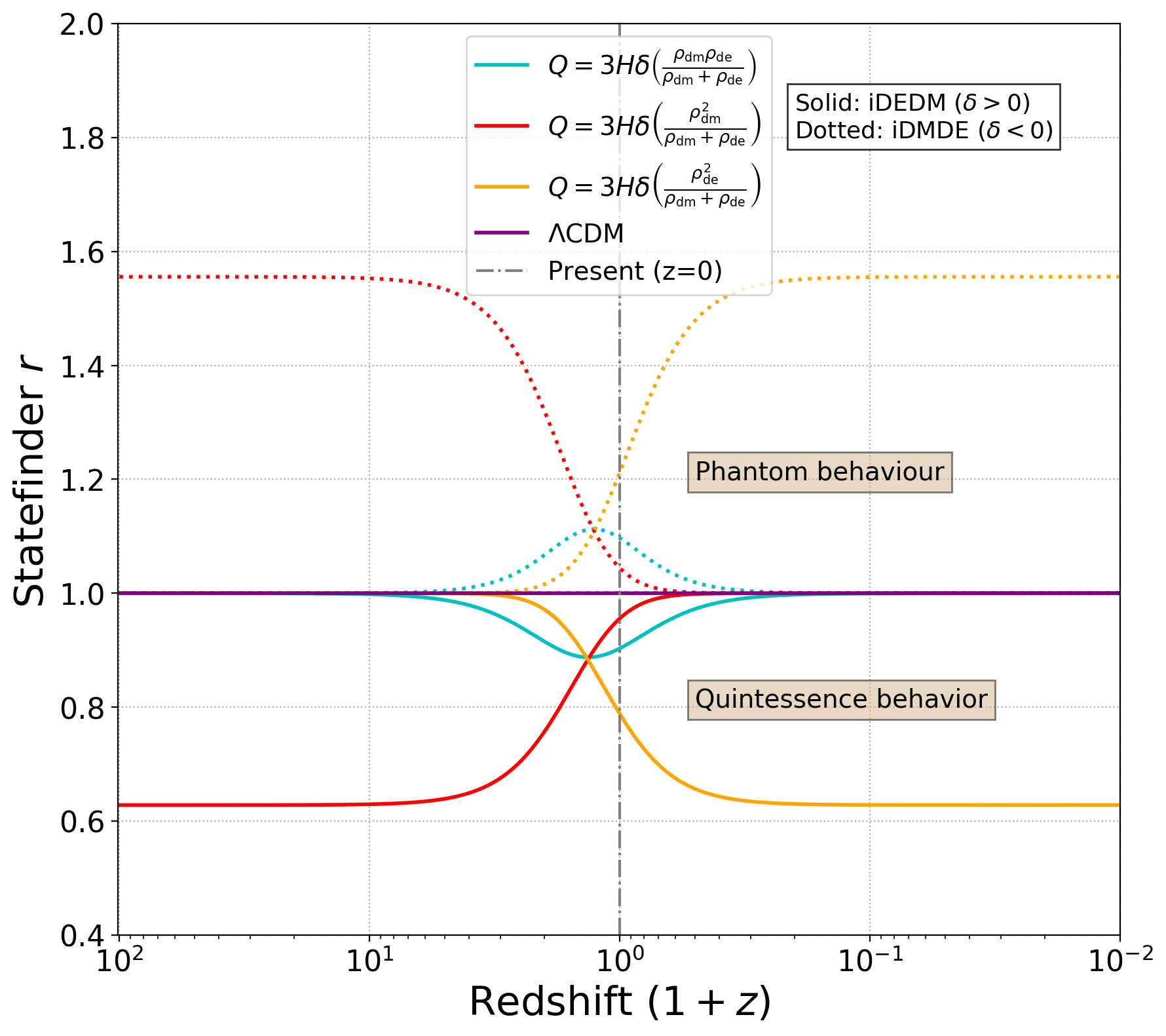}
        \label{fig:Statefinder_linear_rz_iDEDM}
    \end{subfigure}%
    \hspace{0pt} 
    \begin{subfigure}[b]{0.49\linewidth}
        \centering
        \includegraphics[width=\linewidth]{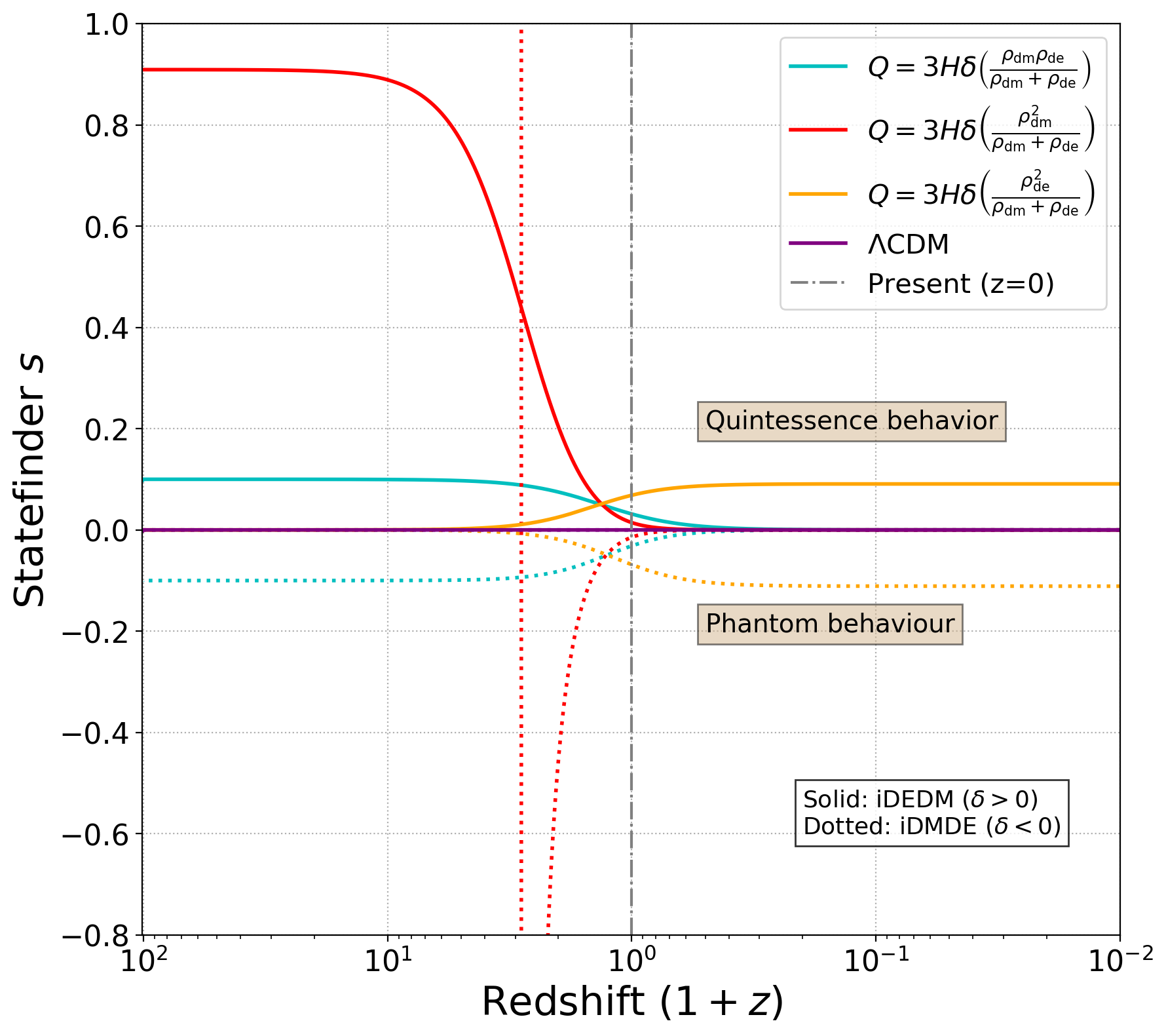}
        \label{fig:Statefinder_linear_sz_iDEDM}
    \end{subfigure} 
    \centering
    \begin{subfigure}[b]{0.49\linewidth}
        \centering
        \includegraphics[width=\linewidth]{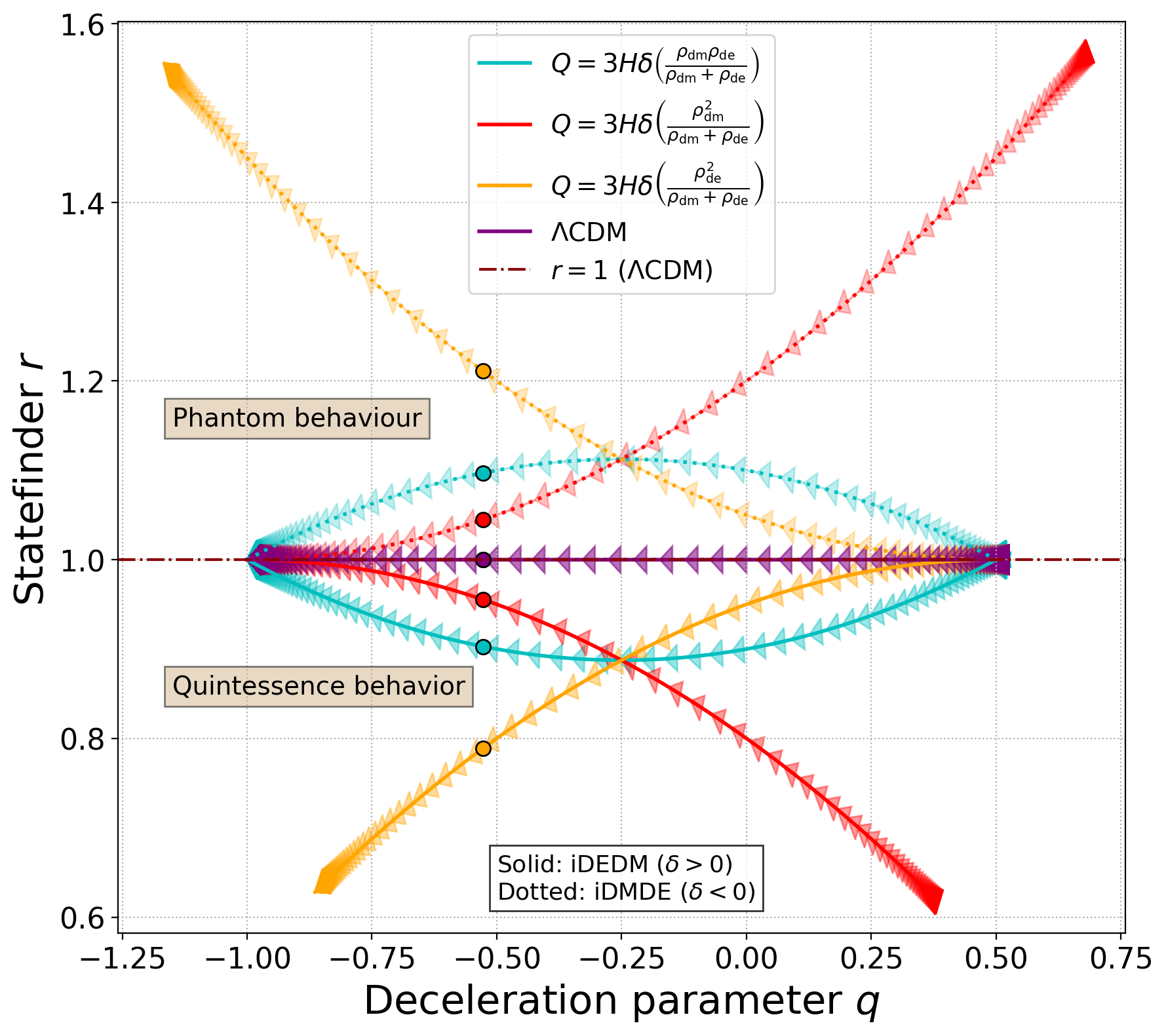}
        \label{fig:Statefinder_linear_qr_iDEDM}
    \end{subfigure}%
    \hspace{0pt} 
    \begin{subfigure}[b]{0.49\linewidth}
        \centering
        \includegraphics[width=\linewidth]{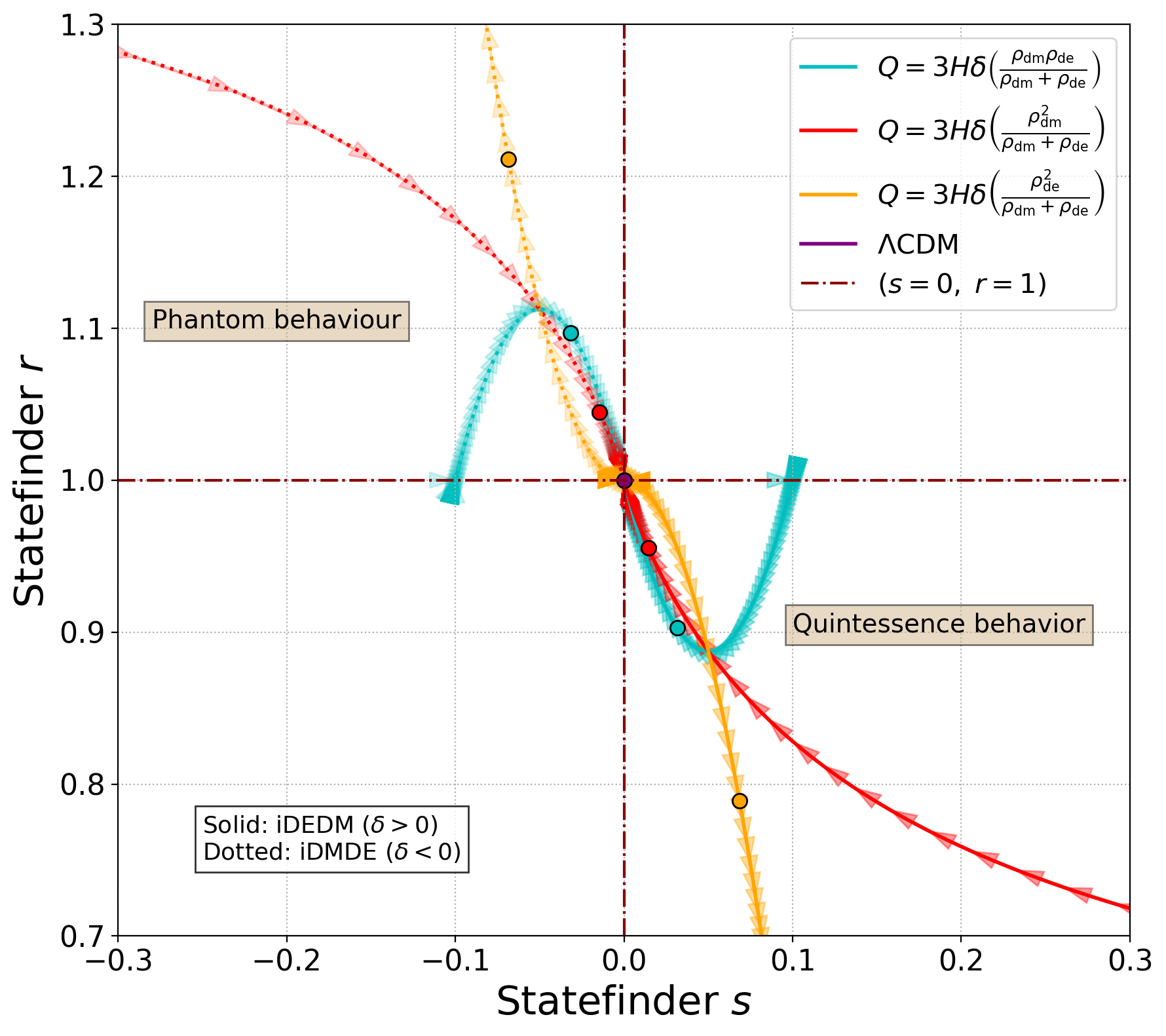}
        \label{fig:Statefinder_linear_sr_iDEDM}
    \end{subfigure}
    \caption{Statefinder parameters — $r$ vs.\ $z$ (top left), $s$ vs.\ $z$ (top right), $q$ vs.\ $r$ (bottom left) and $s$ vs.\ $r$ (bottom right) for non-linear IDE models in the iDEDM and iDMDE regimes with $\delta=\pm0.1$. Circles indicate present coordinates. Phantom DE behavior is defined by the parameter space $(q<-1 \; ; \; r_{\rm{sf}}>1 \; ; \; s_{\rm{sf}}<0 )$, while quintessence behavior is found in $(q>-1 \; ; \; r_{\rm{sf}}<1 \; ; \; s_{\rm{sf}}>0)$. Both interactions $Q=3\delta H  \left(\tfrac{\rho_{\text{dm}}\rho_{\text{de}}}{\rho_{\text{dm}}+\rho_{\text{de}}} \right)$ and $Q=3\delta H \left(\tfrac{\rho_{\text{dm}}^2}{\rho_{\text{dm}}+\rho_{\text{de}}} \right)$ have different past trajectories, but in the future converge to uncoupled $w$CDM behavior when the interaction strength becomes subdominant ($Q\rightarrow0$ as $\rho_{\text{dm}} \rightarrow 0$). Conversely, $Q=3\delta H 
\left(\tfrac{\rho_{\text{de}}^2}{\rho_{\text{dm}}+\rho_{\text{de}}} \right)$ shares the same past origin as $w$CDM, while diverging away as the effect of the interaction becomes dominant during DE domination. In the asymptotic future, this interaction exhibits phantom DE behavior in the iDEDM regime, while quintessence behavior is observed in the iDMDE regime.}
    \label{fig:qr+sr_NLID}
\end{figure}

To obtain the evolution of the statefinder parameters for these models, we substitute the expressions for $\rho_{\text{dm}}$ and $\rho_{\text{de}}$ (using the conversion $ \tfrac{8 \pi G }{3H^2} \rho_i= \Omega_i$) into the expression for $q$, and subsequently into the expressions for $r_{\text{sf}}$ and $s_{\text{sf}}$ given in \eqref{DSA.H}. The expressions for $r_{\text{sf}}$ and $s_{\text{sf}}$ are valid only during late-time expansion, approximately for $z<10^4$. From these, we obtain the evolution of the statefinder parameters in the $s_{\text{sf}}$--$r_{\text{sf}}$ and $r_{\text{sf}}$--$q$ planes, plotted in Figure~\ref{fig:qr+sr_NLID} using $w=-1$. The statefinder parameters also converge to the expressions found in Table~\ref{tab:Statefinder_past_nonlinear} and Table~\ref{tab:Statefinder_future_nonlinear} in the past and future, while expressions for the present are listed in Table~\ref{tab:Statefinder_present_nonlinear}. For ease of comparison, we also include the coordinates for the $\Lambda$CDM model, the non-interacting $w$CDM models (sometimes referred to as \emph{Quiessence}), and a model without DE, the Standard Cold Dark Matter (SCDM) model. For both $\Lambda$CDM and SCDM, the coordinates are fixed points throughout cosmic evolution, while for $w$CDM, $s_{\text{sf}}$ is fixed at $(1+w)$ and $r_{\rm{sf}}$ asymptotically decreases to $1+\tfrac{9}{2}w\left(1+w \right)$~\cite{Sahni_2003, Alam_2003}.

\begin{table}[H]
\centering
\renewcommand{\arraystretch}{1.2} 
\setlength{\tabcolsep}{10pt}     
\begin{tabular}{|c|c|c|c|}
\hline
\textbf{Model}   
 & $q_{\text{(past)}}$
 & $r_{\text{(sf, past)}}$
 & $s_{\text{(sf, past)}}$

\\ \hline \hline

$Q=3\delta H  
\left(\frac{\rho_{\text{dm}}\rho_{\text{de}}}{\rho_{\text{dm}}+\rho_{\text{de}}} \right)$
& $\frac{1}{2} $ & $1$ & $\delta+w+1 $ 
 
\\ \hline

$Q=3\delta H  \left(\frac{\rho_{\text{dm}}^2}{\rho_{\text{dm}}+\rho_{\text{de}}} \right)$
& $\frac{1}{2}(1+3 \frac{\delta w}{\delta-w})$ &$1+\frac{9}{2}\left(\frac{\delta w}{\delta-w} \right)\left(1+\frac{\delta w}{\delta-w} \right)$& $1+\frac{\delta w}{\delta-w}$ 
\\ \hline

$Q=3\delta H  \left(\frac{\rho_{\text{de}}^2}{\rho_{\text{dm}}+\rho_{\text{de}}} \right)$
&$\frac{1}{2}$ & $1$ & $1+w $ 

\\ \hline

$w$CDM
& $\frac{1}{2}$ & $1$ & $1 $ 
\\ \hline

$\Lambda$CDM
& $\frac{1}{2}$ & $1$ & $0 $ 
\\ \hline

SCDM
& $\frac{1}{2}$ & $1$ & $1 $ 
\\ \hline

\end{tabular}
\caption{Comparison of the deceleration parameter $q$ and the statefinder parameters $r_{\rm{sf}}$ and $s_{\rm{sf}}$ during past DM domination for non-linear IDE models.}
\label{tab:Statefinder_past_nonlinear}
\end{table}

\begin{table}[H]
\centering
\renewcommand{\arraystretch}{1.2} 
\setlength{\tabcolsep}{10pt}     
\begin{tabular}{|c|c|c|c|}
\hline
\textbf{Model} & $q_{\rm{(present)}}$  & $r_{\text{(sf, present)}}$
 & $s_{\text{(sf, present)}}$
\\ \hline \hline

$Q=3\delta H  
\left(\frac{\rho_{\text{dm}}\rho_{\text{de}}}{\rho_{\text{dm}}+\rho_{\text{de}}} \right)$
& $\frac{1}{2} \left[\Omega_{\rm{(dm,0)}}   +  \Omega_{\rm{(de,0)}}  \left(1 +3 w \right) \right]$ & $1+\frac{9}{2} \Omega_\text{(de,0)} w\left[ 1+w + \delta \left(\frac{r_0}{1+r_0}\right)\right]$ 
 & $1+w + \delta \left(\frac{r_0}{1+r_0}\right)$
\\ \hline

$Q=3\delta H  \left(\frac{\rho_{\text{dm}}^2}{\rho_{\text{dm}}+\rho_{\text{de}}} \right)$
&$\frac{1}{2} \left[\Omega_{\rm{(dm,0)}}   +  \Omega_{\rm{(de,0)}}  \left(1 +3 w \right) \right]$ & $1+\frac{9}{2} \Omega_\text{(de,0)} w\left[ 1+w + \delta \left(\frac{r_0}{1+\frac{1}{r_0}}\right)\right]$ 
 & $1+w + \delta \left(\frac{r_0}{1+\frac{1}{r_0}}\right)$
\\ \hline

$Q=3\delta H  \left(\frac{\rho_{\text{de}}^2}{\rho_{\text{dm}}+\rho_{\text{de}}} \right)$
&$\frac{1}{2} \left[\Omega_{\rm{(dm,0)}}   +  \Omega_{\rm{(de,0)}}  \left(1 +3 w \right) \right]$ & $1+\frac{9}{2} \Omega_\text{(de,0)} w\left[ 1+w + \delta \left(\frac{1}{1+r_0}\right)\right]$ 
 & $1+w + \delta \left(\frac{1}{1+r_0}\right)$

\\ \hline

$w$CDM
& $\frac{1}{2} \left[\Omega_{\rm{(dm,0)}}   +  \Omega_{\rm{(de,0)}}  \left(1 +3 w \right) \right]$ &$1+\frac{9}{2} \Omega_\text{(de,0)} w\left[ 1+w\right]$
 & $1+w $
\\ \hline

$\Lambda$CDM
&$\frac{1}{2} \left[\Omega_{\rm{(dm,0)}}   -2  \Omega_{\rm{(de,0)}}  \right]$ & $1$ & $0 $ 
\\ \hline

SCDM
&$\frac{1}{2} \Omega_{\rm{(dm,0)}}$ & $1$ & $1 $
\\ \hline

\end{tabular}
\caption{Comparison of the deceleration parameter $q$ and the statefinder parameters $r_{\rm{sf}}$ and $s_{\rm{sf}}$ at present $(a=1 ; z=0)$ for non-linear IDE models.}
\label{tab:Statefinder_present_nonlinear}
\end{table}

\begin{table}[H]
\centering
\renewcommand{\arraystretch}{1.2} 
\setlength{\tabcolsep}{10pt}     
\begin{tabular}{|c|c|c|c|}
\hline
\textbf{Model} & $q_{\rm{(future)}}$ 
 & $r_{\text{(sf, future)}}$
 & $s_{\text{(sf, future)}}$
\\ \hline \hline

$Q=3\delta H  
\left(\frac{\rho_{\text{dm}}\rho_{\text{de}}}{\rho_{\text{dm}}+\rho_{\text{de}}} \right)$
 & $\frac{1}{2}\left( 1+3w \right)$ & $1+\frac{9}{2}w\left(1+w \right)$
 & $1+w $
\\ \hline

$Q=3\delta H  \left(\frac{\rho_{\text{dm}}^2}{\rho_{\text{dm}}+\rho_{\text{de}}} \right)$
&$\frac{1}{2}\left( 1+3w \right)$ & $1+\frac{9}{2}w\left(1+w \right)$
 & $1+w $
\\ \hline

$Q=3\delta H  \left(\frac{\rho_{\text{de}}^2}{\rho_{\text{dm}}+\rho_{\text{de}}} \right)$
& $\frac{1}{2}\left(1-3\frac{w^2}{\delta-w} \right)$ & $1-\frac{9}{2}\left(\frac{w^2}{\delta-w} \right)\left(1-\frac{w^2}{\delta-w} \right)$
 & $1+\frac{w^2}{\delta-w} $

\\ \hline

$w$CDM
& $\frac{1}{2}\left( 1+3w \right)$ & $1+\frac{9}{2}w\left(1+w \right)$
 & $1+w $
\\ \hline

$\Lambda$CDM
& $-1$ & $1$
 & $0$
\\ \hline

SCDM
& $\frac{1}{2}$ & $1$
 & $1$
\\ \hline

\end{tabular}
\caption{Comparison of the deceleration parameter $q$ and the statefinder parameters $r_{\rm{sf}}$ and $s_{\rm{sf}}$ during future DE domination for non-linear IDE models.}
\label{tab:Statefinder_future_nonlinear}
\end{table}

From Figure~\ref{fig:qr+sr_NLID} and Tables~\ref{tab:Statefinder_future_nonlinear},~\ref{tab:Statefinder_present_nonlinear} it can be seen that both interactions 
$Q=3\delta H \left(\frac{\rho_{\text{dm}}\rho_{\text{de}}}{\rho_{\text{dm}}+\rho_{\text{de}}} \right)$ 
and 
$Q=3\delta H \left(\frac{\rho_{\text{dm}}^2}{\rho_{\text{dm}}+\rho_{\text{de}}} \right)$ 
(which also shows divergent behavior for $s_{\rm{sf}}$ in the iDMDE regime due to a zero crossing of the DE density, causing divergence of $w_{\rm{de}}^{\rm{eff}}$, as seen in Figure~\ref{fig:CP+omega_dmde_NLID2}) have different past trajectories, but in the future both models converge to uncoupled $w$CDM behavior when the interaction strength becomes subdominant ($Q\rightarrow0$ as $\rho_{\text{dm}} \rightarrow 0$).  
The evolution for the model 
$Q=3\delta H \left(\frac{\rho_{\text{dm}}\rho_{\text{de}}}{\rho_{\text{dm}}+\rho_{\text{de}}} \right)$ 
in the iDEDM regime lies to the right of $\Lambda$CDM and qualitatively shows quintessence behavior, while in the iDMDE regime it instead exhibits phantom or Chaplygin gas behavior to the left of $\Lambda$CDM, similar to what was seen in Figure~1 of~\cite{Alam_2003}. This close connection between this interaction and the decomposed new generalized Chaplygin gas (NGCG) model has been discussed in more detail in~\cite{Zhang_2006, Li_2014, Wang_2013}.  
Conversely, the interaction 
$Q=3\delta H \left(\frac{\rho_{\text{de}}^2}{\rho_{\text{dm}}+\rho_{\text{de}}} \right)$ 
shares the same past origin as $w$CDM, while diverging away as the effect of the interaction becomes dominant during DE domination. For the future in the iDEDM regime, this interaction exhibits phantom DE behavior $(q<-1 \; ; \;  r_{\rm{sf}}>1 \; ; \; s_{\rm{sf}}<0 )$, while in the iDMDE regime quintessence behavior $(q>-1\; ; \; r_{\rm{sf}}<1 \; ; \;  s_{\rm{sf}}>0)$ is seen.  
The qualitative behavior shown in Figure~\ref{fig:qr+sr_NLID} matches the relevant figures found in~\cite{carrasco2023discriminatinginteractingdarkenergy}, such that 
$Q=3\delta H \left(\frac{\rho_{\text{dm}}\rho_{\text{de}}}{\rho_{\text{dm}}+\rho_{\text{de}}} \right)$ matches FIG.~5(b)--(c), 
$Q=3\delta H \left(\frac{\rho_{\text{dm}}^2}{\rho_{\text{dm}}+\rho_{\text{de}}} \right)$ matches FIG.~6(a)--(b), 
and 
$Q=3\delta H \left(\frac{\rho_{\text{de}}^2}{\rho_{\text{dm}}+\rho_{\text{de}}} \right)$ matches FIG.~7(a)--(b).

For ease of comparison, we have ignored the contribution of baryons, which are separately conserved. Including baryons as a separate fluid will change the past expressions for $q$ and $r_{\rm{sf}}$ for the model $Q=3\delta H  \left(\frac{\rho_{\text{de}}^2}{\rho_{\text{dm}}+\rho_{\text{de}}} \right)$ given in Table~\ref{tab:Statefinder_present_nonlinear}, and subsequently the trajectories in the bottom two panels of Figure~\ref{fig:qr+sr_NLID}. In this case, these expressions will not converge to any stable value in the past, but will remain dynamic. The inclusion of baryons will also modify the expressions for $q_{\rm{present}}$ in Table~\ref{tab:Statefinder_present_nonlinear}, such that $\Omega_{\text{(dm,0)}}$ should be replaced with $\Omega_{\text{(m,0)}}=\Omega_{\text{(dm,0)}}+\Omega_{\text{(bm,0)}}$.

\section{Summary of main results and discussions} \label{summary}

In this study we investigated three IDE models with phenomenological interaction kernels that are non-linear, specifically 
$Q_{1}=3H \delta \left(\frac{\rho_{\rm{dm}}\rho_{\rm{de}}}{\rho_{\rm{dm}}+\rho_{\rm{de}}} \right)$, 
$Q_{2}=3H \delta \left(\frac{\rho_{\rm{dm}}^2}{\rho_{\rm{dm}}+\rho_{\rm{de}}} \right)$, 
and $Q_{3}=3H \delta \left(\frac{\rho_{\rm{de}}^2}{\rho_{\rm{dm}}+\rho_{\rm{de}}} \right)$. 
In Section~\ref{Sec.DSA} a dynamical system analysis of each model was performed, while in Section~\ref{Background_cosmology} the background cosmology was studied using newly derived analytical solutions. 
The expressions obtained in both sections using different methods converge and reduce to the relevant expressions for the $\Lambda$CDM model when $\delta=0$ and $w=-1$, thus validating the results found in both sections. 

In summary, we found that for the first interaction kernel $Q_{1}$, all energies are always positive since $Q=0$ if $\rho_{\text{dm}}=0$ or $\rho_{\text{de}}=0$, preventing the energy densities from crossing into negative values, as illustrated in Figures~\ref{fig:3D_QNLdmde_phase_portraits} and~\ref{fig:Omega_NLID1}. 
Conversely, if energy flows from DM to DE, for interaction $Q_{2}$ we find negative DE in the past, as shown in Figures~\ref{fig:2D_QNLdm_phase_portraits} and~\ref{fig:Omega_NLID2}, while for interaction $Q_{3}$ we find negative DM in the future, as seen in Figures~\ref{fig:2D_QNLde_phase_portraits} and~\ref{fig:Omega_NLID3}. 
We also showed that for $Q_{1}$ and $Q_{2}$, where the effect of the interaction diminishes in the future, a big rip singularity will always occur in the phantom regime ($w<-1$). 
For $Q_{3}$, the interaction remains dominant into the distant future; therefore, a big rip may be avoided if there is sufficient energy flow from DE to DM. 

The main results of this study are the new analytical solutions in Section \ref{Background_cosmology}, and of special importance are the new expressions for the evolution of $\rho_{\rm{dm}}$ and $\rho_{\rm{de}}$ in \eqref{NLID2_dm_de_BG} and \eqref{NLID3_dm_de_BG} for $Q_2$ and $Q_3$. Similarly, the expression for the reconstructed dynamical dark energy equation of state $\tilde{w}(z)$ in \eqref{wz_general} is also useful, as it holds for any IDE model with the same conservation equation given in \eqref{eq:conservation}. We would also like to highlight the new constraints summarized in Tables~\ref{tab:Com_real},~\ref{tab:Com_PEC},~\ref{tab:Com_AE_BR}, and~\ref{tab:Com_CP}. 
From these results, we can see that the behavior of $Q_{2}$ and $Q_{3}$ strongly correlates with those of the more widely studied linear interactions $Q=3H\delta\rho_{\rm{dm}}$ and $Q=3H\delta\rho_{\rm{de}}$, respectively. 
For comparison with these linear models, see our companion papers \cite{vanderWesthuizen:2025I, vanderWesthuizen:2025III} and earlier work~\cite{vanderWesthuizen:2023hcl}.

To get an intuition of the range of the allowed parameters in our tables, we have also included examples where we substitute $\Omega_{\rm{(dm,0)}}=0.266$, $\Omega_{\rm{(de,0)}}=0.685$ (which implies $r_0=0.388$), and $w=-1$.

\begin{table}[H]
\centering
\renewcommand{\arraystretch}{1.2} 
\setlength{\tabcolsep}{10pt}     
\begin{tabular}{|c|c|c|}
\hline
\textbf{Interaction $Q$} 
 & Conditions to avoid imaginary $\rho_{\text{dm/de}}$
 & Conditions to avoid undefined $\rho_{\text{dm/de}}$ \\ \hline \hline

$3H\delta \left( \frac{\rho_{\text{dm}} \rho_{\text{de}} }{\rho_{\text{dm}}+\rho_{\text{de}}} \right)$
 & $\rho_{\text{dm/de}}$ \text{ always real}
 & $\delta\ne -w$
\\ \hline

$3H\delta \left( \frac{\rho^2_{\text{dm}} }{\rho_{\text{dm}}+\rho_{\text{de}}} \right)$
 & $\rho_{\text{dm/de}}$ \text{ always real}
 & $w<0$ ; $w<\delta\le-\frac{w}{r_0}$
\\ \hline

$3H\delta \left( \frac{\rho^2_{\text{de}} }{\rho_{\text{dm}}+\rho_{\text{de}}} \right)$
 & $\rho_{\text{dm/de}}$ \text{ always real}
 & $w<0$ ; $w<\delta\le-w r_0$
\\ \hline

\end{tabular}
\caption{Constraints required to avoid imaginary or undefined energy densities for non-linear interaction kernels.}
\label{tab:Com_real}
\end{table}

\begin{table}[H]
\centering
\renewcommand{\arraystretch}{1.2} 
\setlength{\tabcolsep}{10pt}     
\begin{tabular}{|c|c|c|c|}
\hline
\textbf{Interaction $Q$} 
 & \text{$\rho_{\text{dm/de}}>0$ domain} 
 & \text{$\rho_{\text{dm/de}}>0$ conditions} 
 & \text{Example values} 
\\ \hline \hline

$3H\delta \left( \frac{\rho_{\text{dm}} \rho_{\text{de}} }{\rho_{\text{dm}}+\rho_{\text{de}}} \right)$
 & \text{DE $\leftrightarrow$ DM}
 & $\forall \delta$
 & $\forall \delta$
\\ \hline

$3H\delta \left( \frac{\rho^2_{\text{dm}} }{\rho_{\text{dm}}+\rho_{\text{de}}} \right)$
 & \text{DE $\rightarrow$ DM}
 & $0 \leq \delta \leq -\frac{w}{r_0}$
& $0 \leq \delta \leq 2.575$
\\ \hline

$3H\delta \left( \frac{\rho^2_{\text{de}} }{\rho_{\text{dm}}+\rho_{\text{de}}} \right)$
 & \text{DE $\rightarrow$ DM}
 & $0 \leq \delta \leq -w r_0$
& $0 \leq \delta \leq 0.388$
\\ \hline

\end{tabular}
\caption{Positive energy conditions for non-linear interaction kernels.}
\label{tab:Com_PEC}
\end{table}

\begin{table}[H]
\centering
\renewcommand{\arraystretch}{1.2} 
\setlength{\tabcolsep}{10pt}     
\begin{tabular}{|c|c|c|}
\hline
\textbf{Model} 
 & \text{Accelerated expansion $\left[w=-1 \right]$} 
 & \text{No big rip if $w<-1$ $\left[w=-1.1 \right]$ } 
\\ \hline \hline

$Q_1=3H\delta \left( \frac{\rho_{\text{dm}} \rho_{\text{de}} }{\rho_{\text{dm}}+\rho_{\text{de}}} \right)$
 &  $\forall \delta$ if $w\le-\frac{1}{3}$ 
 & Big rip Inevitable
\\ \hline

$Q_2=3H\delta \left( \frac{\rho^2_{\text{dm}} }{\rho_{\text{dm}}+\rho_{\text{de}}} \right)$
 & $\forall \delta$ if $w\le-\frac{1}{3}$ 
 & Big rip Inevitable
\\ \hline

$Q_3=3H\delta \left( \frac{\rho^2_{\text{de}} }{\rho_{\text{dm}}+\rho_{\text{de}}} \right)$
 & $\delta \leq w(3w+1) $ ; $\left[\delta \leq 2 \right]$
 & $\delta \geq  w(w+1)  $ ; $\left[\delta \geq 0.11 \right]$
\\ \hline

$w$CDM 
 & $w\le-\frac{1}{3}$ 
 & Big rip Inevitable
\\ \hline

\end{tabular}
\caption{Conditions for accelerated expansion and avoidance of a big rip for non-linear interaction kernels.}
\label{tab:Com_AE_BR}
\end{table}

\begin{table}[H]
\centering
\renewcommand{\arraystretch}{1.2} 
\setlength{\tabcolsep}{8pt}      
\begin{tabular}{|c|c|c|}
\hline 
\textbf{Model with $\delta>0$ (iDEDM)} & \text{Coincidence problem (Past)} & \text{Coindence problem (Future)}  \\ \hline\hline

$Q_1=3H\delta \left( \frac{\rho_{\text{dm}} \rho_{\text{de}}}{\rho_{\text{dm}}+\rho_{\text{de}}}\right)$  
  & \text{Alleviated } [$\zeta=-3(w+\delta)$]  
  & \text{Alleviated } [$\zeta=-3(w+\delta)$]   \\ \hline

$Q_2=3H\delta \left( \frac{\rho^2_{\text{dm}} }{\rho_{\text{dm}}+\rho_{\text{de}}} \right)$ 
  & \text{Solved } [$\zeta=0$]  
  & \text{No change } [$\zeta=-3w$]  \\ \hline

$Q_3=3H\delta \left(  \frac{\rho^2_{\text{de}} }{\rho_{\text{dm}}+\rho_{\text{de}}}\right)$  
  & \text{No change } [$\zeta=-3w$]  
  & \text{Solved } [$\zeta=0$]  \\ \hline

$w\text{CDM}$  
  & $\zeta=-3w$  
  & $\zeta=-3w$ \\ \hline

$\Lambda\text{CDM}$  
  & $\zeta=-3$  
  & $\zeta=-3$ \\ \hline

\end{tabular}
\caption{Potential to address the coincidence problem for non-linear interaction kernels.}
\label{tab:Com_CP}
\end{table}

Table~\ref{tab:Com_CP} provides a summary of how each interaction in the iDEDM regime addresses the coincidence problem in the past and the future, with the deviation from
$\zeta=0$ indicating the magnitude of the problem and $\zeta=-3w$ serving as the baseline for non-interacting models. Interestingly, the results differ slightly from the corresponding results for linear IDE models found in Table XII of our companion paper \cite{vanderWesthuizen:2025I}. For interactions proportional to both dark components, we find that the non-linear model $Q_1 \propto \rho_{\rm{dm}}\rho_{\rm{de}}$ only alleviates the coincidence problem, while the linear model $Q \propto \rho_{\rm{dm}}+\rho_{\rm{de}}$ solves it. Similarly, when the interaction is proportional to only one dark component, both $Q\propto \rho_{\rm{dm}}$ and $Q\propto \rho_{\rm{dm}}^2$ solve the coincidence problem in the past during DM domination, but while the linear model alleviates the problem in the past, the non-linear model has no effect. The same holds for $Q\propto \rho_{\rm{de}}$ and $Q\propto \rho_{\rm{de}}^2$, but with the roles of past and future reversed. This highlights a qualitative difference between the linear and non-linear interactions considered here: for any IDE model, the effect of the interaction is strongest during the domination of the fluid present in the kernel $Q$, but for linear models there remains a significant, albeit smaller, effect when the fluid is subdominant. By contrast, in the non-linear models studied here, the effect almost completely vanishes once the relevant fluid becomes subdominant, effectively recovering the non-interacting case in such circumstances.

The conclusion from this study is similar to what we found when investigating linear interaction kernels. We find that allowing for a small interaction with energy flowing from DE to DM appears to favor positive energy densities, helps alleviate the coincidence problem, and may avoid a big rip. Nevertheless, the ultimate test will come from observational data, which will determine whether these theoretical features correspond to physical reality. Additionally, for both $Q_1$ and $Q_3$ in the iDEDM regime, we observe the presence of an effective DE phantom crossing $w^{\rm{eff}}_{\rm{de}}=-1$ from the phantom to the quintessence regime (shown in Figure~\ref{fig:CP+omega_dmde_NLID1} and~\ref{fig:CP+omega_dmde_NLID3}), consistent with the recent results from DESI~\cite{DESI:2025fii}. 

Lastly, the only case that allows bidirectional energy transfer while maintaining positive energies is $Q_1$, which offers a wider parameter space with fewer complications. We therefore argue that $Q_1$ warrants the most attention among the models studied here for further observational constraints, as recently done in~\cite{vanderwesthuizen2025compartmentalizationdarksectoruniverse}.

\textbf{Data Availability Statements:} Data sharing is not applicable to this article as no datasets were generated or analyzed during the current study. More detailed calculations for any section can be provided by the authors on reasonable request.

\begin{acknowledgments}
EDV is supported by a Royal Society Dorothy Hodgkin Research Fellowship. This article is based upon work from the COST Action CA21136 - ``Addressing observational tensions in cosmology with systematics and fundamental physics (CosmoVerse)'', supported by COST - ``European Cooperation in Science and Technology''.
\end{acknowledgments}

\appendix 
\section{Finding the expressions of $\rho_{\rm{dm}}$ and $\rho_{\rm{de}}$ for non-linear IDE model 2: $Q_2=3\delta H  \left(\frac{\rho_{\text{dm}}^2}{\rho_{\text{dm}}+\rho_{\text{de}}} \right)$} \label{Appendix_A}
Our derivation starts with the expression for the evolution of the ratio $r=\frac{\rho_{\text{dm}}}{\rho_{\text{de}}}$ and the total dark sector density $\rho_{\text{t}}$, which are derived and found in equations (21) and (22)  from~\cite{Arevalo:2011hh}.
\begin{equation}
\begin{split}
 r &= r_0\frac{w }{\left( w+ \delta r_0\right) a^{-3w} -\delta r_0}.  
\label{NLID_r'_Q2_4} 
\end{split}
\end{equation}
\begin{equation}
\begin{split}
\rho_{\text{t}} &=\rho_{\text{(t,0)}} a^{-3\left(1- \frac{w\delta }{w -\delta} \right)}   \left[\frac{(w+\delta r_0)a^{-3w } + r_0(w -\delta) }{ w (1+r_0 )}    \right]^{\frac{w }{w -\delta}}.
\label{NLID2_rhoT_der_13} 
\end{split}
\end{equation}
In order to obtain expressions for $\rho_{\text{dm}}$ and $\rho_{\text{de}}$, we need to consider the relationship of $\rho_{\text{dm}}$ and $\rho_{\text{de}}$, with $r$ and $\rho_{\text{t}}$, which are given by the following expressions:
The DM density $\rho_{\text{dm}}$, as well as its present value $\rho_{\text{(dm,0)}}$ are related to $r$ and $\rho_{\text{t}}$  as:
\begin{equation}
\begin{split} 
\rho_{\text{dm}} = \rho_{\text{t}}\left(\frac{r}{1+r} \right)  \quad ; \quad \rho_{\text{(dm,0)}} = \rho_{\text{(t,0)}}\left(\frac{r_0}{1+r_0} \right) .
\label{NLID1_dm_der_1}
\end{split} 
\end{equation}
Similarly, the DE density $\rho_{\text{de}}$, as well as its present value $\rho_{\text{(de,0)}}$ are given by expressions:
\begin{equation}
\begin{split} 
\rho_{\text{de}} = \rho_{\text{t}}\left(\frac{1}{1+r} \right)  \quad ; \quad \rho_{\text{(de,0)}} = \rho_{\text{(t,0)}}\left(\frac{1}{1+r_0} \right) .
\label{NLID1_de_der_1}
\end{split} 
\end{equation}
Thus, we will need an expression for $1/(1+r)$, which we obtain using \eqref{NLID_r'_Q2_4}:
\begin{equation}
\begin{split}
\frac{1}{1+r} &= \frac{\left( w+ \delta r_0\right)a^{-3w} -\delta r_0} {\left( w+ \delta r_0\right)a^{-3w} + r_0(w - \delta) },  
\label{NLID2_rhoT_der_2} 
\end{split}
\end{equation}
The DM density $\rho_{\text{dm}}$ may be obtained for this model by first substituting the derived expressions for $r$ from \eqref{NLID_r'_Q2_4} and $1+r$ from \eqref{NLID2_rhoT_der_2} into \eqref{NLID1_dm_der_1}, which gives:
\begin{equation}
\begin{split} 
\rho_{\text{dm}} &= \rho_{\text{t}}\left(\frac{r}{1+r} \right)  =  \rho_{\text{t}}\left(\frac{\frac{r_0 w }{\left( w+ \delta r_0\right) a^{-3w} -\delta r_0}}{\frac{\left( w+ \delta r_0\right)a^{-3w} +  r_0(w - \delta) }{\left( w+ \delta r_0\right)a^{-3w} -\delta r_0}} \right)
=  \rho_{\text{t}}\left( \frac{r_0 w }{\left( w+ \delta r_0\right)a^{-3w} +  r_0(w - \delta)} \right).\\
\label{NLID2_dm_der_1}
\end{split} 
\end{equation}
We can now substitute $\rho_t$ from \eqref{NLID2_rhoT_der_13} into  \eqref{NLID2_dm_der_1} to obtain:
\begin{equation}
\begin{split} 
\rho_{\text{dm}} &=  \left[ \rho_{\text{(t,0)}} a^{-3\left(1- \frac{w\delta }{w -\delta} \right)}   \left[\frac{(w+\delta r_0)a^{-3w } + r_0(w -\delta) }{ w (1+r_0 )}    \right]^{\frac{w }{w -\delta}} \right]\left( \frac{r_0 w }{\left( w+ \delta r_0\right)a^{-3w} +  r_0(w - \delta)} \right)\\
&=   \rho_{\text{(t,0)}} \frac{1}{1+r_0} a^{-3\left(1- \frac{w\delta }{w -\delta} \right)}   \left[\frac{(w+\delta r_0)a^{-3w } + r_0(w -\delta) }{ w (1+r_0 )}    \right]^{\frac{w }{w -\delta}-1}\\
\label{NLID2_dm_der_2}
\end{split} 
\end{equation}
Substituting the initial DE density from \eqref{NLID1_de_der_1} into \eqref{NLID2_dm_der_2} gives the final expression for the DE density of this non-linear interaction model:\begin{equation}
\begin{split} 
\boxed{ \rho_{\text{dm}} =  \rho_{\text{(dm,0)}}  a^{-3\left(1- \frac{w\delta }{w -\delta} \right)} \left[\frac{(w+\delta r_0)a^{-3w } + r_0(w -\delta) }{ w (1+r_0 )}    \right]^{\frac{\delta }{w -\delta}}}     . 
\label{NLID2_dm_der_3}
\end{split} 
\end{equation}
It should be noted that for the non-interacting case $\delta=0$, \eqref{NLID2_dm_der_3} reduces back to the $\Lambda$CDM case where $\rho_{\text{dm}} = \rho_{\text{(dm,0)}}  a^{-3}$. 
The DM density can also be obtained by substituting the derived expression for $1/(1+r)$ from \eqref{NLID2_rhoT_der_2} and $\rho_{\text{t}}$ from \eqref{NLID2_rhoT_der_13}  into \eqref{NLID1_de_der_1} above:
\begin{equation}
\begin{split} 
\rho_{\text{de}} =\rho_{\text{t}}\left(\frac{1}{1+r} \right) &=\rho_{\text{(t,0)}} a^{-3\left(1- \frac{w\delta }{w -\delta} \right)}   \left[\frac{(w+\delta r_0)a^{-3w } + r_0(w -\delta) }{ w (1+r_0 )}    \right]^{\frac{w }{w -\delta}}\left(\frac{\left( w+ \delta r_0\right)a^{-3w} -\delta r_0} {\left( w+ \delta r_0\right)a^{-3w} + r_0(w - \delta) } \right)  \\
 &=\rho_{\text{(t,0)}} \frac{1}{1+r_0}  a^{-3\left(1- \frac{w\delta }{w -\delta} \right)}   \left[\frac{(w+\delta r_0)a^{-3w } + r_0(w -\delta) }{ w (1+r_0 )}    \right]^{\frac{w }{w -\delta}-1} \left(\frac{\left( w+ \delta r_0\right)a^{-3w} -\delta r_0} {w } \right).
\label{NLID2_de_der_1}
\end{split} 
\end{equation}
Substituting the initial DE density from \eqref{NLID1_de_der_1} into \eqref{NLID2_de_der_1} gives the final expression for the DM density of this non-linear interaction model: 
\begin{equation}
\begin{split} 
\boxed{ \rho_{\text{de}} =  \rho_{\text{(de,0)}}   a^{-3\left(1- \frac{w\delta }{w -\delta}  \right)} \left(\frac{\left( w+ \delta r_0\right)a^{-3w} -\delta r_0} {w } \right)  \left[\frac{(w+\delta r_0)a^{-3w } + r_0(w -\delta) }{ w (1+r_0 )}    \right]^{\frac{\delta }{w -\delta}} } . 
\label{NLID2_de_der_2}
\end{split} 
\end{equation}
It should be noted that for the non-interacting case $\delta=0$, \eqref{NLID2_dm_der_2} reduces back to the $\Lambda$CDM case where $\rho_{\text{de}} = \rho_{\text{(de,0)}} $

\section{Finding the expressions of $\rho_{\rm{dm}}$ and $\rho_{\rm{de}}$ for non-linear IDE model 3: $Q_3=3\delta H  \left(\frac{\rho_{\text{de}}^2}{\rho_{\text{dm}}+\rho_{\text{de}}} \right)$} \label{Appendix_B}
Similarly, this derivation starts with the expression for the evolution of the ratio $r=\frac{\rho_{\text{dm}}}{\rho_{\text{de}}}$ and the total dark sector density $\rho_{\text{t}}$, which are derived and found in equations (25) and (26)  from~\cite{Arevalo:2011hh}. We have followed a similar derivation, but without taking $w=-|w|$, leading to the equivalent expressions below:
\begin{equation}
\begin{split}
r &= \left(r_0 +\frac{\delta}{w} \right) a^{3 w} -\frac{\delta}{w}  .  
\label{NLID_r'_Q3_4} 
\end{split}
\end{equation}
\begin{equation}
\begin{split}
\rho_{\text{t}}&=\rho_{\text{(t,0)}}   a^{-3\left(1+ \frac{w^2 }{w -\delta} \right)}   \left[\frac{(w r_0+\delta )a^{3w } + (w -\delta) }{w (1+r_0)}\right]^{\frac{w}{w -\delta}  }. 
\label{NLID3_rhoT_der_13} 
\end{split}
\end{equation}
The DM density $\rho_{\text{dm}}$ may be obtained for this model by first substituting the derived expressions for $r$ from \eqref{NLID_r'_Q3_4} into \eqref{NLID1_dm_der_1}, which gives:
\begin{equation}
\begin{split} 
\rho_{\text{dm}} = \rho_{\text{t}}\left(\frac{r}{1+r} \right)  
=  \rho_{\text{t}}\left(\frac{\left(r_0 +\frac{\delta}{w} \right) a^{3 w} -\frac{\delta}{w}}{\left(r_0 +\frac{\delta}{w} \right) a^{3 w} +\left(1 -\frac{\delta}{w}\right)}  \right)=  \rho_{\text{t}}\left(\frac{(w r_0+\delta )a^{3w }   -\delta }{(w r_0+\delta )a^{3w } + (w -\delta) }  \right).
\label{NLID3_dm_der_1}
\end{split} 
\end{equation}
We can now substitute $\rho_t$ from \eqref{NLID3_rhoT_der_13} into  \eqref{NLID3_dm_der_1} to obtain:
\begin{equation}
\begin{split} 
\rho_{\text{dm}} 
&=  \rho_{\text{(t,0)}}   a^{-3\left(1+ \frac{w^2 }{w -\delta} \right)}   \left[\frac{(w r_0+\delta )a^{3w } + (w -\delta) }{w (1+r_0)}\right]^{\frac{w}{w -\delta}  }\left(\frac{(w r_0+\delta )a^{3w }   -\delta }{(w r_0+\delta )a^{3w } + (w -\delta) }  \right)\\
&=  \rho_{\text{(t,0)}} \frac{r_0}{1+r_0}   a^{-3\left(1+ \frac{w^2 }{w -\delta} \right)}   \left[\frac{(w r_0+\delta )a^{3w } + (w -\delta) }{w (1+r_0)}\right]^{\frac{w}{w -\delta} -1 }\left(\frac{(w r_0+\delta )a^{3w }   -\delta }{w r_0 }  \right).\\
\label{NLID3_dm_der_2}
\end{split} 
\end{equation}
Substituting the initial DE density from \eqref{NLID1_de_der_1} into \eqref{NLID2_dm_der_2} gives the final expression for the DE density of this non-linear interaction model:
\begin{equation}
\begin{split} 
\boxed{\rho_{\text{dm}} 
=  \rho_{\text{(dm,0)}}a^{-3\left(1+ \frac{w^2 }{w -\delta} \right)}  \left(\frac{(w r_0+\delta )a^{3w }   -\delta }{w r_0 }  \right) \left[\frac{(w r_0+\delta )a^{3w } + (w -\delta) }{w (1+r_0)}\right]^{\frac{\delta}{w -\delta} } }.
\label{NLID3_dm_der_3}
\end{split} 
\end{equation}
It should be noted that for the non-interacting case $\delta=0$, \eqref{NLID3_dm_der_3} reduces back to the $\Lambda$CDM case where $\rho_{\text{dm}} = \rho_{\text{(dm,0)}}  a^{-3}$. The DE density can also be obtained by substituting $r$ from \eqref{NLID_r'_Q2_4} and $\rho_{\text{t}}$ from \eqref{NLID3_rhoT_der_13}  into \eqref{NLID3_de_der_1} above:
\begin{equation}
\begin{split} 
\rho_{\text{de}} =\rho_{\text{t}}\left(\frac{1}{1+r} \right)  
 &=\rho_{\text{(t,0)}}   a^{-3\left(1+ \frac{w^2 }{w -\delta} \right)}   \left[\frac{(w r_0+\delta )a^{3w } + (w -\delta) }{w (1+r_0)}\right]^{\frac{w}{w -\delta}  } \left( \frac{1}{\left(r_0 +\frac{\delta}{w} \right) a^{3 w} +\left(1 -\frac{\delta}{w}\right)} \right) \\
 &=\rho_{\text{(t,0)}} \frac{1}{1+r_0} a^{-3\left(1+ \frac{w^2 }{w -\delta} \right)}   \left[\frac{(w r_0+\delta )a^{3w } + (w -\delta) }{ w (1+r_0 )}    \right]^{\frac{w }{w -\delta}-1}. 
\label{NLID3_de_der_1}
\end{split} 
\end{equation}
Substituting the initial DM density from \eqref{NLID1_de_der_1} into \eqref{NLID3_de_der_1} gives the final expression for the DM density of this non-linear interaction model: 
\begin{equation}
\begin{split} 
\boxed{ \rho_{\text{de}} =  \rho_{\text{(de,0)}}   a^{-3\left(1+ \frac{w^2 }{w -\delta} \right)}   \left[\frac{(w r_0+\delta )a^{3w } + (w -\delta) }{ w (1+r_0 )}    \right]^{\frac{\delta }{w -\delta}}}. 
\label{NLID3_de_der_2}
\end{split} 
\end{equation}
It should be noted that for the non-interacting case $\delta=0$, \eqref{NLID2_dm_der_2} reduces back to the $\Lambda$CDM case where $\rho_{\text{de}} = \rho_{\text{(de,0)}} $

\section{Derivation of an expression for the reconstructed DE equation of state $\tilde{w}(z)$ for any IDE model} \label{Appendix_C}
In order to reconstruct  $\tilde{w}(z)$, we start by noting that the normalized Hubble parameter $h(z)$ for a dynamical dark energy model in a flat universe without any interactions in the dark sector is given by: \\
\begin{gather} \label{wz_1}
\begin{split}
h^2(z)&= \Omega_{\text{(r,0)}}(1+z)^{4}+ \Omega_{\text{(bm,0)}}(1+z)^{3}+ \Omega_{\text{(dm,0)}}(1+z)^{3}+  \underbrace{ \Omega_{\text{(de,0)}} \text{exp}\left[ 3 \int_0^z dz' \frac{1+\tilde{w}(z')}{1 + z'} \right]}_{X_{\rm{de}}(z)}, \\
X_{\rm{de}}(z)&= h^2(z)- \Omega_{\text{(r,0)}}(1+z)^{4}- \Omega_{\text{(bm,0)}}(1+z)^{3}- \Omega_{\text{(dm,0)}}(1+z)^{3},
\end{split}
\end{gather}
where $X_{\rm{de}}(z)$ is the apparent DE density. We can differentiate and invert $X_{\rm{de}}(z)$ to obtain an expression for $\tilde{w}(z')$: \\
\begin{gather} \label{wz_2}
\begin{split}
\tilde{w}(z)&=   \frac{(1+z) }{3 X_{\rm{de}}(z) } \frac{d X_{\rm{de}}}{dz}  -1.
\end{split}
\end{gather}
Thus, we can see that we only require expressions for $ X_{\rm{de}}(z)$ and $\frac{d X_{\rm{de}}}{dz}$ to find $\tilde{w}(z)$. It will be useful to convert the derivative with redshift to that of time, using the transformation:
\begin{gather} \label{wz_3}
\begin{split}
\frac{d}{dz}=-\frac{1}{H (1+z) }\frac{d}{dt}
\end{split}\end{gather}
Using \eqref{wz_3}, we can write \eqref{wz_2} as:
\begin{gather} \label{wz_4}
\begin{split}
\tilde{w}(z)&=  - \frac{1}{3H X_{\rm{de}}(z) } \frac{d X_{\rm{de}}}{dt}  -1.
\end{split}
\end{gather}
Using the Hubble function \eqref{DSA.H} for a IDE model with radiation and baryons, we can find a expression for $X_{\rm{de}}$ from \eqref{wz_1}:
\begin{gather} \label{wz_5}
\begin{split}
h^2&= \Omega_{\text{(r,0)}}(1+z)^{4}+ \Omega_{\text{(bm,0)}}(1+z)^{3}+\frac{\rho_{\text{dm}}+\rho_{\text{de}}}{\rho_{(c,0)}},  \\
\rightarrow X_{\rm{de}}(z)&=\frac{\rho_{\text{dm}}+\rho_{\text{de}}-\rho_{\text{(dm,0)}}(1+z)^{3}}{\rho_{(c,0)}},
\end{split}
\end{gather}
where the present critical density is given by $\rho_{(c,0)}=\frac{3H_0^2}{8 \pi G}$. Taking the time derivative of $X_{\rm{de}}(z)$ \eqref{wz_5}, while noting that $\frac{d}{dt}(1+z)^3=-3H(1+z)^3$ gives:
\begin{gather} \label{wz_6}
\begin{split}
\frac{d X_{\rm{de}}(z)}{dt} &=\frac{\dot\rho_{\text{dm}}+\dot\rho_{\text{de}}+3H\rho_{\text{(dm,0)}}(1+z)^{3}}{\rho_{(c,0)}}.
\end{split}
\end{gather}
From the conservation equations \eqref{eq:conservation}, we can get the following relations for $\dot\rho_{\text{dm}}$ and $\dot\rho_{\text{de}}$:
\begin{gather} \label{wz_7}
\begin{split}
\dot{\rho}_{\text{dm}}  = Q- 3H \rho_{\text{dm}} \quad &; \quad  \dot{\rho}_{\text{de}}  = -Q-3H [1 + w(z)] \rho_{\text{de}}.\\
\end{split}
\end{gather}
Substituting \eqref{wz_7} back into \eqref{wz_6}, gives:
\begin{gather} \label{wz_8}
\begin{split}
\frac{d X_{\rm{de}}(z)}{dt} &=-\frac{3H\left[\rho_{\text{dm}}+[1+w(z)]\rho_{\text{de}}-\rho_{\text{(dm,0)}}(1+z)^{3} \right]}{\rho_{(c,0)}}.
\end{split}
\end{gather}
Substituting $ X_{\rm{de}}(z)$ from \eqref{wz_5} and $\frac{d X_{\rm{de}}(z)}{dt}$ from \eqref{wz_8} back into $\tilde{w}(z)$ from \eqref{wz_4}, gives:
\begin{gather} \label{wz_9}
\begin{split}
\tilde{w}(z)&=  - \frac{1}{3H \left[ \frac{\rho_{\text{dm}}+\rho_{\text{de}}-\rho_{\text{(dm,0)}}(1+z)^{3}}{\rho_{(c,0)}} \right] } \left[-\frac{3H\left[\rho_{\text{dm}}+[1+w(z)]\rho_{\text{de}}-\rho_{\text{(dm,0)}}(1+z)^{3} \right]}{\rho_{(c,0)}} \right]  -1 \\
\tilde{w}(z)&=  \frac{\rho_{\text{dm}}+[1+w(z)]\rho_{\text{de}}-\rho_{\text{(dm,0)}}(1+z)^{3} }{\rho_{\text{dm}}+\rho_{\text{de}}-\rho_{\text{(dm,0)}}(1+z)^{3}}   -1 \\
\tilde{w}(z)&=  \frac{w(z)\rho_{\text{de}}}{\rho_{\text{dm}}+\rho_{\text{de}}-\rho_{\text{(dm,0)}}(1+z)^{3}}  .
\end{split}
\end{gather}
This may also be written in terms of the ratio of DM to DE, such that we get the final expression:
\begin{gather} \label{wz_10}
\begin{split}
\boxed{\tilde{w}(z) =\frac{w(z) }{1+r-\frac{\rho_{\rm{(dm,0)}}(1+z)^3}{\rho_{\rm{de}}}}}. \end{split}
\end{gather}
This holds for any IDE model, even with a inherent dynamical equation of state $w(z)$, though we have assumed in this paper a constant $w(z)=w$. We may also note that in the case where $\delta=0$, we get the uncoupled $\tilde{w}(z)=w(z)$. 

We may use expression \eqref{wz_9} to understand why divergent behavior of $\tilde{w}(z)$ occurs for the iDEDM regime, but not the iDMDE regime. The divergence occurs when the denominator in \eqref{wz_9} becomes zero, which occurs when:
\begin{gather} \label{wz_11}
\begin{split}
\rho_{\text{dm}}+\rho_{\text{de}}=\rho_{\text{(dm,0)}}(1+z)^{3}.
\end{split}
\end{gather}
Thus, we can see that divergence will occur in the past if the sum of DM and DE is equal to the amount of DM in an uncoupled scenario at any point. Lets look at this possibility for the two directions of energy transfer.

\begin{itemize}
    \item For the iDMDE regime, we have more DM and less DE in the past, in comparison to the uncoupled case. This implies that at all times $\rho_{\rm{dm}}>\rho_{\text{(dm,0)}}(1+z)^{3}$ and therefore $\rho_{\text{dm}}+\rho_{\text{de}}\neq\rho_{\text{(dm,0)}}(1+z)^{3}$ at any point in the past. \textit{Thus, no divergence of $\tilde{w}(z)$ is expected in the iDMDE regime}.
    \item Conversely, for the iDEDM regime, we have less DM and more DE in the past, in comparison to the uncoupled case. This implies that at all times $\rho_{\rm{dm}}<\rho_{\text{(dm,0)}}(1+z)^{3}$ and since DE is small in the past, we can expect at some point in past that $\rho_{\text{dm}}+\rho_{\text{de}}=\rho_{\text{(dm,0)}}(1+z)^{3}$. \textit{Thus, divergence of $\tilde{w}(z)$ is expected in the iDEDM regime}. 
\end{itemize}
Lastly, it is important to remember that this \textit{divergence is merely an artefact of this parametrization of DE, and does not indicate a real pathology of the underlying dynamics}, as both DM and DE have been shown to be well behaved with a small coupling in the iDEDM regime.

\bibliographystyle{apsrev4-2}
\bibliography{References2,biblio}  

\end{document}